\documentclass[twocolumn,tighten]{aastex62}

\usepackage[english]{babel}
\usepackage{boldline}
\usepackage[intlimits]{amsmath}
\usepackage{bm}
\usepackage{gensymb}
\usepackage[caption=false]{subfig}
\usepackage{txfonts}
\usepackage{dsfont}
\usepackage{verbatim}
\usepackage{etoolbox}
\usepackage{float}
\usepackage{varioref}				
\usepackage{hyperref}				

\usepackage{array}
\renewcommand{\arraystretch}{1.2}

\labelformat{chapter}{Chap.~#1}
\labelformat{section}{Sec.~#1}
\labelformat{appendix}{App.~#1}
\labelformat{subsection}{Sec.~#1}
\labelformat{subsubsection}{Sec.~#1}
\labelformat{figure}{Fig.~#1}
\labelformat{subfigure}{Fig.~\thefigure #1}
\labelformat{table}{Tab.~#1}
\labelformat{equation}{Eq.~(#1)}

\usepackage[procnames]{listings}
\usepackage{textcomp}
\usepackage{numprint}

\definecolor{text}{HTML}{000000}
\definecolor{keyword}{HTML}{0000FF}
\definecolor{builtin}{HTML}{900090}
\definecolor{definition}{HTML}{000000} 
\definecolor{comment}{HTML}{ADADAD} 
\definecolor{string}{HTML}{00AA00}
\definecolor{number}{HTML}{800000}
\definecolor{instance}{HTML}{924900} 
\definecolor{linenumber}{HTML}{ADADAD} 

\definecolor{green}{rgb}{0,1,00}
\definecolor{lightgreen}{rgb}{0,0.5,0}
\definecolor{red}{rgb}{1,0,0}
\definecolor{lightred}{rgb}{0.5,0,0}
\newcounter{inCounter}[section]
\newcommand{\iin}{\refstepcounter{inCounter}{\color{lightgreen}In [{\color{green}\arabic{inCounter}}]:}}
\newcommand{\out}{{\color{lightred}Out[{\color{red}\arabic{inCounter}}]:}}

\makeatletter
\newif\iffirstchar\firstchartrue
\newif\ifstartedbyadigit

\newcommand\ProcessLetter
{%
	\ifnum\lst@mode=\lst@Pmode%
		\iffirstchar%
				\global\startedbyadigitfalse%
			\fi
			\global\firstcharfalse%
		\fi
}
\newcommand\ProcessDigit
{%
	\ifnum\lst@mode=\lst@Pmode%
		\iffirstchar%
				\global\startedbyadigittrue%
			\fi
			\global\firstcharfalse%
  \fi
}
\lst@AddToHook{Output}%
{%
	\ifstartedbyadigit%
		\def\lst@thestyle{\color{number}}%
	\fi
	\global\firstchartrue%
	\global\startedbyadigitfalse%
}
\newtoks\python@toks
\python@toks={language=Python,tabsize=4,frame=none,commentstyle=\itshape\color{comment},basicstyle=\footnotesize\ttfamily\color{text},keywordstyle=[1]\color{keyword},keywordstyle=[2]\color{builtin},keywordstyle=[3]\itshape\color{instance},stringstyle=\color{string},identifierstyle=\color{text},morestring=[d]{"""},showspaces=false,sensitive=true,showstringspaces=false,numbers=none,numbersep=5pt,numberstyle=\tiny\color{linenumber},breaklines=true,gobble=4,upquote=true,framexleftmargin=0mm,xleftmargin=12pt,escapechar=|,rulecolor=\color{black},keepspaces=true,procnamekeys={def,class},procnamestyle=\bfseries\color{definition},morekeywords=[1]{as},morekeywords=[2]{True,False},morekeywords=[3]{self,@property},alsoletter={0123456789.},alsodigit={.},SelectCharTable=%
}
\def\add@savedef#1#2{%
  \begingroup\lccode`?=#1\relax
  \lowercase{\endgroup
  \edef\@temp{%
    \noexpand\lst@DefSaveDef{\number#1}%
    \expandafter\noexpand\csname lsts@?\endcsname{%
      \expandafter\noexpand\csname lsts@?\endcsname\noexpand#2}%
  }}%
  \python@toks=\expandafter{\the\expandafter\python@toks\@temp}%
}
\count@=`0
\loop
  \add@savedef\count@\ProcessDigit
  \ifnum\count@<`9
  \advance\count@\@ne
\repeat
\count@=`A
\loop
  \add@savedef\count@\ProcessLetter
  \ifnum\count@<`Z
  \advance\count@\@ne
\repeat
\count@=`a
\loop
  \add@savedef\count@\ProcessLetter
  \ifnum\count@<`z
  \advance\count@\@ne
\repeat
\begingroup\edef\x{\endgroup
  \noexpand\lstdefinestyle{defaultpython}{\the\python@toks}
}\x
\makeatother

\definecolor{azure}{rgb}{0.0, 0.5, 1.0}
\definecolor{indiagreen}{rgb}{0.07, 0.53, 0.03}

\newcommand{\evnote}[1]{#1}                                        

\newcommand{\textsw}[1]{\textsc{#1}}								
\newcommand{\textcl}[1]{\textsc{#1}}								

\newcommand{\python}{\textsw{Python}}
\newcommand{\numpy}{\textsw{NumPy}}
\newcommand{\prism}{\textsw{Prism}}
\newcommand{\mypackage}{\textsw{e13Tools}}
\newcommand{\pipeline}{\textcl{Pipeline}}
\newcommand{\emulator}{\textcl{Emulator}}

\newcommand{\modellink}{\textcl{ModelLink}}
\newcommand{\gaussianlink}{\textcl{GaussianLink}}
\newcommand{\mlxtend}{\textsw{Mlxtend}}
\newcommand{\sklearn}{\textsw{Scikit-learn}}

\newcommand{\meraxes}{\textsw{Meraxes}}
\newcommand{\emcee}{\textsw{emcee}}

\newcommand{\BLA}{Bayes linear approach}

\newcommand{\prob}{\mathrm{P}}							
\newcommand{\cov}{\mathrm{Cov}}							
\newcommand{\var}{\mathrm{Var}}							
\newcommand{\E}{\mathrm{E}}								

\renewcommand{\vec}[1]{\bm{\mathrm{#1}}}

\makeatletter
\preto{\@verbatim}{\topsep=0.5\baselineskip  \partopsep=0.5\baselineskip}
\makeatother

\hypersetup{
	pdfauthor={van der Velden et al.},%
    pdftitle={Model dispersion with PRISM},
}

\received{24/01/2019}
\revised{29/04/2019}
\accepted{03/05/2019}
\submitjournal{ApJS}

\shorttitle{Model dispersion with PRISM}
\shortauthors{van der Velden et al.}

\begin{document}

\title{Model dispersion with PRISM; an alternative to MCMC for rapid analysis of models}

\correspondingauthor{Ellert van der Velden}
\email{evandervelden@swin.edu.au}

\author[0000-0002-1559-9832]{Ellert van der Velden}
\affiliation{Centre for Astrophysics and Supercomputing, Swinburne University of Technology, PO Box 218, Hawthorn, VIC 3122, Australia}
\affiliation{ARC Centre of Excellence for All Sky Astrophysics in 3 Dimensions (ASTRO 3D)}

\author[0000-0002-9636-1809]{Alan R. Duffy}
\affiliation{Centre for Astrophysics and Supercomputing, Swinburne University of Technology, PO Box 218, Hawthorn, VIC 3122, Australia}
\affiliation{ARC Centre of Excellence for All Sky Astrophysics in 3 Dimensions (ASTRO 3D)}

\author[0000-0002-5009-512X]{Darren Croton}
\affiliation{Centre for Astrophysics and Supercomputing, Swinburne University of Technology, PO Box 218, Hawthorn, VIC 3122, Australia}
\affiliation{ARC Centre of Excellence for All Sky Astrophysics in 3 Dimensions (ASTRO 3D)}

\author[0000-0002-3166-4614]{Simon J. Mutch}
\affiliation{ARC Centre of Excellence for All Sky Astrophysics in 3 Dimensions (ASTRO 3D)}
\affiliation{School of Physics, University of Melbourne, Parkville, VIC 3010, Australia}

\author[0000-0002-4845-1228]{Manodeep Sinha}
\affiliation{Centre for Astrophysics and Supercomputing, Swinburne University of Technology, PO Box 218, Hawthorn, VIC 3122, Australia}
\affiliation{ARC Centre of Excellence for All Sky Astrophysics in 3 Dimensions (ASTRO 3D)}




\begin{abstract}
    We have built \prism, a \textit{Probabilistic Regression Instrument for Simulating Models}.
   	\prism\ uses the Bayes linear approach and history matching to construct an approximation (`emulator') of any given model, by combining limited model evaluations with advanced regression techniques, covariances and probability calculations.
   	It is designed to easily facilitate and enhance existing Markov chain Monte Carlo (MCMC) methods by restricting plausible regions and exploring parameter space efficiently.
   	However, \prism\ can additionally be used as a standalone alternative to MCMC for model analysis, providing insight into the behavior of complex scientific models.
   	With \prism, the time spent on evaluating a model is minimized, providing developers with an advanced model analysis for a fraction of the time required by more traditional methods.

    This paper provides an overview of the different techniques and algorithms that are used within \prism.
	We demonstrate the advantage of using the Bayes linear approach over a full Bayesian analysis when analyzing complex models.
    Our results show how much information can be captured by \prism\ and how one can combine it with MCMC methods to significantly speed up calibration processes (${>}15$ times faster).
    \prism\ is an open-source \python\ package that is available under the BSD 3-Clause License (BSD-3) at \url{https://github.com/1313e/PRISM} and hosted at \url{https://prism-tool.readthedocs.io}.
    \evnote{\prism\ has also been reviewed by \textit{The Journal of Open Source Software} \citep{PRISM_JOSS}.}
\end{abstract}




\section{Introduction}
\label{sec:Introduction}
Rapid technological advancements allow for both computational resources and observational/experimental instruments to become better, faster and more precise with every passing year.
This leads to an ever-increasing amount of scientific data being available and more research questions being raised.
As a result, scientific models that attempt to address these questions are becoming more abundant, and are pushing the available resources to the limit as these models incorporate more complex science and more closely resemble reality.

However, as the number of available models increases, they also tend to become more distinct.
This causes scientific phenomena to be described in multiple different ways.
For example, there are many models in existence that attempt to describe the Galactic magnetic field (\evnote{GMF,} e.g., \citealt{Sun08,Jaffe10,Pshirkov11,VanEck11,JF12a,JF12b,Jaffe13,Terral16,UngerFarrar17}) or study the formation of galaxies (e.g., \citealt{SAGE,Meraxes,Shark}).
Due to the complexity and diversity of such models, \evnote{it is already a difficult task to compare them with each other (as demonstrated for GMF models by the IMAGINE pipeline, \citealt{IMAGINE}), but} it can \evnote{easily} become \evnote{just as} difficult to keep track of their individual qualities.
Therefore, a full analysis of every model would be required in order to recognize these qualities.

We commonly employ Markov chain Monte Carlo (MCMC) methods \citep{BayesianBook,gelman2014bayesian} when performing this task.
\evnote{Monte Carlo methods are a class of algorithms that use the randomness of a system in order to solve problems, given that the problem is deterministic in principle.
By repeatedly sampling randomly over the system, the idea of Monte Carlo methods is that with enough samples, the problem can be solved by a combination of these samples.
If the problem can be parametrized, which is often the case with scientific models, this process can be sped up by adding one or several Markov chains to it.
A Markov chain is (in this context) a sequence of evaluations where the outcome of every evaluation solely depends on the state the system was in right before it, creating a memoryless process.
Doing so gives the class of MCMC methods.}

When combined with a full Bayesian analysis, MCMC has the potential to create an accurate approximation of the posterior probability distribution function (PDF).
This PDF can be used to identify regions of parameter space that compare well with the available data.
Currently, there are many different MCMC methods available, like Metropolis-Hastings \citep{Metropolis,Hastings}; Gibbs \citep{Gibbs}; Hamiltonian/Hybrid Monte Carlo \citep{brooks2011handbook,Betancourt}; nested \citep{Skilling}; no-U-turn \citep{NUTS}; and affine invariant \citep{affine_invariant} sampling to name but a few.
Some of these are completely unique, while others are simply extensions of already existing methods to make them suitable for specific tasks.

However, as specialized as some MCMC methods are, they have a couple of drawbacks:
\begin{enumerate}
	\item Random walk MCMC methods (methods based on the Metropolis-Hastings algorithm) tend to move around the equilibrium distribution of the model in relatively small steps or get stuck in local extrema, causing them to sample parts of parameter space that have been visited before.
	Not only does this make the process very slow, but it also causes the model to be reevaluated unnecessarily.
	Several MCMC methods, like Hamiltonian Monte Carlo, try to circumvent this by not using a random walk algorithm, but instead use the gradient field of the desired distribution.
	This however does require one to have additional knowledge about the model and that the gradient field exists, which might be difficult to obtain due to the number of degrees-of-freedom involved;

	\item Given the discontinuous and irregular natures some models tend to have, MCMC methods can completely fail to converge due to irregularities in the probability distribution.
	This also increases the difficulty of obtaining a description of the gradient field, since it is not defined everywhere;

	\item The rate at which an MCMC method converges depends on the initial number of samples and their locations.
	Not using enough may cause it to never converge, while using too many will make the process move forward very slowly.
\end{enumerate}

Given the reasons above, it may not look appealing to some to perform an extensive model analysis.
However, we believe that analyzing a model does not necessarily require the full posterior PDF to be accurately known.
Instead, the converging process towards this PDF provides insight on the workings of a model.
Therefore, we propose to use a combination of the \textit{\BLA} and the \textit{emulation technique} with \textit{history matching}, which can be seen as special cases of Bayesian statistics.

The emulation technique can be used to create an \textit{emulator}, which is an approximate system of \evnote{mainly} polynomial functions \evnote{and Gaussian processes} based on limited knowledge about the model.
Using the \BLA\ and history matching, an emulator can quickly make predictions on the expected relevance (`implausibility') of all parts in model parameter space, taking into account the variance that is introduced by doing so.
By imposing cutoffs on this relevance, parts of parameter space can be excluded, defining a smaller parameter space over which an improved emulator can be defined.
Performing this process iteratively allows one to quickly reduce the size of relevant parameter space while being provided with several snapshots of the convergence process, giving insight on the model's \evnote{behavior}.

These techniques have been applied to various systems several times in the past (either together or separately), including the study of whales \citep{Raftery95}, oil reservoirs \citep{Craig96,Craig97}, galaxy formation \evnote{\citep{Bower10,Vernon10,Vernon14,Rodriques17}}\evnote{, disease studies \citep{Andrianakis15,Andrianakis16,Andrianakis17}}, biological systems \citep{Vernon18} and simply described in general \citep{Sacks89,Currin91,Oakley02,O'Hagan06}.
Many of these works show how much information can be extracted from an emulator, and the advantages \evnote{the combination of the \BLA, emulation technique and history matching} has over using a full Bayesian analysis.
However, the algorithms used are often focused on a specific application and are typically not publicly available.

In this work, we introduce \prism, a \textit{Probabilistic Regression Instrument for Simulating Models}.
\prism\ is a publicly available framework that uses the \BLA\ to combine the power of emulation with the process of history matching.
Although this has been done before (e.g., \evnote{\citealt{Bower10,Vernon10,Vernon14,Rodriques17,Vernon18}}), \prism\ is unique in that it is not built for a specific application.
Instead, \prism\ provides a universal, versatile framework in which both simple and sophisticated models can be analyzed with minimal effort.
Additionally, its implementation is highly modular, allowing for extensions to be added easily.

In \ref{sec:Model analysis}, we describe the \BLA, emulation technique and history matching, which are the main methods behind analyzing models with \prism.
Then, using this knowledge, we give an overview of the \prism\ framework in \ref{sec:PRISM} and its various components.
We show in \ref{sec:Basic usage} what \prism\ can do when combined with and compared to normal MCMC methods.
Finally, we give a short introduction to the larger applications that \prism\ will be used for in \ref{sec:Conclusions}.

\section{Model analysis}
\label{sec:Model analysis}
In this section, we describe the various different techniques that are used in \prism, including a discussion on variances, the \BLA\ and the emulation technique, history matching and implausibility measures.
Note that this section is meant for those seeking a general understanding of the used methodology.
\evnote{See \citet{Vernon10} for further details and justification.}
For those looking for a description of \prism\ itself, we refer to \ref{sec:PRISM}.

\subsection{Uncertainty analysis}
\label{subsec:Uncertainty analysis}
When evaluating a model, it is required to have knowledge of all the individual components that influence the outcome of the model before performing the actual evaluation.
This allows one to interpret the meanings behind the constructed model realization, determine its accuracy by comparing it to the physical system one is describing and study its behavior.
Due to a variety of reasons, these components each contribute to how certain (or uncertain) one is that the output given by the model is correct given the observational data.
When performing an analysis on a scientific model, it is important to know what all the individual contributions to the uncertainty are.
Below, we give an overview of the most common uncertainty contributions, describe the underlying process and how they are treated in this work.

\paragraph*{Observational uncertainties}
\texttt{ }\\
In order to explain the Universe, one attempts to make a connection between observations and modelling.
However, observational instruments are not perfect, giving rise to uncertainties in the true value of the observed data.
Additionally, measurements are often obtained by performing conversions/integrations on the observations, increasing the uncertainty.
In this work, the uncertainties described above are collected into one term called the \textit{observational variance}, which for a single data point $i$ is denoted by $\var(\epsilon_{\mathrm{obs}, i})$.

\paragraph*{Model output uncertainties}
\texttt{ }\\
The output of a model is usually determined by individual contributions from a large number of deterministic functions, which combined can be seen as a single deterministic computer function.
Because evaluating this single function is often very expensive, it is appropriate to say that the output of the function is unknown for \evnote{almost all} combinations of input parameter values.
In this work, we attempt to solve this problem by creating a statistical representation of this function by constructing something called an \textit{emulator} (see \ref{subsec:Emulation technique}).
While the model discrepancy variance discussed below gives the contribution to the overall uncertainty of the known model realizations, the \textit{emulator variance} $\var(f_i(\vec{x}))$ describes the uncertainty of the unknown realizations.

\paragraph*{Knowledge uncertainties}
\texttt{ }\\
Since the observational data that we have at our disposal is exhaustive, the possibility exists that we have overlooked a certain dynamic, mechanic or phenomenon that is important for the model.
Additionally, limited computational resources often force us to use approximations of the involved physics or use other models, inevitably introducing uncertainties to the problem.
Although very small, these are all contributions to the knowledge uncertainty of the model, causing a discrepancy between reality and the model.
This so-called \textit{model discrepancy variance} $\var(\epsilon_{\mathrm{md}, i})$ \evnote{\citep{Craig96,Craig97,Kennedy01,Vernon10}} is usually the most challenging of all uncertainties and it is very important to not underestimate its contribution to the total variance involved.
Given its importance, we give a more detailed description of this in \ref{subsec:Variances}, describing how \prism\ treats different variances.

\subsection{Bayes linear approach}
\label{subsec:BLA}
As discussed in \ref{sec:Introduction}, performing a full Bayesian analysis on a complex model can be quite complicated.
In order to undertake such a Bayesian analysis, one uses Bayes' equation given as:
\begin{align}
\label{eq:Bayes}
    \prob(M|D,I) &= \frac{\prob(D|M,I)\cdot\prob(M|I)}{\prob(D|I)},
\end{align}
where $M$ is the model (realization), $D$ the comparison data and $I$ the remaining background information.
Bayes' equation gives one the probability that a given model realization $M$ can explain the comparison data $D$, taking into account the background information $I$.

Although this gives one an absolute answer, it is required for the user to parametrize everything about the model in order to quantify its \textit{prior}, $\prob(M|I)$.
This includes that one knows all uncertainties that are related to the model in question.
\evnote{Not only that, but it is also required to specify a meaningful \textit{likelihood}, which is often very difficult to do properly, as this requires realistic assumptions and robustness (slight changes in the likelihood form should not massively alter its outcome).}
\evnote{And, finally,} the same applies for the comparison data, since it is required to know the \textit{evidence} value, $\prob(D|I)$, as well when dealing with model comparisons.
It is therefore necessary to have full knowledge about the model and the data one wants to use, which may be hard or even impossible to obtain.

Another complication here is how one finds the parameter set that can explain the comparison data \evnote{the} best.
As mentioned before, we commonly employ MCMC methods to perform this task.
Although MCMC can narrow down parameter space where one can find the exact model realization that can explain the comparison data the best, it requires many evaluations of the model in order to get there.
When dealing with complex models, one usually does not want to unnecessarily evaluate the computationally expensive model.
It is therefore not seen as ideal to evaluate the model in parts of parameter space where the probability of finding that exact parameter set, is very low.
This does however happen fairly often due to the nature of MCMC, which tries to find its path through parameter space by sampling in random directions and accepting new samples with certain probabilities (which is commonly known as \textit{Metropolis-Hastings sampling}, \citealt{Metropolis,Hastings}).

There is a more appropriate method to analyze a complex scientific model.
Instead of using full Bayesian analysis and MCMC methods, we use the \BLA\ \citep{Goldstein99,Goldstein00,BLA}.
The difference between the \BLA\ and a full Bayesian analysis, is that the \BLA\ uses expectations as its primary output instead of probabilities.
The advantage of this is that the \BLA\ does not require the \evnote{full joint specification of an appropriate probabilistic model for the prior and all associated data points} (which is required for determining probabilities), allowing one to use it even when not all details are known.
This makes the \BLA\ much simpler in terms of belief and analysis, since it is based on only the mean/expectation, variance and covariance specifications, for which we follow the definitions by \citet{DeFinetti74,DeFinetti75}.
The two main equations of the \BLA\ are the updating equations for the expectations and variances of a vector $M$ (the model realization), given a vector $D$ (the comparison data):
\begin{align}
\label{eq:adj_exp_BLA}
    \E_D(M) &= \E(M)+\cov(M, D)\cdot\var(D)^{-1}\cdot(D-\E(D)),\\
\label{eq:adj_var_BLA}
    \var_D(M) &= \var(M)-\cov(M, D)\cdot\var(D)^{-1}\cdot\cov(D, M),
\end{align}
where $\E_D(M)$ and $\var_D(M)$ are known as the \textit{adjusted expectation and adjusted variance of $M$ given $D$} \citep{Goldstein99,BLA}.

The equation for the adjusted expectation is fairly similar in meaning to Bayes' equation.
In the \BLA, the expectation of the model realization $M$ is the equivalent of the probability that $M$ is correct in full Bayesian analysis.
In particular, if one would use a full Gaussian specification for all of the relevant quantities, one would end up with similar updating formulas.
An overview of the \BLA\ is given in \citet{Goldstein99}, and we direct readers to \citet{BLA} for a detailed description of it.

There are two general reasons for \evnote{why} one may choose \evnote{the} \BLA\ over a full Bayesian analysis.
We discussed earlier that performing a full Bayesian analysis can take a significant amount of time, and may not be possible due to constraints on knowledge and computational resources.
In this scenario, one could view the \BLA\ as a special case of full Bayesian analysis.
The \BLA\ is simplified compared to a full Bayesian analysis, since we only require the expectations, variances and covariances of all random quantities that are involved in the problem.
Therefore, instead of carrying out a full posterior calculation using Bayes' equation given in \ref{eq:Bayes}, we carry out a Bayes linear update by using \ref{eq:adj_exp_BLA} and \ref{eq:adj_var_BLA}.
The adjusted expectation value of $M$ given $D$ ($\E_D(M)$) can be seen as the best linear fit for $M$, given the elements in the data $D$, with the adjusted variance value being the minimized squared error.
This best linear fit depends solely on the collection of functions of the elements in the data $D$, which can be chosen to be whatever we like it to be.
If we would choose all functions that are possible in the data $D$ (up to infinite order polynomials), we are effectively recreating the full Bayesian analysis.

The second, more fundamental reason comes from the meaning of a full Bayesian analysis.
A full Bayesian analysis has a lot of value, since it allows others to include their own expertise and knowledge in the analysis of their model.
However, at some point, models become so complex and sophisticated, that it becomes hard to make expert judgments on the prior specifications of the model, which are required for doing a full Bayesian analysis.
The \BLA\ however, does not require \evnote{the specifications of any prior distribution (which can be very complex), but instead requires the simpler prior expectations and (co)variances} of a model realization\evnote{, which can still be used for including expert prior knowledge}.
For a more detailed overview of these two reasons, see \citet{Goldstein06}.

The benefit of \evnote{the \BLA\ combined with the emulation technique and history matching} is that it does not require a high number of model evaluations in order to say something useful about the model.
For example, one can usually already exclude certain parts of parameter space based on minimal knowledge of the model, as long as taking into account all related variances still does not make it likely that this part is important.
Therefore, using the \BLA\ \evnote{in this way} allows one to decrease the part of parameter space that is considered important and only evaluate the model there, by only having a limited amount of knowledge.
This is a big advantage over using a traditional full Bayesian analysis with MCMC methods, since MCMC will not have this knowledge (due to its Markov chain nature) and therefore can potentially have a large \textit{burn-in phase}.

Another benefit of the \BLA\ \evnote{(when combined with the emulation technique and history matching)} that can be crucial to model analysis, is its ability to accept additional constraints while already analyzing a model.
When performing a full Bayesian analysis, it is required that it already contains all constraints (given by data, previously acquired results, etc.) that one wants to put on the model.
If one would attempt to introduce additional constraints during a full Bayesian analysis, it could potentially corrupt the results or confuse the process.
However, since the expectation value of the model realization ($\E(M)$) is important rather than the model realization itself, additional comparison data can be easily added to the analysis at a later stage, without requiring the model to be reevaluated at all previously evaluated samples.
This allows for one to update/improve their analysis of a model when new data becomes available or to only incrementally add data constraints in order to minimize the amount of time spent on evaluating.

\subsection{Emulation technique}
\label{subsec:Emulation technique}
When analyzing a model, it is usually desirable to cover its full parameter space.
This would allow one to study the dependencies between all parameters and their behaviors in general.
However, this approach rapidly becomes unfeasible when one increases the number of model parameters.
If one does not make any assumptions about the dependencies between model parameters, then the easiest way of model exploration is achieved by direct sampling, which means that every single combination of model parameters needs to be checked.
In other words, choosing $10$ different values for every of the $N$ model parameters, will yield $10^N$ different combinations.
Anything more than $5$ model parameters will already give a very large number of model realizations to be evaluated, while the density of evaluated samples in parameter space is extremely low.
In order to avoid this, a different approach is required.

We propose to use the technique of emulation, by constructing a so-called \textit{emulator system} for every output of a given model.
An emulator system can be described as a stochastic belief specification for a deterministic function, which has been successfully applied to various systems several times in the past (e.g., \citealt{Sacks89,Currin91,Craig96,Craig97,Oakley02,O'Hagan06,Bower10,Vernon10,Vernon14,Vernon18}).
Unlike a model, an emulator system can be evaluated very quickly, allowing one to explore its parameter space in a more efficient way.
Knowing this, we substitute a collection of emulator systems (called an \textit{emulator}) for the model evaluations, while taking into account the additional variances that are introduced in doing so.

\evnote{Following the form used in \citet{Vernon10}, an emulator system is constructed in the following way.}
Suppose that for a vector of input parameters $\vec{x}$, the output of a model is given by the function $f(\vec{x})$.
If the output of this model would have no variance, then we can say that $f_i(\vec{x})$ (output $i$ of $f(\vec{x})$) is given by a function $r_i(\vec{x})$, which has the form
\begin{align}
\label{eq:r}
    r_i(\vec{x}) &= \sum_j\beta_{ij}g_{ij}(\vec{x}),
\end{align}
where $\beta_{ij}$ are unknown scalars/coefficients, $g_{ij}$ are known deterministic functions of $\vec{x}$ and $r_i(\vec{x})$ is known as the \textit{regression term}.

Realistically speaking, \evnote{we are uncertain if $f_i(\vec{x})$ follows the above description for every $\vec{x}$, as we have not evaluated $f_i(\vec{x})$ everywhere and thus we do not know all of its corresponding outputs}.
Therefore, it is required to add a second term to $f_i(\vec{x})$ that describes this \evnote{uncertainty}, which gives
\begin{align}
\label{eq:f1}
    f_i(\vec{x}) &= r_i(\vec{x})+u_i(\vec{x}),
\end{align}
with $u_i(\vec{x})$ a weakly stochastic process with constant variance and uncorrelated with the regression term $r_i(\vec{x})$. 
The regression term describes the general behavior of the function, while $u_i(\vec{x})$ represents the local variations from this behavior near $\vec{x}$.

When dealing with high dimensional parameter spaces, it is not unusual to find that a subset of the model parameters $\vec{x}$ can explain a significant part of the variance in the model output.
Therefore, we introduce the so-called \textit{active parameters}, $\vec{x}_{\mathrm{A},i}$, which is a subset of $\vec{x}$ and varies with $i$.
However, due to the introduction of active parameters, one final term needs to be added to $f_i(\vec{x})$, changing it to
\begin{align}
\label{eq:f2}
    f_i(\vec{x}) &= r_i(\vec{x}_{\mathrm{A},i})+u_i(\vec{x}_{\mathrm{A},i})+w_i(\vec{x}),
\end{align}
where $w_i(\vec{x})$ is the variance in $f_i$ caused by the \textit{passive parameters}. 

When using the \BLA, the emulator system requires the definition of the prior expectation and the prior (co)variance of all three terms in \ref{eq:f2}.
Commonly, localized deviations in a function are assumed to be of Gaussian form, which has a covariance that is defined as
\begin{align}
\label{eq:cov_u}
    \cov\left(u_i(\vec{x}_{\mathrm{A},i}), u_i(\vec{x}'_{\mathrm{A},i})\right) &= \sigma_{u_i}^2\exp\left(-\left\|\vec{x}_{\mathrm{A},i}-\vec{x}'_{\mathrm{A},i}\right\|^2/\theta_i^2\right),
\end{align}
with $\sigma_{u_i}^2$ and $\theta_i$ being the Gaussian variance and correlation length, respectively.
The Gaussian correlation length is defined as the maximum distance between two values of a specific model parameter within which the Gaussian contribution to the correlation between the values is still significant.

Here, $u_i(\vec{x}_{\mathrm{A},i})$ is called the \textit{Gaussian term} and has an expectation of zero.
The other variation term, the so-called \textit{passive term}, $w_i(\vec{x})$, describes the variance that is caused by passive parameters and therefore has no expectation value and a constant variance $\sigma_{w_i}^2$:
\begin{align}
\label{eq:cov_w}
    \cov\left(w_i(\vec{x}), w_i(\vec{x}')\right) &= \sigma_{w_i}^2\delta_{\vec{x}, \vec{x}'},
\end{align}
with $\delta_{\vec{x}, \vec{x}'}$ the \textit{Kronecker delta} of $\vec{x}$ and $\vec{x}'$.

We can now use the emulator system to evaluate and calculate the expectation and variance values of the function for any input $\vec{x}$, and the covariance between values of $f_i$ for any pair of inputs $\vec{x}, \vec{x}'$.
From \ref{eq:f2}, we get that the prior expectation is given by
\begin{align}
\label{eq:prior_exp}
\E(f_i(\vec{x})) &= \sum_j\E(\beta_{ij})g_{ij}(\vec{x}_{\mathrm{A},i}),
\end{align}
and that the prior covariance is given by
\begin{align}
\nonumber
    c_i(\vec{x}, \vec{x}') &= \cov\left(f_i(\vec{x}), f_i(\vec{x}')\right),\\
\label{eq:prior_cov}
    \begin{split}
    &= \sum_j\sum_k\cov(\beta_{ij},\beta_{ik})\cdot g_{ij}(\vec{x}_{\mathrm{A},i})g_{ik}(\vec{x}'_{\mathrm{A},i})\\
    &\quad+\sigma_{u_i}^2\exp\left(-\left\|\vec{x}_{\mathrm{A},i}-\vec{x}'_{\mathrm{A},i}\right\|^2/\theta_i^2\right)+\sigma_{w_i}^2\delta_{\vec{x}, \vec{x}'},
    \end{split}
\end{align}
where all cross-covariances are equal to zero (as all three terms are uncorrelated).
Here the first term is the covariance of the regression term, which can be derived using the relation $\cov\left(X,Y\right)=\E\left(X\cdot Y\right)-\E\left(X\right)\cdot\E\left(Y\right)$:
\begin{align}
\nonumber
    \cov(r_i(\vec{x}), r_i(\vec{x}')) &= \E(r_i(\vec{x})\cdot r_i(\vec{x}'))-\E(r_i(\vec{x}))\cdot\E(r_i(\vec{x}')),\\
\nonumber
    &=\E\left(\sum_j\sum_k\beta_{ij}\beta_{ik}\cdot g_{ij}(\vec{x})g_{ik}(\vec{x}')\right)\\
\nonumber
    &\quad-\E\left(\sum_j\beta_{ij}g_{ij}(\vec{x})\right)\cdot\E\left(\sum_k\beta_{ik}g_{ik}(\vec{x}')\right),\\
\nonumber
    &=\sum_j\sum_k\Big(\E(\beta_{ij}\beta_{ik})-\E(\beta_{ij})\E(\beta_{ik})\Big)\\
\nonumber
    &\quad\cdot g_{ij}(\vec{x})g_{ik}(\vec{x}'),\\
\label{eq:cov_r}
    &= \sum_j\sum_k\cov(\beta_{ij}, \beta_{ik})\cdot g_{ij}(\vec{x})g_{ik}(\vec{x}').
\end{align}
This is all that is required for the \BLA\ in order to update our beliefs in terms of expectation and variances about the model output $f_i(\vec{x})$ at an unevaluated input $\vec{x}$, given a set of known model realizations.
This process is described below.

Suppose that we have evaluated the model for $n$ different input parameter sets.
Then we can write the individual inputs as $\vec{x}^{(k)}$ with $k = 1, \ldots, n$, where each $\vec{x}^{(k)}$ represents the vector of parameter values in parameter space for the $k^{\mathrm{th}}$ evaluation.
Using the same notation, $\vec{x}_{\mathrm{A},i}^{(k)}$ gives the vector of active parameter values for the $k^{\mathrm{th}}$ evaluation.
We define the vector of model outputs for output $i$ of known model realizations as $D_i = \left(f_i(\vec{x}^{(1)}), f_i(\vec{x}^{(2)}), \ldots, f_i(\vec{x}^{(n)})\right)$.
If we now replace the vector $M$ in the \BLA\ update equation of the expectation value (\ref{eq:adj_exp_BLA}) with the model output $f_i(\vec{x})$ and vector $D$ with $D_i$, we obtain the adjusted expectation for a given emulator system $i$:
\begin{align}
\nonumber
    \E_{D_i}(f_i(\vec{x})) &= \E(f_i(\vec{x}))\\
\nonumber
    &\quad+\cov\left(f_i(\vec{x}), D_i\right)\cdot\var(D_i)^{-1}\cdot\left(D_i-\E(D_i)\right),\\
\intertext{which combined with \ref{eq:prior_exp} and \ref{eq:prior_cov} gives}
\label{eq:adj_exp}
    \begin{split}
    \E_{D_i}(f_i(\vec{x})) &= \sum_j\E(\beta_{ij})g_{ij}(\vec{x}_{\mathrm{A},i})\\
    &\quad+t(\vec{x})\cdot A^{-1}\cdot\left(D_i-\E(D_i)\right),
    \end{split}
\end{align}
where $t(\vec{x}) = \left(c_i(\vec{x}, \vec{x}^{(1)}), c_i(\vec{x}, \vec{x}^{(2)}), \ldots, c_i(\vec{x}, \vec{x}^{(n)})\right) = \linebreak \cov\left(f_i(\vec{x}), D_i\right)$ is the vector of covariances between the new and known points, and $A$ is the $n\times n$ matrix of covariances between known points with elements $A_{jk} = c_i(\vec{x}^{(j)}, \vec{x}^{(k)})$.

Similarly, the adjusted variance can be found by combining \ref{eq:prior_exp} and \ref{eq:prior_cov} with \ref{eq:adj_var_BLA}:
\begin{align}
\nonumber
    \var_{D_i}(f_i(\vec{x})) &= \var(f_i(\vec{x}))\\
\nonumber
    &\quad-\cov(f_i(\vec{x}), D_i)\cdot \var(D_i)^{-1}\cdot \cov(D_i, f_i(\vec{x})),\\
\label{eq:adj_var}
    \begin{split}
    &= \var\left(\sum_j\beta_{ij}g_{ij}(\vec{x}_{\mathrm{A},i})\right)+\sigma_{u_i}^2+\sigma_{w_i}^2\\
    &\quad -t(\vec{x})\cdot A^{-1}\cdot t(\vec{x})^T.
    \end{split}
\end{align}

The adjusted expectation, $\E_{D_i}(f_i(\vec{x}))$, and adjusted variance, $\var_{D_i}(f_i(\vec{x}))$, represent our updated beliefs about the output of the model function $f_i(\vec{x})$ for model parameter set $\vec{x}$, given a set of $n$ model evaluations with output $D_i$.
These \textit{adjusted values} are used in the implausibility measures for the process of history matching, which are described in \ref{subsec:History matching}.
Note that the adjusted values for any known model evaluation are equal to $\E_{D_i}(f_i(\vec{x}^{(k)})) = f_i(\vec{x}^{(k)})$ and $\var_{D_i}(f_i(\vec{x}^{(k)})) = 0$.
\evnote{For further discussion, we point interested readers to \citet{Vernon10,Vernon18}}.

From \ref{eq:adj_exp} and \ref{eq:adj_var}, it is clear that the regression term and the Gaussian term are strongly connected to each other: increasing the variance that the regression term can explain will decrease the remaining variance that is described by the Gaussian term and vice verse.
Since the regression term is much more complicated than the Gaussian term, one could argue here whether it is preferable to put a lot of time, effort and resources in constructing the regression term for every emulator system, while one can also place more weight on the Gaussian term.
By default, \prism\ aims to explain as much variance as possible with the regression term, but this balance is provided as a user-chosen parameter.
The reasons why we prefer putting a lot of emphasis on the regression term are:
\begin{enumerate}
	\item We aim to make \prism\ suited for analyzing any given model, but in particular those that are highly complex.
	Such models often have strong physical interpretations and interactions driving them, influencing the results that the model gives.
	The embedded physical laws \evnote{are often reasonably well-behaved and therefore} usually come in the form of polynomial functions, making it only natural to express them in the same way through the regression term;

	\item Studying the behavior of the model according to the emulator system, is much easier to do when one is given the polynomial terms and their expected coefficients.
	This allows the user to check whether the emulator system is consistent with expectation, while also allowing for new physical interactions to be discovered;

	\item The more information is contained within the regression term, the less information remains for the Gaussian term.
	This makes it easier to compare emulator systems with each other, while also allowing one to remove the calculation of the Gaussian term if the remaining (Gaussian) variance drops below a certain threshold;

	\item And most importantly, the Gaussian term, $u_i(\vec{x}_{\mathrm{A},i})$, is an approximation in itself (of Gaussian form) and therefore might have trouble explaining the smoothness of a complex model \evnote{on both large and small scales simultaneously}.
	A Gaussian correlation (or any other correlation form) makes certain assumptions about the behavior of the model.
	If the model does not follow these assumptions, a correlation can have trouble coming up with a good fit.
	\evnote{Therefore, the Gaussian term is mainly used for explaining local behavior, whereas the regression term captures the global behavior.}
\end{enumerate}

In summary, for each model output, one is required to:
\begin{itemize}
	\item have a collection of known model realizations $D_i$;
    \item identify the set of active parameters $\vec{x}_{\mathrm{A},i}$;
	\item select the polynomial regression terms $g_{ij}$;
    \item determine the coefficients of these terms $\beta_{ij}$;
    \item obtain the residual variance $\sigma_i^2=\sigma_{u_i}^2+\sigma_{w_i}^2$ from the regression term;
    \item and, if required, calculate the covariance of the polynomial coefficients $\cov(\beta_{ij}, \beta_{ik})$.
\end{itemize}
Then, we can use \ref{eq:adj_exp} and \ref{eq:adj_var} to update our beliefs on the model output function $f_i(\vec{x})$, by obtaining the adjusted expectation and covariance values for any given parameter set $\vec{x}$.
Afterward, we have to carry out a diagnostic analysis on the emulator, or we can alternatively decide to study the properties and behavior of the emulator by making a \textit{projection} first (see \ref{subsec:Projection}).
This diagnostic analyzing of the emulator is called \textit{history matching} and is explained below.

\subsection{History matching}
\label{subsec:History matching}
The idea behind \prism\ is to provide the user with the collection of parameter sets $\mathcal{X}^*$ that gives an acceptable match to the observations $z$ when evaluated in $f(\vec{x})$.
This collection contains the best parameter set $\vec{x}^*$ as well as parameter sets that yield acceptable matches and can be used for studying the properties of the emulated model.
The process of obtaining this collection $\mathcal{X}^*$ is usually referred to as \textit{history matching}. 
This terminology is common in various different fields \citep{Raftery95,Craig96,Craig97}, although one rarely tries to find all matches instead of just a few.
To give the user more flexibility and more information, we think it is better to try to find as many matches as possible.

The process of history matching can be compared to \textit{model calibration} \evnote{(discussed in more detail in \citealt{Vernon10,Vernon18})}, where we assume that there is a single true but unknown parameter set $\vec{x}^*$ and our goal is to make probabilistic statements about $\vec{x}^*$ based on a prior specification, the collection of model evaluations and the observed history \citep{Kennedy01,Goldsteinetal06}.
Although history matching and model calibration look alike and are certainly related, they are very different in terms of approach.
Whereas model calibration gives one a posterior PDF of parameter space that can be used to evaluate various parameter sets, history matching can conclude that no best parameter set $\vec{x}^*$ exists even if it should.
If this is the case, this might be an indication that there are some serious issues with the model, while model calibration can have trouble coming to the same conclusion.
Because of this, we think that history matching is very important for analyzing complex models.

The way we approach history matching is by evaluating so-called \textit{implausibility measures} \citep{Craig96,Craig97}\evnote{, for which we use the same form as in \citet{Vernon10}}.
An implausibility measure is a function that is defined over parameter space which gives a measure of our tolerance of finding a match between the model and the modelled system.
When the implausibility measure is high, it suggests that such a match would exceed our stated tolerance limits, and that we therefore should not consider the corresponding parameter set $\vec{x}$ to be part of $\mathcal{X}^*$.
If we again consider $f_i$ to be a single model output, then we would like to know for a given parameter set $\vec{x}$ whether the output $f_i(\vec{x})$ is within tolerance limits when compared to the system's true value $y_i$.
In order to do this, we would have to evaluate the standardized distance given as \[\frac{\left(f_i(\vec{x})-y_i\right)^2}{\var(\epsilon_{\mathrm{md}, i})}.\]
In reality, we do not know $y_i$ and instead have to use its observed value $z_i$, which has its own measurement error and converts the standardized distance to \[\frac{\left(f_i(\vec{x})-z_i\right)^2}{\var(\epsilon_{\mathrm{md}, i})+\var(\epsilon_{\mathrm{obs}, i})}.\]
However, for most parameter sets $\vec{x}$, we are not able to evaluate the model and obtain $f_i(\vec{x})$.
Therefore, we have to use the emulated value $\E(f_i(\vec{x}))$ and compare this with $z_i$.
This defines the implausibility measure as
\begin{align}
\label{eq:f_impl}
    I_i^2(\vec{x}) &= \frac{\left(\E(f_i(\vec{x}))-z_i\right)^2}{\var\left(\E(f_i(\vec{x}))-z_i\right)}.
\end{align}
Using \ref{eq:f_impl} and taking into account that the data, model and emulator system have a variance, we obtain the implausibility measure for the emulator system:
\begin{align}
\label{eq:impl_sq}
    I_i^2(\vec{x}) &= \frac{\left(\E_{D_i}(f_i(\vec{x}))-z_i\right)^2}{\var_{D_i}(f_i(\vec{x}))+\var(\epsilon_{\mathrm{md}, i})+\var(\epsilon_{\mathrm{obs}, i})},
\end{align}
with $\E_{D_i}(f_i(\vec{x}))$ the adjusted emulator expectation (\ref{eq:adj_exp}), $\var_{D_i}(f_i(\vec{x}))$ the adjusted emulator variance (\ref{eq:adj_var}), $\var(\epsilon_{\mathrm{md}, i})$ the model discrepancy variance and $\var(\epsilon_{\mathrm{obs}, i})$ the observational variance.

When, for a given parameter set $\vec{x}$, the corresponding implausibility value $I_i(\vec{x})$ is large, it suggests that it would be unlikely that we would view the match between the model output and the comparison data as acceptable, if we would evaluate the model at this parameter set.
Therefore, whenever this happens, we can say that any parameter set $\vec{x}$ for which the implausibility value $I_i(\vec{x})$ is large, should not be considered part of the potential parameter sets in the collection $\mathcal{X}^*$.
By imposing certain maximum values for the implausibility measure, we can ensure that only those parameter sets that give low implausibility values are not discarded.
Seeing that the distribution of the function $\left(\E_{D_i}(f_i(\vec{x}'))-z_i\right)$ is both unimodal (\evnote{the two terms in \ref{eq:adj_exp} are independent of each other and $z_i$}) and continuous (the emulator system solely contains deterministic functions) for a fixed parameter set $\vec{x}$, we can use the $3\sigma$-rule given by \citet{Pukelsheim94}.
\evnote{This rule implies that for any continuous, unimodal distribution, $95\%$ of its probability must lie within $\pm 3\sigma$ ($I_i(\vec{x}) \leq 3$), which even applies for asymmetric, skewed, tailed or heavily varying distributions.}
Values higher than $3$ would usually mean that the proposed parameter set $\vec{x}$ should be discarded, but \prism\ allows the user full control over this.

\begin{figure*}
\begin{center}
	\subfloat[Initial Gaussian emulator with $5$ model evaluations.]{\label{subfig:gaussian_0D_1}\includegraphics[width=\textwidth]{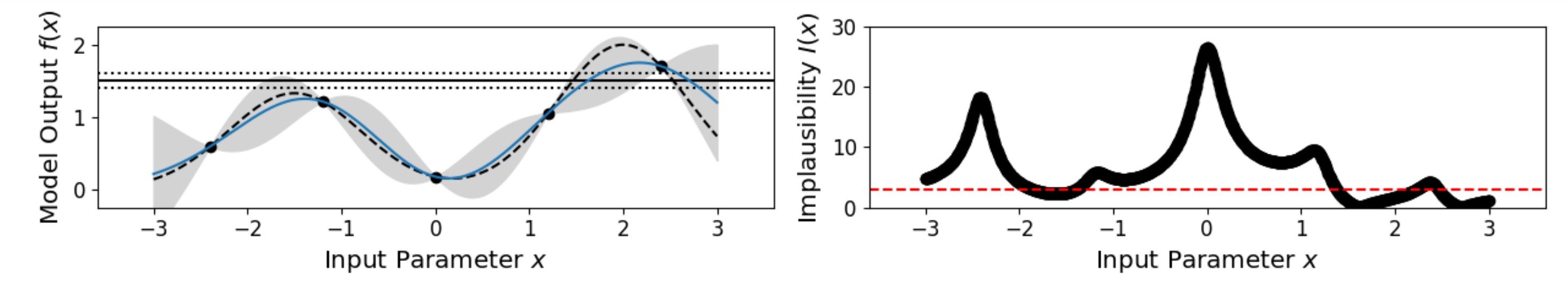}}\\
    \subfloat[Updated Gaussian emulator with $11$ model evaluations.]{\label{subfig:gaussian_0D_2}\includegraphics[width=\textwidth]{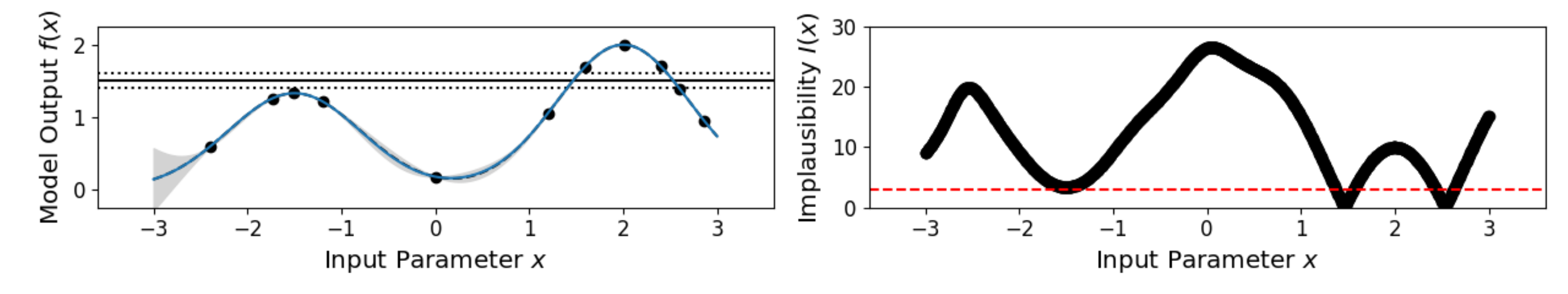}}
	\caption{Emulator of two simple Gaussians, defined as $f(x)=2\cdot\exp\left(-(2-x)^2\right)+1.33\cdot\exp\left(-(-1.5-x)^2\right)$.
	\textbf{Left column:} Gaussian model $f(x)$ (\textbf{dashed}), model evaluations $D$ (\textbf{dots}), emulator $\E_D(f(x))$ (\textbf{solid}), emulator uncertainty $\E_D(f(x))\pm 3\sqrt{\var_D(f(x))}$ (\textbf{shaded}), observational data with $2\sigma$ errors (\textbf{horizontal lines}).
	\textbf{Right column:} Implausibility values $I(x)$ (\textbf{dots}) with cut-off at $I(x)=3$ (\textbf{dashed}).
    \textbf{Top row:} Initial emulator.
    \textbf{Bottom row:} Updated emulator.
    Note that the updated emulator focused solely on the parts of parameter space that were considered plausible.}
    \label{fig:gaussian_0D}
\end{center}
\end{figure*}
To illustrate how the theory above can be applied to a model, we have used the emulation method on a simple double Gaussian model, given in \autoref{fig:gaussian_0D}.
Here, we have made an emulator of a model defined as
\begin{align*}
    f(x) &= 2\cdot\exp\left(-(2-x)^2\right)+1.33\cdot\exp\left(-(-1.5-x)^2\right),
\end{align*}
which are two Gaussians with different mean and amplitude, and a standard deviation $\sigma$ chosen such that $2\sigma^2=1$.
On the left in \ref{subfig:gaussian_0D_1}, the real model function $f(x)$ is shown in dashed black (which usually is not known, but displayed for convenience), which has been evaluated a total of $5$ times as indicated by the black dots.
Using these $5$ evaluations, we can construct an emulator using \ref{eq:adj_exp} and \ref{eq:adj_var}, where the adjusted expectation value $\E_D(f(x))$ is given by the solid (light blue) line and its $3\sigma$ confidence interval by the shaded area\evnote{, where $\sigma$ here is defined as $\sqrt{\var_D(f(\vec{x})}$}.

Now suppose that we also have a comparison data point, given as the black horizontal line with its $2\sigma$ confidence interval.
Using the constructed emulator, we can determine for what values of $x$ we expect a model realization that might be within our stated tolerance of $I(x)\leq3$, whose results are shown on the right in \ref{subfig:gaussian_0D_1}.
From the plotted implausibility values, we can see that there are small parts of parameter space that are likely to yield such model realizations and therefore require further analysis.
By evaluating the model $6$ more times in the plausible region and defining a new emulator using these evaluations, we obtain the plots in \ref{subfig:gaussian_0D_2}.
We can now see that the emulator has been greatly improved in the regions of parameter space that were still interesting, and that the implausibility values are only low enough around $x=1.5$ and $x=2.5$, which is as expected.

Obviously, the example given in \autoref{fig:gaussian_0D} is highly simplified and does not even require emulation to solve.
For more complex models with more parameters and outputs, the problem quickly becomes too complicated to cover by a single emulator system.
In \prism, every model output $f_i$ has its own implausibility measure defined, since every emulator system is vastly different from another.
However, all emulator systems (the \textit{emulator}) share the same parameter space and a parameter set $\vec{x}$ must give acceptable results in every emulator system to be considered part of the collection $\mathcal{X}^*$.
Therefore, it is required that we combine the various different implausibility measures together in order to know which parts of parameter space are definitely not contained within $\mathcal{X}^*$.
By maximizing over $I_i(\vec{x})$, one obtains the highest implausibility value that is reached for all these outputs.
This so-called \textit{maximum implausibility measure} is given by
\begin{align*}
    I_{\mathrm{max}, 1}(\vec{x}) &= \max_i \left(I_i(\vec{x})\right).
\end{align*}
This measure can be used to rate the emulated outputs $\E(f(\vec{x}))$ in terms of how well they compare to the comparison data $z$.

However, early on in the emulation process, the emulator systems are still fairly inaccurate due to a low density of model evaluation samples.
This causes these emulator systems to have a high probability of excluding a part of parameter space that should not be excluded or at least currently still contains acceptable choices for $\vec{x}$.
Therefore, one should not select the highest implausibility value, but the second (or third) highest implausibility value as a safety measure early on.
This is then given by
\begin{align*}
    I_{\mathrm{max}, 2}(\vec{x}) &= \max_i \left(I_i(\vec{x}) \backslash \left\lbrace I_{\mathrm{max}, 1}(\vec{x})\right\rbrace\right),\\
    I_{\mathrm{max}, 3}(\vec{x}) &= \max_i \left(I_i(\vec{x}) \backslash \left\lbrace I_{\mathrm{max}, 1}(\vec{x}), I_{\mathrm{max}, 2}(\vec{x})\right\rbrace\right),
\end{align*}
with $I_{\mathrm{max}, 1}(\vec{x})$ ($I_{\mathrm{max}, 2}(\vec{x})$) being the highest (second-highest) implausibility value and `$\backslash$' meaning `except/without'.
Generalizing the functions above, gives the function for the so-called \textit{implausibility cut-off} as
\begin{align}
\label{eq:impl_cut}
    I_{\mathrm{max}, n}(\vec{x}) &= \max_i \left(I_i(\vec{x}) \backslash \left\lbrace I_{\mathrm{max}, 1}(\vec{x}), I_{\mathrm{max}, 2}(\vec{x}), \ldots, I_{\mathrm{max}, n-1}(\vec{x})\right\rbrace\right).
\end{align}

History matching is an iterative process, in which parts of parameter space are removed based on the implausibility values of evaluated parameter sets, which in turn leads to a smaller parameter space to evaluate the model in.
This process is known as \textit{refocusing} \evnote{(described in \citealt{Vernon10})}, and \prism\ uses this process to shrink down parameter space with every step.
At each iteration $I$, the algorithm can be summarized in the following way:
\begin{enumerate}
	\item A Latin-Hypercube design \citep{LHS} of parameter sets $\vec{x}$ is created over the current plausible region $\mathcal{X}_I$, which is based on all preceding implausibility measures (or covers full parameter space if no previous iterations exist);
    \item Each parameter set $\vec{x}$ in this Latin-Hypercube design is evaluated in the model;
    \item The active parameters $\vec{x}_{\mathrm{A},i}$ are determined for every model output (see \ref{subsec:Active parameters and regression});
    \item The corresponding model outputs $f(\vec{x})$ are used to construct a more accurate emulator which is solely defined over the current plausible region $\mathcal{X}_I$;
    \item The implausibility measures are then recalculated over $\mathcal{X}_I$ by using the new emulator systems;
    \item Cut-offs are imposed on the implausibility measures, which defines a new and smaller plausible region $\mathcal{X}_{I+1}$ which should satisfy $\mathcal{X}^* \subset \mathcal{X}_{I+1} \subset \mathcal{X}_I$;
    \item Depending on the user, repeat step 1 to 6 unless a certain condition is met;
    \item \prism\ stops improving the emulator and gives back its final results.
\end{enumerate}

We have multiple different reasons to believe that every iteration will decrease the part of parameter space that is still considered plausible:
\begin{enumerate}
	\item Increasing the density of model evaluations in parameter space gives the emulator systems more information and therefore increases their accuracy.
	However, since the emulator systems are not required to have a high accuracy in order to exclude parts of parameter space, it is important to only gradually increase the density to allow for higher evaluation rates.
	The algorithm above ensures that this happens;
	\item Reducing the plausible region of parameter space might make it easier for the emulator systems to emulate the model output and therefore make the function smoother;
    \item All parameters that were not considered active in earlier iterations due to them not accounting for much of the variance, may become active in the current one, allowing more variance to be captured by the emulator systems.
\end{enumerate}

The use of continued refocusing is very useful, but it also has its complications.
For example, the only way of knowing if a parameter set $\vec{x}$ is contained within the plausible region, is by calculating the implausibility values for all its model outputs and then using \ref{eq:impl_cut} to see if it satisfies all imposed implausibility cut-offs.
Although evaluating a single parameter set can be done very quickly, obtaining a reasonably sized Latin-Hypercube design for step 1 in the algorithm requires the evaluation of thousands, tens of thousands and maybe even hundreds of thousands of parameter sets.
This means that \prism\ must be fast and efficient in evaluating model parameter sets.
A full detailed description of \prism\ is given in \ref{sec:PRISM}.

\section{PRISM pipeline}
\label{sec:PRISM}
\subsection{Structure}
\label{subsec:Structure}
\begin{figure*}
\begin{center}
	\includegraphics[width=\textwidth]{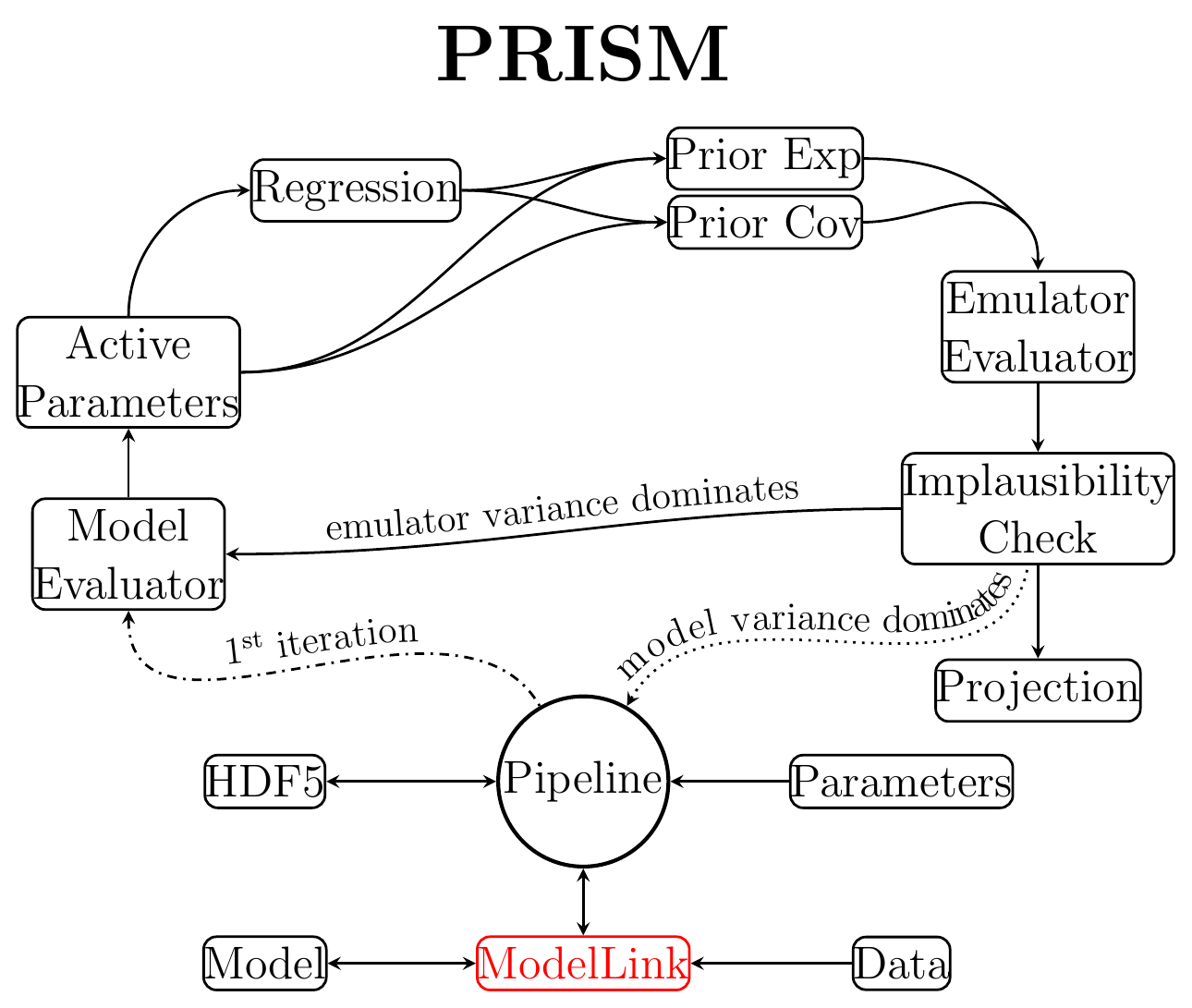}
    \caption{The structure of the \prism\ pipeline.}
    \label{fig:PRISM}
\end{center}
\end{figure*}

The overall structure of \prism\ (as of version $1.0.0$) can be seen in \autoref{fig:PRISM} and will be discussed below.
The \texttt{Pipeline} object plays a key-role in the \prism\ framework as it governs all other objects and orchestrates their communications and method calls.
It also performs the process of history matching and refocusing (as explained in \ref{subsec:History matching}).
It is linked to the model by a user-written \texttt{ModelLink} object (see \ref{subsec:ModelLink}), allowing the \texttt{Pipeline} object to extract all necessary model information and call the model.
In order to ensure flexibility and clarity, the \prism\ framework writes all of its data to one or several HDF5-files\footnote{\url{https://portal.hdfgroup.org/display/HDF5/HDF5}} using \textsw{h5py} \citep{h5py}, as well as \numpy\ \citep{NumPy}.

The analysis of a provided model and the construction of the emulator systems for every output value, start and end with the \texttt{Pipeline} object.
When a new emulator is requested, the \texttt{Pipeline} object creates a large Latin-Hypercube design \citep{LHS} of model evaluation samples to get the construction of the first iteration of the emulator systems started.
To ensure that the maximum amount of information can be obtained from evaluating these samples, we have written a custom Latin-Hypercube sampling code based on the work by \citet{JH08}.
This produces Latin-Hypercube designs that attempt to satisfy both the \textit{maximin} criterion \citep{JMY90,MM95} as well as the \textit{correlation} criterion \citep{IC82,Owen94,Tang98}.
This code is customizable through \prism\ and publicly available in the \mypackage\footnote{\url{https://github.com/1313e/e13Tools}} \python\ package.

This Latin-Hypercube design is then given to the \texttt{Model Evaluator}, which through the provided \texttt{ModelLink} object evaluates every sample.
Using the resulting model outputs, the \texttt{Active Parameters} for every emulator system can now be determined.
Next, depending on the user, polynomial functions will be constructed by performing an extensive \texttt{Regression} process (see \ref{subsec:Active parameters and regression}) for every emulator system, or this can be skipped in favor of a sole Gaussian analysis (faster, but less accurate).
No matter the choice, the emulator systems now have all the required information to be constructed, which is done by calculating the \texttt{Prior Expectation} and \texttt{Prior Covariance} values for all evaluated model samples ($\E(D_i)$ and $\var(D_i)$ in \ref{eq:adj_exp} and \ref{eq:adj_var}).

Afterward, the emulator systems are fully constructed and are ready to be evaluated and analyzed.
Depending on whether the user wants to prepare for the next emulator iteration or create a projection (see \ref{subsec:Projection}), the \texttt{Emulator Evaluator} creates one or several Latin-Hypercube designs of emulator evaluation samples, and evaluates them in all emulator systems, after which an \texttt{Implausibility Check} is carried out.
The samples that survive the check can then either be used to construct the new iteration of emulator systems by sending them to the \texttt{Model Evaluator}, or they can be analyzed further by performing a \texttt{Projection}.
The \texttt{Pipeline} object performs a single cycle by default (to allow for user-defined analysis algorithms), but can be easily set to continuously cycle.

In addition to the above, \prism\ also features a high-level \textit{Message Passing Interface} (MPI, \citealt{mpi-1,mpi-2}) implementation using the \python\ package \textsw{mpi4py} \citep{mpi4py}.
All emulator systems in \prism\ can be constructed independently from each other, in any order, and only require to communicate when performing the implausibility cut-off checks during history matching.
Additionally, since different models and/or architectures require different amounts of computational resources, \prism\ can run on any number of MPI processes (including a single one in serial to accommodate for OpenMP codes) and the same emulator can be used on a different number of MPI processes than it was constructed on (e.g., constructing an emulator using $8$ MPI processes and reloading it with $6$).
More details on the MPI implementation and its scaling can be found in \ref{app:MPI implementation}.

In the following, we discuss some of the components of \prism\ in more detail.

\subsection{Making the model-link}
\label{subsec:ModelLink}
In order to start analyzing models, the most important ingredient is the model itself.
This model must be connected to the \prism\ framework such that it is capable of distilling all the information it requires about the model.
And, this will allow the framework to call the model for a certain combination of model parameter values, after which \prism\ is provided with all the model outputs that require emulation.
The difficulty here is that every model is different: they can be written in different languages; must be called/executed differently; have different computational requirements; their I/O has a different format; and some will require certain external files that others do not need.
As everybody has the most knowledge about their own model, and knows best how the model works, we think that this should always be the only real necessary requirement for using \prism\ in addition to basic \python\ knowledge, without having to perform large conversions or transformations to make it work.

To help users with the task of connecting their model to the \prism\ framework, we have written the \modellink\ abstract base class.
An abstract base class in \python\ provides the general structure that the subclass that ``links'' the model to the \prism\ framework, needs to have in order to allow \prism\ to work with the model.
An example of a \modellink\ subclass can be found in \ref{app:Writing a ModelLink subclass}.

The \modellink\ class has several important properties to make it as easy as possible for the user of \prism\ to use it:
\begin{enumerate}
	\item It treats the linked model as a ``black box'': it takes a set of values for all model parameters, performs some unknown operations and returns the model outputs for which emulator systems need to be constructed;
    \item Several flags can be set to customize \prism\ how to call the model (some models require to be called in MPI, while others require to load in data first and are therefore best called for all samples at once);
	\item The \modellink\ class provides multiple different options for importing the details about the model parameters and the observational data that needs to be compared to the emulated model outputs;
\end{enumerate}

The reason for treating a model as a ``black box'' is that the \prism\ framework does not require knowledge of the actual structure of the linked model in order to perform the analysis, since its purpose is to identify the structure and behavior of the model and show this to the user.
By not knowing how the model works and what its behavior is, we ensure that the outcome of the analysis is in no way influenced by any knowledge that the framework has obtained from the model besides that which is required.
We therefore also guarantee to provide the user with an unbiased analysis result, which describes what the framework has identified as the behavior of the model and what characteristics it has recognized and extracted.
\evnote{This also greatly increases the generality of \prism, making it more universally applicable.}

Additionally, it may be required for an emulator construction or analysis process to be interrupted mid-way, and restored later.
Since a constructed emulator describes the workings of a specific instance of a model (and therefore a specific \modellink\ subclass), mismatches could occur if a different model were used.
In order to prevent this from happening, \prism\ supports the use of user-defined naming schemes for \modellink\ subclasses as part of the emulator's meta-data.
This can be used to give a different name to each \modellink\ subclass (assigned automatically if not done manually), which ensures that \prism\ does not link a constructed emulator to the wrong model(link).

\subsection{Active parameters and regression}
\label{subsec:Active parameters and regression}
\subsubsection{Active parameters}
\label{subsubsec:Active parameters}
In \ref{subsec:Emulation technique}, we discussed that during the early iterations of the emulator systems, it is not unreasonable to find that a subset of the model parameters can explain most of the variance in the model function.
Therefore, we introduced a set of \textit{active parameters} $\vec{x}_{\mathrm{A},i}$ for each model output $i$, where we try to explain as large an amount of variance in the model function $f_i(\vec{x})$ using as few parameters as possible.
For each of the selected model outputs, \prism\ uses the set of model evaluation samples for the specific iteration, to reduce the set of potentially active model parameters $\vec{x}_{\mathrm{A,pot}}$ by \textit{backwards step-wise elimination} (cf., starting with all polynomial terms and repeating the fit for less terms every time).
The set of potentially active model parameters is defined by the user, and determines which model parameters are allowed to become active for a specific iteration, thereby forcing all other model parameters to be passive regardless of their actual importance.
We have three reasons for why we want to divide the model parameters into \textit{active} and \textit{passive parameters}.

\begin{enumerate}
    \item For early iterations of the emulator systems (e.g., the first iteration), the density of model evaluation samples in parameter space is very low.
    All model parameters have a different influence on the outcome of the model, where some can easily swing the outcome from one end of the scale to the other, while others only make very small contributions.
    Parameters with small contributions cannot be resolved in parameter space when only a low density of evaluated model samples is available.
    Including these (passive) parameters in the analysis can only introduce more uncertainties that are hard to parametrize, as opposed to the predictable contribution to the uncertainty when they are excluded;

    \item It is not necessary to include those parameters that make small contributions to the model output when the plausible region of parameter space $\mathcal{X}$ is still large.
    The contributing parameters can be used to initially start reducing the relevant part of parameter space, making it smaller with every iteration while increasing the density of evaluated model samples in the plausible region.
    \evnote{This may also remove the possibly erratic behavior of uninteresting parts of parameter space from the emulator, which could be hard but unnecessary to emulate accurately.}
    Doing so will allow \prism\ to accurately pinpoint the values that these parameters should take, at which point the parameters with smaller contributions become important.
    This in turn increases the number of important model parameters and thus the number of active parameters;

    \item The more parameters for which polynomial terms $g_{ij}$ and their coefficients $\beta_{ij}$ need to be determined, the slower the regression process (see \ref{subsubsec:Regression}) will be, given that the total number of polynomial terms is
    \evnote{\begin{align*}
        N_{\mathrm{terms}} &= \sum_{n=0}^{N_{\mathrm{order}}}\left({N_{\mathrm{par}}\choose n}\right) = {N_{\mathrm{par}}+N_{\mathrm{order}}\choose N_{\mathrm{order}}},
    \end{align*}}
    where $\choose$ is the binomial coefficient.
    Even though one might argue that this only slows down the construction of an emulator system (which is done once), it will also increase the amount of time it takes to evaluate that emulator system.
    Since more polynomial terms need to be calculated and scaled by their coefficients, it takes longer to carry out a single evaluation.
    This is important, because the earlier iterations are evaluated much more than the later ones, given that a parameter set is only evaluated in an iteration if it was found plausible in the previous iteration.
    Because of this, we put heavy emphasis on determining which model parameters should be considered active and which ones should not.
\end{enumerate}

The active parameters for each model output $i$ are determined in the following way.
Suppose that we want to fit a function $h_i(\vec{x})$ to the $i^{\mathrm{th}}$ output of a model.
As a start, \prism\ takes all linear terms in the potentially active parameters $\vec{x}_{\mathrm{A,pot}}$, which defines $h_i(\vec{x})$ as
\begin{align*}
    h_i(\vec{x}) &= \vec{\alpha}\cdot\vec{x}_{\mathrm{A,pot}},\\
    &= \alpha_1x_{\mathrm{A,pot},1} + \alpha_2x_{\mathrm{A,pot},2} + \ldots + \alpha_nx_{\mathrm{A,pot},n},
\end{align*}
where $\vec{\alpha}$ is the vector of polynomial coefficients and $n$ is the number of potentially active parameters.

This function $h_i(\vec{x})$ is then fitted to the set of known model realizations $D_i$ using an ordinary least-squares fit, which yields the vector $\vec{\alpha}$.
Every polynomial term (model parameter) that has a non-zero $\alpha$ value, is automatically considered to be an active parameter, which defines a vector of linearly active parameters $\vec{x}_{\mathrm{A,lin}}$.

If a parameter $x_j$ was not selected, then \prism\ will fit all corresponding polynomial terms up to third order (by default) plus all linear terms that were selected, to the model samples $D_i$ again.
This changes $h_i(\vec{x})$ to
\begin{align*}
    h_i(\vec{x}) &= \vec{\alpha}\cdot\vec{x}_{\mathrm{A,lin}} + \left(\vec{\beta}\cdot\vec{x}_{\mathrm{A,pot}}\right)x_j + \left(\vec{\gamma}\cdot\vec{x}_{\mathrm{A,pot}}\otimes\vec{x}_{\mathrm{A,pot}}\right)x_j + \ldots,
\end{align*}
with $\otimes$ being the \textit{outer product}.
If any value of $\vec{\beta}$ or $\vec{\gamma}$ (or others if higher order terms are included) after the fit is non-zero, parameter $x_j$ will be considered active as well.
This will be done for all parameters in $\vec{x}_{\mathrm{A,pot}}$ that are not in $\vec{x}_{\mathrm{A,lin}}$, which in the end yields the set of active parameters $\vec{x}_{\mathrm{A},i}$ for output $i$.

By performing an ordinary least-squares fit to only the linear terms, we allow those parameters that have high contributions to the model output, given the current plausible region $\mathcal{X}$, to be easily recognized as \textit{active}.
However, since it is perfectly possible that a certain model parameter only has a significant influence on the model output if it is scaled by another parameter, \prism\ performs the fit again with all relevant polynomial terms to check if this is the case.
The method described above cannot possibly extract \textit{only} those parameters that should be considered active, but it does ensure that at least all parameters that should be considered as such are.
Since not extracting an active parameter bears a higher cost than considering a passive parameter active, we wrote the algorithm for the active parameters in this way.

When the active parameters $\vec{x}_{\mathrm{A},i}$ have been determined for all outputs, a full third order polynomial function can be fitted to all active parameters.
This is done during the (optional) \textit{regression} process, described in the following section.
For performing the above mentioned operations, we make use of the \python\ packages \sklearn\ \citep{Sklearn} and \mlxtend\ \citep{Mlxtend}.

\subsubsection{Regression}
\label{subsubsec:Regression}
Following the active parameter determination step, the surviving parameters $\vec{x}_{\mathrm{A},i}$ are then analyzed by the regression process and the full regression function $r_i(\vec{x})$ (\ref{eq:r}) can be determined for every emulator system.
This process is entirely optional and can be skipped in favor of a sole Gaussian analysis, which is much faster, but also less accurate.
Obtaining the regression function also allows the user to receive more information on the workings of the model, given that it provides the polynomial terms and their coefficients that lead to a specific data point (and should therefore have a logical form).
For this reason, the regression function and the process of obtaining it are considered to be of vital importance.

In order to obtain this regression function, we have to determine which deterministic functions of the active parameters $\vec{x}_{\mathrm{A},i}$ we are going to use and what their polynomial coefficients are.
For this, we make use of \textit{forward step-wise linear regression}.
Forward step-wise linear regression works by first determining all polynomial terms for all active parameters $\vec{x}_{\mathrm{A},i}$ of an emulator system.
This gives the set of deterministic functions $\vec{g}_i$ as
\begin{align*}
    \vec{g}_i &= \lbrace\vec{x}_{\mathrm{A},i},\vec{x}_{\mathrm{A},i}\otimes\vec{x}_{\mathrm{A},i},\vec{x}_{\mathrm{A},i}\otimes\vec{x}_{\mathrm{A},i}\otimes\vec{x}_{\mathrm{A},i},\ldots\rbrace.
\end{align*}

Then, an ordinary least-squares fit is performed for every individual term in $\vec{g}_i$, after which the corresponding mean squared errors are determined.
Out of the $N_{\mathrm{terms}}$ fits, the polynomial term that yielded the lowest mean squared error is then considered to be a part of the final regression function.
After this step, $N_{\mathrm{terms}}-1$ fits are performed using the current regression function and every individual non-chosen term in $\vec{g}_i$.
From these fits, the second polynomial term of the regression function can be determined by calculating the mean squared errors of every fit again.

By applying this algorithm repeatedly ($N_{\mathrm{terms}}$ times in total), the regression function improves with every iteration.
However, one runs into the risk of over-fitting the regression function, given that adding more degrees-of-freedom (polynomial terms) generally improves its score while not really improving the fit itself.
In order to make sure that this does not happen, \prism\ additionally requests that the chosen form of the regression function has the best cross-validation performance among all possible forms.
The ($k$-fold) cross-validation \citep{Stone74} of an ordinary least-squares fit is the process of dividing the full training set of samples up into $k$ parts, using all except one as the training set and using the remaining one as the test set (and doing this $k$ times).
Since over-fitting usually causes the fit to become completely different with the slightest changes to the training set, over-fitted regression functions will not perform as well as those that are not, as demonstrated by \citet{Cawley10}.
By using cross-validation, in the end, the regression function will only contain those polynomial terms $g_{ij}$ that make significant contributions to it, which gives us all non-zero coefficients $\beta_{ij}$.

\subsection{Variances}
\label{subsec:Variances}
In order to determine how (un)likely the emulator estimates it is that any given evaluation sample would yield a model realization that would be marked as `acceptable', one has to calculate the implausibility value of this sample for every emulator system (using \ref{eq:impl_sq}) and perform the implausibility cut-off check (with \ref{eq:impl_cut}).
Looking at \ref{eq:impl_sq}, we find that it is required to know what the adjusted expectation and variance values of the emulator are, in addition to the model discrepancy and observational variances.
While the latter two are to be provided externally in the \modellink\ subclass, the adjusted values need to be calculated for every evaluation sample as mentioned in \ref{subsec:History matching}.
Below, we discuss the importance and meanings of the various different variances in determining the adjusted values, as well as how to extract the residual variance $\sigma_i^2=\sigma_{u_i}^2+\sigma_{w_i}^2$ from the regression term.

\subsubsection{Adjusted values}
\label{subsubsec:Adjusted values}
Calculating the adjusted values of a parameter set $\vec{x}$ using \ref{eq:adj_exp} and \ref{eq:adj_var} requires the prior expectation and (co)variance values of various terms.
The adjusted expectation value consists of the prior expectation of the unknown model output $\E(f_i(\vec{x}))$ and its adjustment term.
According to \ref{eq:prior_exp}, the prior expectation only contains contributions from the regression term, given that a Gaussian is always centered around zero and the passive term has a constant variance (and therefore both have an expectation of zero).
Because of this, the prior expectation value is a measure of how much information/variance of the model output is captured by the emulator system, being zero when regression is not used (since no information is captured).
This prior expectation is then adjusted by the expectation adjustment term, in which the emulator system takes into account the knowledge about the behavior of the model and itself.

The corresponding adjusted variance value is similarly obtained, combining the prior variance with its adjustment term.
\ref{eq:prior_cov} shows that the prior variance of a sample $\vec{x}$ is dominated by $\sigma_{u_i}^2+\sigma_{w_i}^2=\sigma_i^2$, which is the residual variance of the regression process (or the square of the user-provided Gaussian error in case no regression is used).
The residual variance $\sigma_i^2$ is all the variance that could not be explained in the regression function (its mean squared error), which is then split up into the Gaussian variance and the passive variance according to \evnote{(for which we follow the form given in \citealt{Vernon10})}
\begin{align*}
\sigma_{u_i}^2 = (1-\omega_{\mathrm{pas},i})\cdot\sigma_i^2\ &&\text{and}&&\ \sigma_{w_i}^2 = \omega_{\mathrm{pas},i}\cdot\sigma_i^2,
\end{align*}
with $\omega_{\mathrm{pas},i}$ being the fraction of model parameters that are passive for model output $i$.
The variance adjustment term accounts for the lack of available knowledge.

Both adjustment terms require the $\cov\left(f_i(\vec{x}), D_i\right)\cdot\var(D_i)^{-1}$ term, which can be described as a measure of the density of the available knowledge and its proximity to $\vec{x}$.
If $\vec{x}$ would be far away from all known model evaluation samples, this term would decrease in value since the relevance of the available knowledge for $\vec{x}$ is low.
Although this makes the calculations for the adjusted values look very similar, their underlying meanings are distinctively different.
The expectation adjustment term describes the emulator system's tendency to either overestimate or underestimate the model output $i$ in parts of parameter space that are currently known ($\left(D_i-\E(D_i)\right)$), which is then scaled by its `relevance' and added to the prior expectation.
On the other hand, the variance adjustment term decreases the adjusted variance with increasing knowledge.

An example of these differences is when one considers the parameter set $\vec{x}$ to be equal to one of the known model evaluation samples, say $\vec{x}^{(1)}$.
In this scenario, the adjusted expectation $\E_{D_i}(f_i(\vec{x}^{(1)}))$ should be equal to $f_i(\vec{x}^{(1)})\equiv D_{i,1}$, since the value is already known.
Given that $\cov(f_i(\vec{x}^{(1)}), D_i)\cdot\var(D_i)^{-1}=\left(1, 0, 0, \ldots, 0\right)$, because $\cov(f_i(\vec{x}^{(1)}), D_i)$ is the first column/row of $\var(D_i)$, we can see that this is true:
\begin{align*}
\E_{D_i}(f_i(\vec{x}^{(1)})) &= \E(f_i(\vec{x}^{(1)}))+\left(1, 0, 0, \ldots, 0\right)\cdot\left(D_i-\E(D_i)\right),\\
&= \E(f_i(\vec{x}^{(1)}))+\left(D_{i,1}-\E(D_{i,1})\right),\\
&= f_i(\vec{x}^{(1)}).
\end{align*}
Using the same approach for calculating the adjusted variance $\var_{D_i}(f_i(\vec{x}^{(1)}))$ gives its value as zero, because the emulator is certain that the adjusted expectation value is correct.
\evnote{Note that this is only true if there are no passive parameters, as otherwise both the adjusted expectation and adjusted variance will be shifted by a value of the order of $\sigma_{w_i}$.}

\subsubsection{Model discrepancy variance}
\label{subsubsec:Model discrepancy variance}
In \ref{subsec:Uncertainty analysis}, we discussed the main contributions to the overall uncertainty of the emulation process, with the most important contribution being the \textit{model discrepancy variance}.
The model discrepancy variance describes all uncertainty about the correctness of the model output that is caused by the model itself.
This includes the accuracy of the code implementation, completeness of the inclusion of the involved physics, made assumptions and the accuracy of the output itself, amongst others.
Here, we would like to describe how the model discrepancy variance is treated and in what ways it affects the results of \prism.

The model discrepancy variance is extremely important for the emulation process, as it is a measure of the quality of the model to emulate.
\prism\ attempts to make a perfect approximation of the emulated model that covers the plausible regions of parameter space, that would be reached if the adjusted emulator variance $\var_{D_i}(f_i(\vec{x}))$ is equal to zero for all $\vec{x}$.
In this case, the emulator and the emulated model should become indistinguishable, which converts the implausibility measure definition given in \ref{eq:impl_sq} to
\begin{align}
\label{eq:red_impl_sq}
I_i^2(\vec{x}) &= \frac{\left(\E_{D_i}(f_i(\vec{x}))-z_i\right)^2}{\var(\epsilon_{\mathrm{md}, i})+\var(\epsilon_{\mathrm{obs}, i})},
\end{align}
where $\E_{D_i}(f_i(\vec{x}))$ should be equal to $f_i(\vec{x})$.

From this, it becomes clear that if the model discrepancy variance $\var(\epsilon_{\mathrm{md}, i})$ is incorrect, evaluating \ref{eq:red_impl_sq} will result in the (im)plausible region of parameter space being described improperly.
This means that the final emulator is defined over a different region of parameter space than desired, $\mathcal{X}^*\neq\mathcal{X}_{\mathrm{final}}$, where $\mathcal{X}$ is the part of parameter space that is still plausible.
When the model discrepancy variance is generally higher than it should be, this will often result in the emulator not converging as far as it could have ($\mathcal{X}^*\subset\mathcal{X}_{\mathrm{final}}$), while the opposite will likely miss out on important information ($\exists\vec{x}\in\mathcal{X}^*: \vec{x}\notin\mathcal{X}_{\mathrm{final}}$).

Because of the above, overestimating the model discrepancy variance is much less costly than underestimating its value.
It is therefore important that this variance is properly described at all times.
However, since the description of the model discrepancy variance can take a large amount of time, \prism\ uses its own default description in case none was provided, which is defined as $\var(\epsilon_{\mathrm{md}, i})=\left(z_i/6\right)^2$.
If one assumes that a model output within half of the data is considered to be acceptable, with acceptable being defined as the $3\sigma$-interval, then the model discrepancy variance is obtained as
\begin{align*}
\left[z_i-3\sigma, z_i+3\sigma\right] &= \left[\frac{1}{2}z_i, \frac{3}{2}z_i\right],\\
6\sigma &= z_i,\\
\sigma &= \frac{z_i}{6},\\
\var(\epsilon_{\mathrm{md}, i}) &= \sigma^2 = \left(\frac{z_i}{6}\right)^2.
\end{align*}

This description of the model discrepancy variance usually works well for simple models, and acts as a starting point within \prism.
When models become bigger and more complex, it is likely that such a description is not enough.
Given that the model discrepancy variance is unique to every model and might even be different for every model output, \prism\ cannot possibly cover all scenarios.
It is therefore advised that the model discrepancy variance is provided externally.

\section{Basic usage and application}
\label{sec:Basic usage}
In this section, we discuss the basic usage of \prism, and give an overview of several applications showcasing what \prism\ can do.
As \prism\ is built to replace MCMC as the standard for analyzing models, but co-exist with MCMC when it comes to constraining and calibrating one, we will show how \prism\ and MCMC methods can be used together.

\subsection{Minimal example}
\label{subsec:Minimal example}
Here, we show a minimal example on how to initialize and use the \prism\ pipeline.
First, we have to import the \pipeline\ class and a \modellink\ subclass:
\begin{lstlisting}[style=defaultpython]
	|\iin| from prism import Pipeline
	|\iin| from prism.modellink import GaussianLink
\end{lstlisting}
Normally, one would import a custom-made \modellink\ subclass, but for this example we use one of the two \modellink\ subclasses that come with the package (see \ref{app:Writing a ModelLink subclass} for the basic structure of writing a custom \modellink\ subclass).

Next, we have to initialize our \modellink\ subclass, the \gaussianlink\ class in this case.
In addition to user-defined arguments, every \modellink\ subclass takes two optional arguments, \verb|model_parameters| and \verb|model_data|.
The use of either one will add the provided parameters/data to the default parameters/data defined in the class.
Since the \gaussianlink\ class does not have default data defined, we have to supply it with some data during initialization:
\begin{lstlisting}[style=defaultpython]
	|\iin| model_data = {
    	3: [3.0, 0.1],   # f(3) = 3.0 +- 0.1
    	5: [5.0, 0.1],   # f(5) = 5.0 +- 0.1
        7: [3.0, 0.1]}   # f(7) = 3.0 +- 0.1
    |\iin| modellink_obj = GaussianLink(model_data=model_data)
\end{lstlisting}
Here, we initialized the \gaussianlink\ class by giving it three data points and using its default parameters.
We can check this by looking at its representation:
\begin{lstlisting}[style=defaultpython]
    |\iin| modellink_obj
    |\out| GaussianLink(
    	model_parameters={'A1': [1.0, 10.0, 5.0],
        				  'B1': [0.0, 10.0, 5.0],
                          'C1': [0.0, 5.0, 2.0]},
        model_data={7: [3.0, 0.1],
        			5: [5.0, 0.1],
                    3: [3.0, 0.1]})
\end{lstlisting}

The \pipeline\ class takes several optional arguments, which are mostly paths and the type of \emulator\ that must be used.
It also takes one mandatory argument, which is an instance of the \modellink\ subclass to use.
We have already initialized it, so we can now initialize the \pipeline\ class:
\begin{lstlisting}[style=defaultpython]
	|\iin| pipe = Pipeline(modellink_obj)
    |\iin| pipe
    |\out| Pipeline(
    	GaussianLink(
        	model_parameters={
            	'A1': [1.0, 10.0, 5.0],
            	'B1': [0.0, 10.0, 5.0],
                'C1': [0.0, 5.0, 2.0]},
            model_data={7: [3.0, 0.1],
            			5: [5.0, 0.1],
                        3: [3.0, 0.1]}),
        working_dir='prism_0')
\end{lstlisting}
Since we did not provide a working directory for the \pipeline\ and none already existed, it automatically created one (\textsc{prism\_0}).

\prism\ is now ready to start emulating the model.
The \pipeline\ allows for all steps in a full cycle shown in \autoref{fig:PRISM} to be executed automatically:
\begin{lstlisting}[style=defaultpython]
	|\iin| pipe.run()
\end{lstlisting}
which is equivalent to:
\begin{lstlisting}[style=defaultpython]
	|\iin| pipe.construct(analyze=False)
    |\iin| pipe.analyze()
    |\iin| pipe.project()
\end{lstlisting}
This will construct the next iteration (first in this case) of the emulator, analyze it to check if it contains plausible regions and then make projections of all active parameters.

The current state of the \pipeline\ object can be viewed by calling the \verb|details()| user-method (called automatically after most user-methods), which gives an overview of many properties that the \pipeline\ object contains.

This is all that is required to construct an emulator of the model of choice.
All user-methods, with one exception\footnote{The \textit{evaluate()}-method of the \pipeline\ class takes a parameter set as an input argument.}, solely take optional arguments and perform the operations that make the most sense given the current state of the \pipeline\ object.
These arguments allow the user to modify the performed operations, like reconstructing/reanalyzing previous iterations, projecting specific parameters, evaluating the emulator and more.

For those interested in a small overview of how to write a \modellink\ subclass, we refer to \ref{app:Writing a ModelLink subclass}.

\subsection{Visualizing model dispersion}
\label{subsec:Projection}
Using the minimal example from \ref{subsec:Minimal example}, we can construct an emulator that constrains the Gaussian model wrapped in the \gaussianlink\ class.
Given that images can provide much more insight into the emulator's performance than numbers, we would like to make some plots showcasing how the emulator is doing.
However, now that we are using a model that uses more than one parameter, we can no longer use the same method as in \autoref{fig:gaussian_0D} for this.
Therefore, a different form of visualization is required.

To solve this problem, we use \textit{projections}; \evnote{a method described by \citet{Vernon10,Vernon18} for visualizing the emulator's performances}.
Projections are three dimensional figures made for active parameters that allow for the behavior of the model to be studied\evnote{, and have been used extensively several times in the past \citep{Vernon10,Andrianakis15,Andrianakis16,Andrianakis17,Vernon18}}.
In the following, we describe how these projections are made and what information can be derived from them.
In order to properly visualize the behavior of the model, we created special colormaps for \prism, which are described in \ref{app:Colormaps}.

\subsubsection{Dispersing model behavior}
\label{subsubsec:Projection figures}
\begin{figure*}
\begin{center}
	\subfloat{\includegraphics[width=0.49\linewidth]{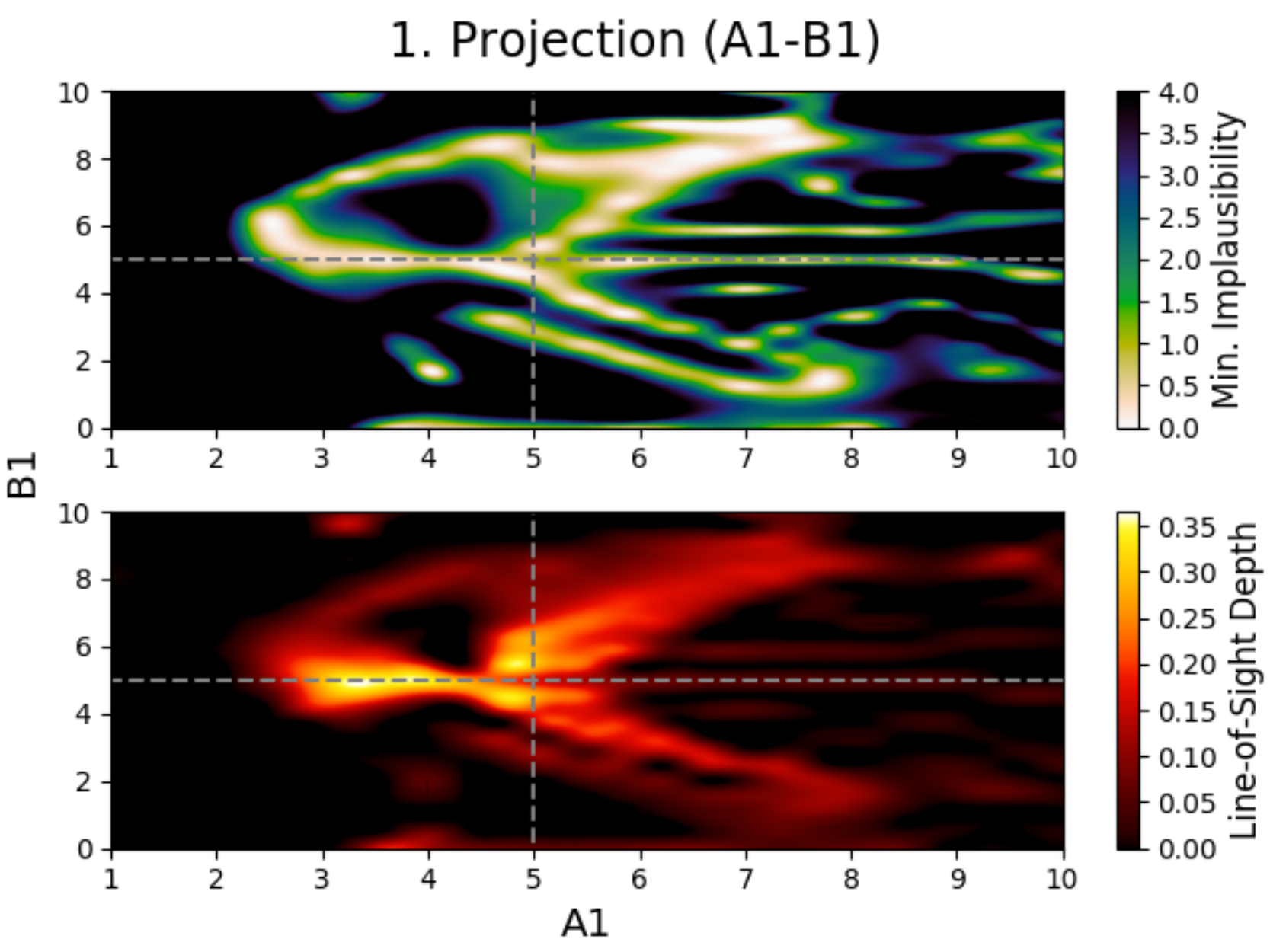}}
    \subfloat{\includegraphics[width=0.49\linewidth]{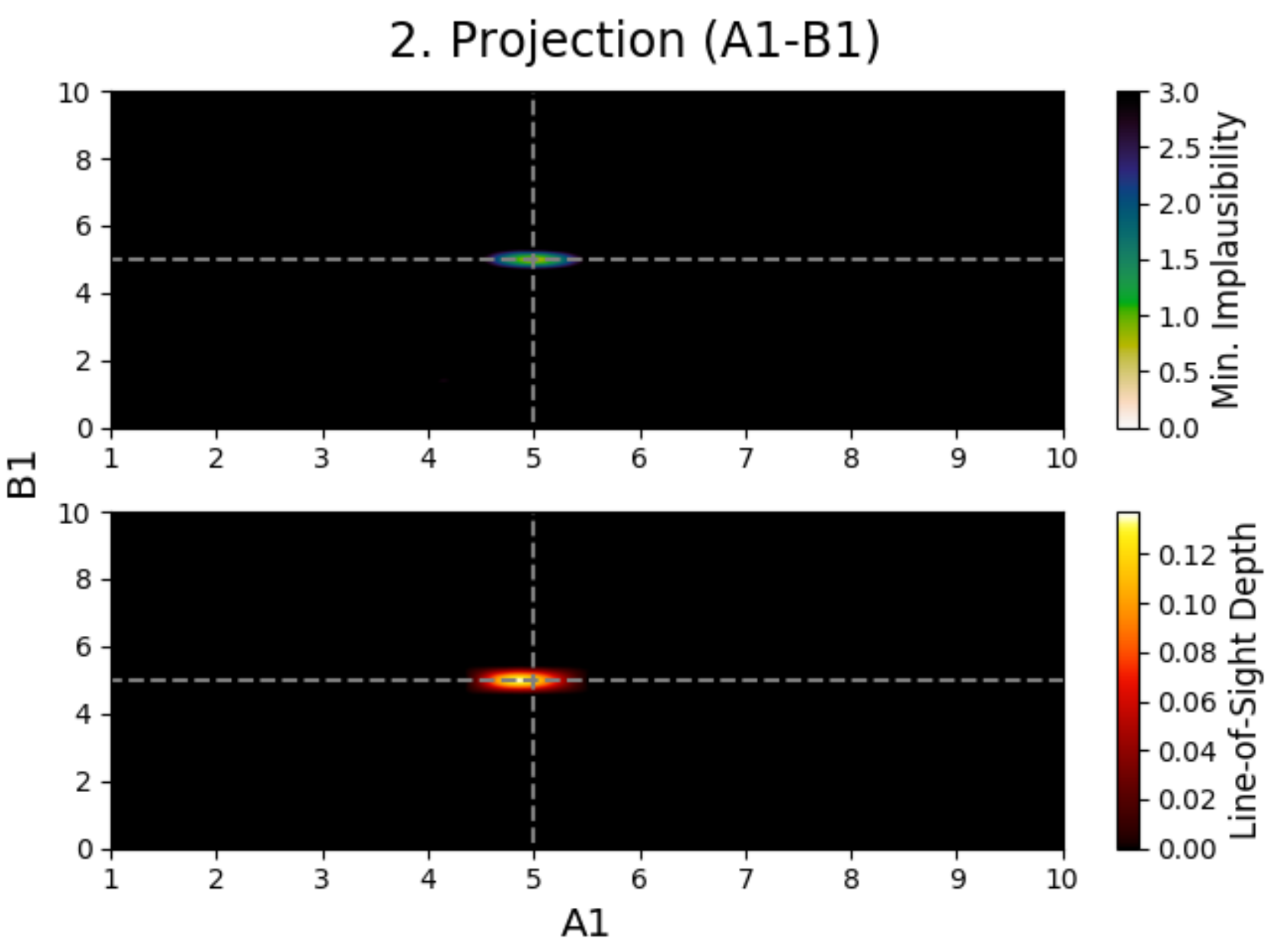}}\\
    \subfloat{\includegraphics[width=0.49\linewidth]{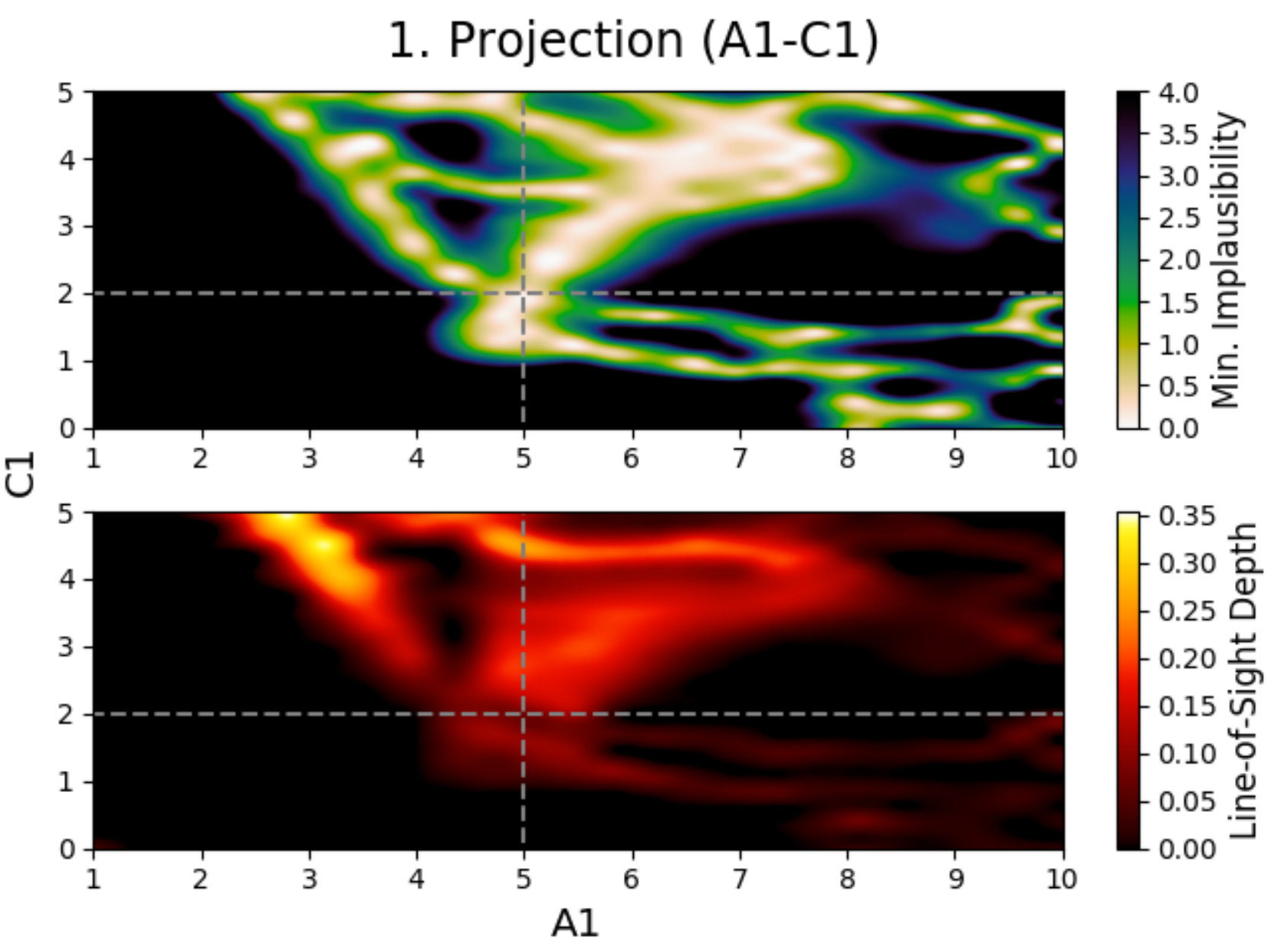}}
    \subfloat{\includegraphics[width=0.49\linewidth]{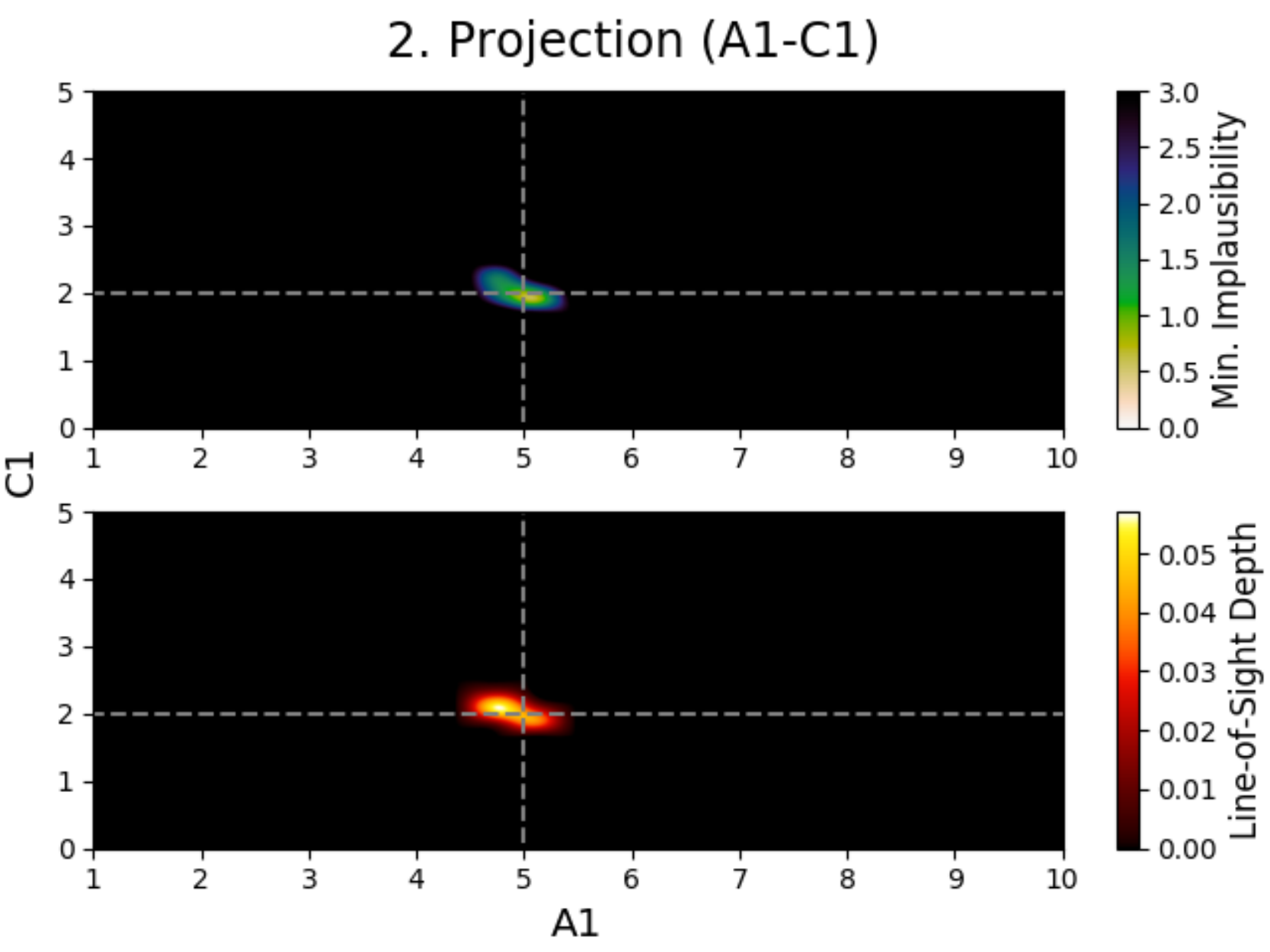}}\\
    \subfloat{\includegraphics[width=0.49\linewidth]{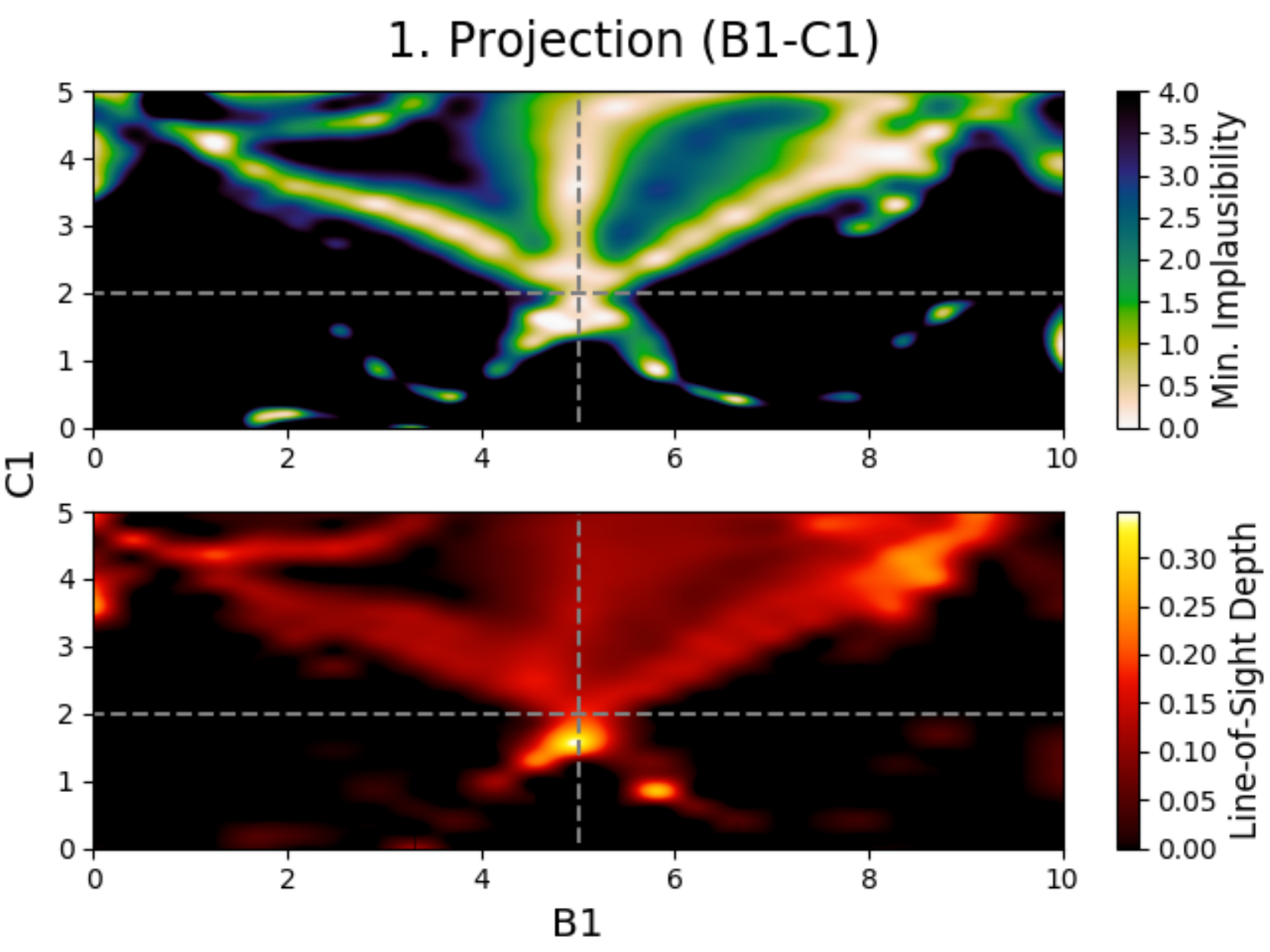}}
    \subfloat{\includegraphics[width=0.49\linewidth]{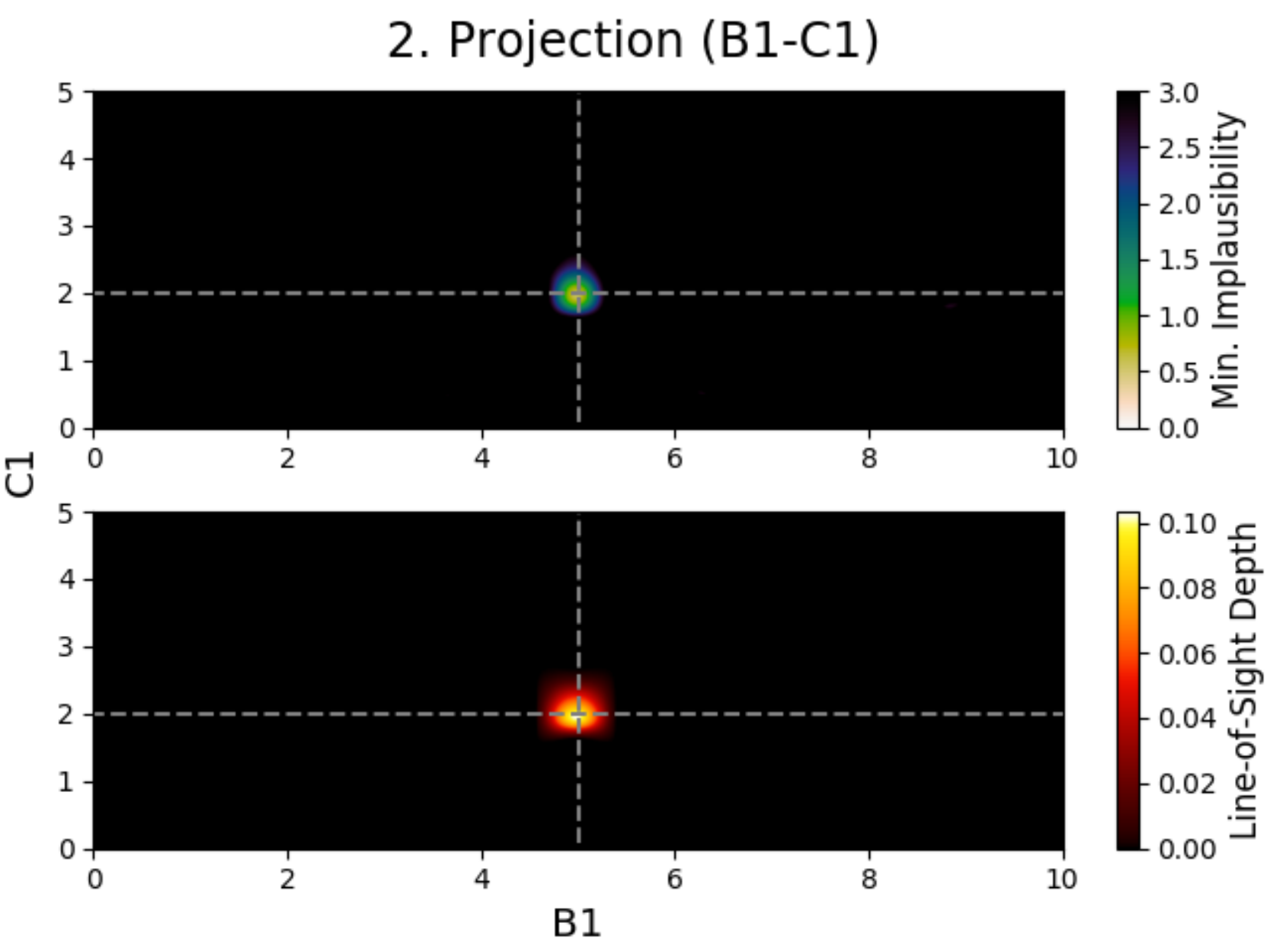}}
	\caption{Projection figures of the emulator of the \gaussianlink\ class used in \ref{subsec:Minimal example}, where the Gaussian is defined as $f(x)=A_1\cdot\exp\left(-\frac{(x-B_1)^2}{2C_1^2}\right)$.
    Every combination of two active model parameters generates a projection figure, which consists of two subplots.
    A projection figure is created by analyzing a grid of $25\times25$ points for the plotted parameters, where the values for the remaining parameters in every individual grid point are given by a Latin-Hypercube design of $250$ samples.
    The results of all samples in every grid point are then processed to yield a single result per grid point, which is independent of the non-plotted parameters.
    \textbf{Left column:} First emulator iteration, $150$ model evaluations, $4.62\%$ of parameter space remaining.
    \textbf{Right column:} Second emulator iteration, $1,110$ model evaluations, $0.0312\%$ of parameter space remaining.
    \textbf{Top subplot:} The minimum implausibility value (at the first cut-off) that can be reached for any given value combination of the plotted parameters.
    \textbf{Bottom subplot:} The fraction of samples (``line-of-sight depth'') that is plausible for any given value combination of the plotted parameters.
    \textbf{Gray lines:} Estimates of the plotted parameters, which only show up if the user provided them.
    Note that the first emulator iteration used a wildcard, while the second one did not.
    See \autoref{fig:2D_gaussian_projections} in \ref{app:Figures and Tables} for the 2D projection figures of every model parameter.}
    \label{fig:3D_gaussian_projections}
\end{center}
\end{figure*}
We introduce \evnote{\textit{projection figures} (as described by \citealt{Vernon10})}, shown in \autoref{fig:3D_gaussian_projections} for the simple Gaussian model discussed earlier.
For every combination of two active model parameters\footnote{``Active parameter'' here indicates any parameter that is active in at least one emulator system in the emulator iteration} in a given emulator iteration, a projection figure can be made.
These figures can be used to derive many properties of the model, the used model comparison data and the performed emulation.

A projection figure is created by generating a square grid of $25\times25$ points for the two active parameters that are plotted.
For each grid point, a Latin-Hypercube design of $250$ samples for the remaining non-plotted parameters is generated.
This gives a Latin-Hypercube design of $250$ samples for every grid point, with the values for the plotted parameters fixed per grid point, and a total of $25\cdot25\cdot250=156,250$ samples for the entire grid.
The grid size/resolution and depth can be chosen freely by the user, with the defaults being the values used here.

Every sample in the grid is then evaluated and analyzed in the emulator, saving whether or not this sample is plausible and what the implausibility value at the first cut-off is (the first $n$ for which $I_{\mathrm{cut}, n}(\vec{x})$ is defined).
This yields $250$ results for every grid point, which can be used to determine the fraction of samples that is plausible and the minimum implausibility value at the first cut-off in this point.
Doing this for the entire grid and interpolating them, creates a map of results that is independent of the values of the non-plotted parameters.

Using these results, a figure consisting out of two subplots can be made.
The first subplot (shown on the top for every panel in \autoref{fig:3D_gaussian_projections}) shows a map of the minimum implausibility value (at the first cut-off) that can be reached for any given value combination of the two plotted parameters.
The second subplot (shown on the bottom) gives a map of the fraction of samples that is plausible in a specified point on the grid (called ``line-of-sight depth'' due to the way it is calculated).
Another way of describing this map is that it gives the probability that a parameter set with given plotted values is plausible.
While the first subplot gives insight into the correlation between the plotted parameters, the second subplot shows where the high-density regions of plausible samples are.
A combination of both allows for many model properties to be derived, as discussed in the following.

\subsubsection{Studying a Gaussian}
\label{subsubsec:Gaussian projections}
Looking at the projection figures of the Gaussian model in \autoref{fig:3D_gaussian_projections}, several features can be noticed.
We can see that the emulator for the first iteration (the projection figures on the left) is very conservative in its approach, which is mainly due to the fact that it used a `wildcard'.
Using a wildcard here means that the worst fitting comparison data point does not have any influence on the implausibility of an evaluation sample, therefore making $I_{\mathrm{cut}, 2}(\vec{x})$ the first implausibility cut-off.
Despite this, the behavior of a Gaussian can still be seen.

This is obvious when looking at the top-left panel.
Combining both subplots together, one can see that there are two/three main relations between the parameters $A_1$ (amplitude) and $B_1$ (mean).
Increasing the amplitude seems to mostly yield plausible samples if the mean changes as well, with some plausible samples being possible if the mean stays the same, while decreasing the amplitude requires the mean to stay the same.
Taking into account that a wildcard was used here and that the comparison data points were taken at $x=\{3, 5, 7\}$, this is expected behavior.
When the amplitude ($A_1$) increases, the mean ($B_1$) has to change in value to make sure that either $x=\{3, 5\}$ or $x=\{5, 7\}$ still yield plausible samples (and the third data point is the wildcard).
Decreasing the amplitude requires the mean to stay the same and the non-plotted standard deviation ($C_1$) to change to allow for $x=\{3, 7\}$ to yield plausible samples.
The top-left panel also shows that this last effect can generate plausible samples when increasing the amplitude, although with much lower yields.
The remaining two projection panels on the left show similar patterns.

From these projections, it is clear that the emulator is not accurate enough yet, which is mostly due to our conservative approach in using a wildcard ($I_{\mathrm{cut}, 2}(\vec{x})=4$).
Therefore, for the second iteration, we remove the wildcard by setting the first implausibility cut-off to $I_{\mathrm{cut}, 1}(\vec{x})=3$, in addition to also improving the emulator in the part of parameter space that is still plausible.
Doing so yield the projection figures on the right in \autoref{fig:3D_gaussian_projections}.
Here, the emulator has basically converged past the point where the correlations between the parameters are visible (only the influence of $C_1$ is still noticeable) and designates a small part ($0.0312\%$) of parameter space as plausible.

Interestingly enough, it does seem that the two subplots do not fully agree with each other in all three projection figures.
The minimum implausibility subplot yields the best values for the parameter estimates, while the line-of-sight depth subplot does not, shifting away a bit from the intersection between the gray lines.
Even though the effect here is small, it does show the importance of having the minimum implausibility, as the line-of-sight depth is very similar in meaning to the walker distribution used in MCMC methods.
Using an MCMC algorithm will yield a marginalized density map of the walker chains that looks the same as the line-of-sight depth.
Although the highest density does not necessarily correspond to the location of the highest posterior probability, the first is more commonly used in results than the latter.
It can therefore \evnote{pinpoint} a result that is slightly off, as can be seen clearly for $A_1$ in \autoref{fig:2D_gaussian_projections} in \ref{app:Figures and Tables}.
As the model discrepancy variance $\var(\epsilon_{\mathrm{md,i}})$ is now much larger than the adjusted emulator variance $\var_{D_i}(f_i(\vec{x}))$, the emulator cannot be optimized any further.

\subsection{Constraining a multi-Gaussian model}
\label{subsec:Application}
Thus far, we have only used \prism\ on simple models with a few parameters.
Although these models are great for showing the basics behind \prism, they do not showcase how it can be used to speed up model parameter estimation.
In the following, we introduce the concept of \textit{hybrid sampling}, where \prism\ and MCMC are combined together, and compare it to normal MCMC sampling using a multi-Gaussian model.
For performing the MCMC operations, we make use of the popular \python\ package \emcee\ \citep{emcee}, which is based on the \textit{affine invariant} sampling method by \citet{affine_invariant}.\footnote{Note that different MCMC methods can yield different results, as their algorithms are often unique.}

\subsubsection{Hybrid sampling}
\label{subsubsec:Hybrid sampling}
During a Bayesian analysis, one or several MCMC chains are created that individually explore parameter space.
Normally, when considering a parameter set, there is no prior information that this parameter set is (un)likely to result into a desirable model realization.
This means that such a parameter set must first be evaluated in the model before any probabilities can be calculated.
However, by constructing an emulator of the model, one can use it as an additional prior for the posterior probability calculation.

This so-called \textit{hybrid sampling} allows one to use \prism\ to first analyze a model's behavior, and later use the gathered information to speed up parameter estimations (by using the emulator as an additional prior in a Bayesian analysis).
Hybrid sampling works in the following way:
\begin{enumerate}
    \item Whenever an MCMC walker proposes a new sample, it is first passed to the emulator of the model;
    \item If the sample is not within the defined parameter space, it automatically receives a prior probability of zero (or $-\infty$ in case of logarithmic probabilities).
    Otherwise, it will be evaluated in the emulator;
    \item If the sample is labeled as \textit{implausible} by the emulator, it also receives a prior probability of zero. If it is plausible, the sample is evaluated in the same way as for normal sampling;
    \item Optionally, a scaled value of the first implausibility cut-off is used as an \evnote{exploratory method by adding an} additional (non-zero) prior probability, which is defined as $P(\vec{x})=1-I_{\mathrm{max}, n}(\vec{x})/I_{\mathrm{cut}, n}$ with $n$ the index of first cut-off.
\end{enumerate}

There are several advantages of using this system over normal sampling:
\begin{itemize}
    \item Acceptable samples are guaranteed to be within plausible space;
    \item This in turn makes sure that the model is only evaluated for plausible samples, which heavily reduces the number of required evaluations;
    \item No \textit{burn-in phase} is required, as the starting positions of the MCMC walkers are chosen to be in plausible space;
    \item As a consequence, varying the number of walkers tends to have a much lower negative impact on the convergence probability and speed;
    \item Samples with low implausibility values can optionally be favored.
\end{itemize}

In the remaining part of this section, we will show this by comparing hybrid sampling and normal sampling with each other using a multi-Gaussian model.

\subsubsection{Parameter estimations}
\label{subsubsec:Parameter estimations}
\begin{figure}
\begin{center}
    \includegraphics[width=\linewidth]{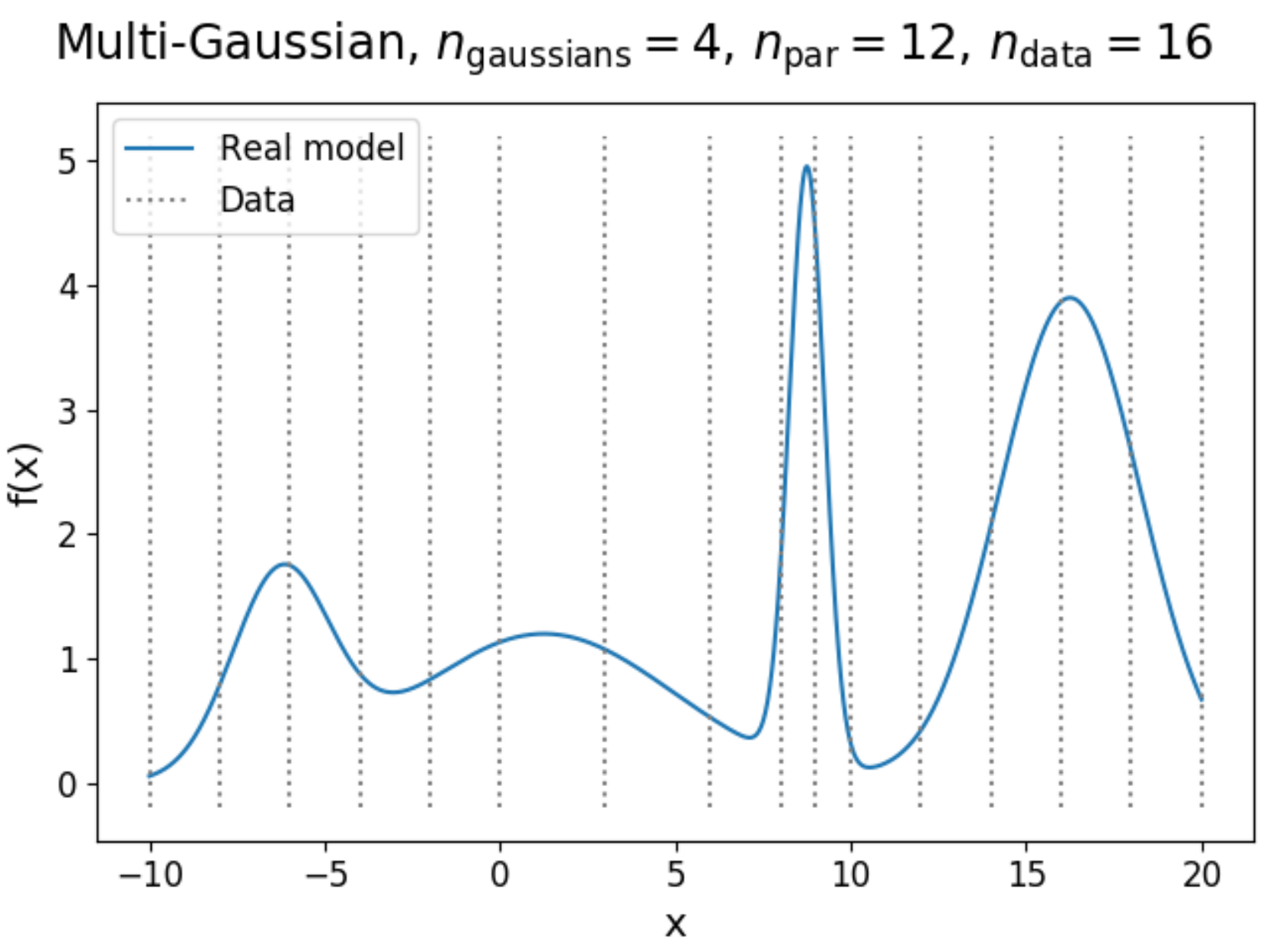}
    \caption{Mock realization of a multi-Gaussian model used for testing hybrid sampling and normal sampling, defined as $f(x)=\sum_{i=1}^4 A_i\cdot\exp\left(-\frac{(x-B_i)^2}{2C_i^2}\right)$.
    The vertical dotted lines show the $16$ known data points.
    The four Gaussians are labeled in order from left to right.
    See \autoref{tab:multi_gaussian_parameters} in \ref{app:Figures and Tables} for the used parameter values.}
    \label{fig:multi_gaussian_model}
\end{center}
\end{figure}
\begin{figure*}
\begin{center}
    \subfloat{\label{subfig:multi_hybrid_100}\includegraphics[width=.37\textwidth]{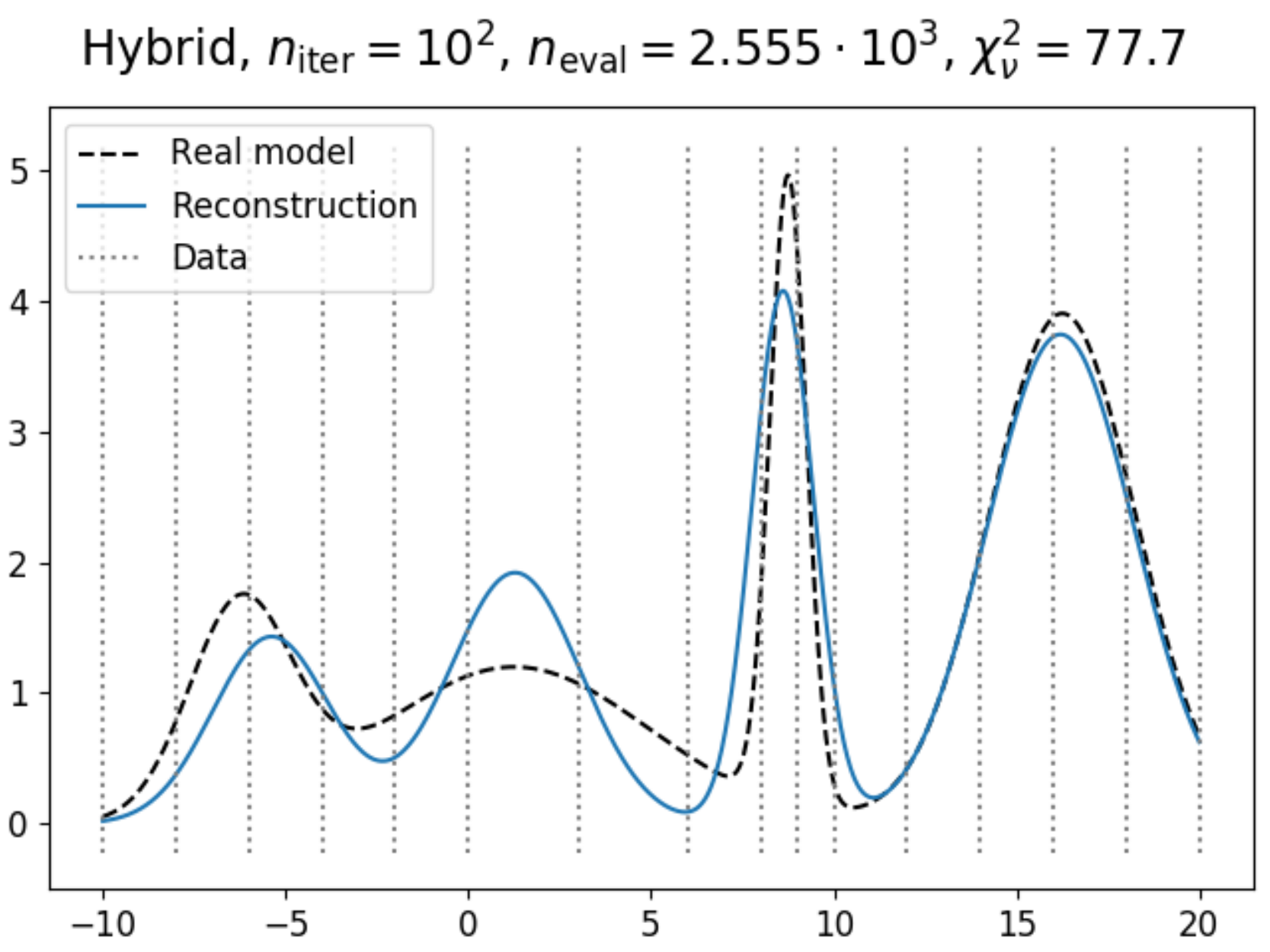}}
    \subfloat{\label{subfig:multi_normal_100}\includegraphics[width=.37\textwidth]{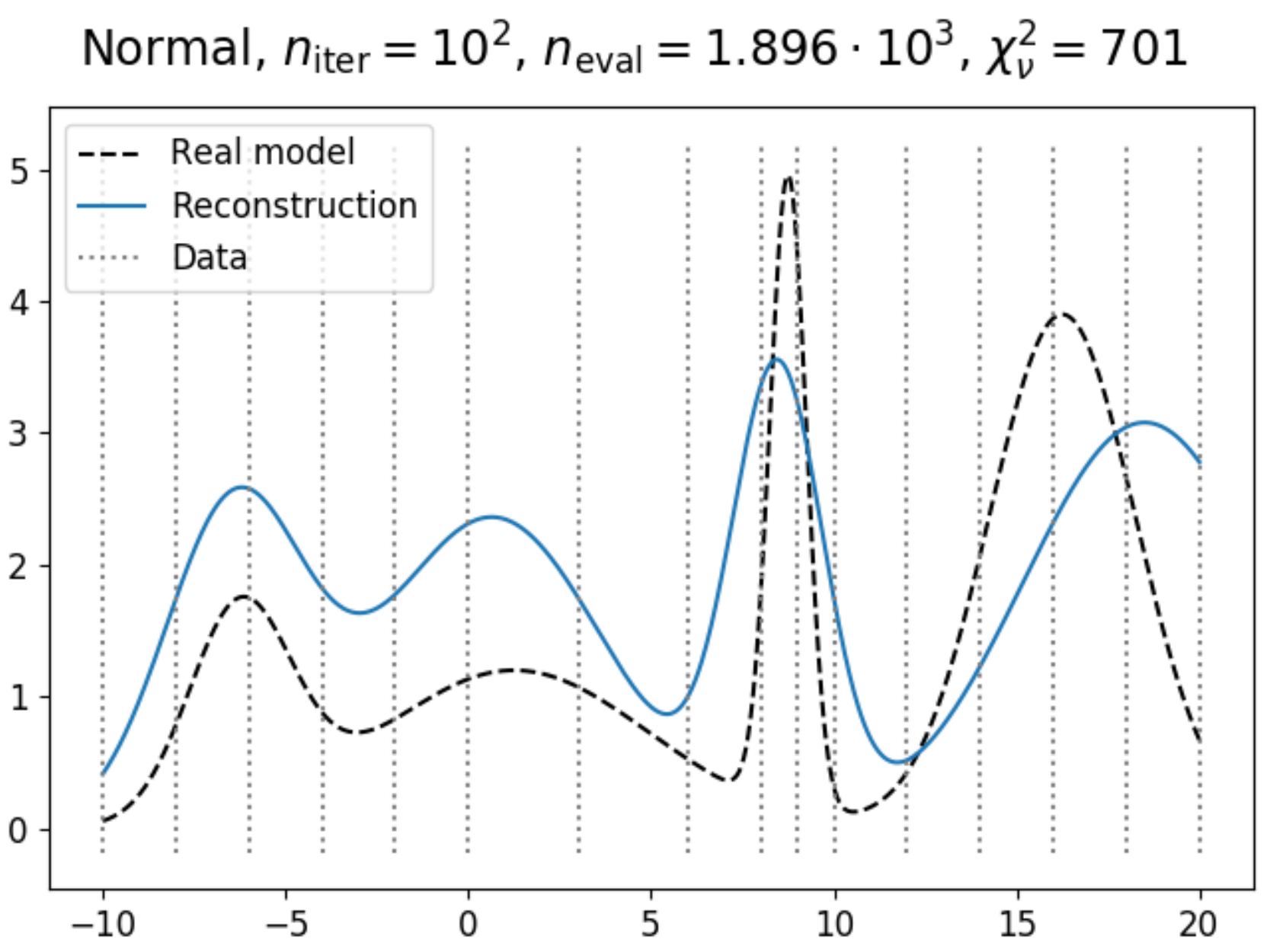}}\\
    \subfloat{\label{subfig:multi_hybrid_1000}\includegraphics[width=.37\textwidth]{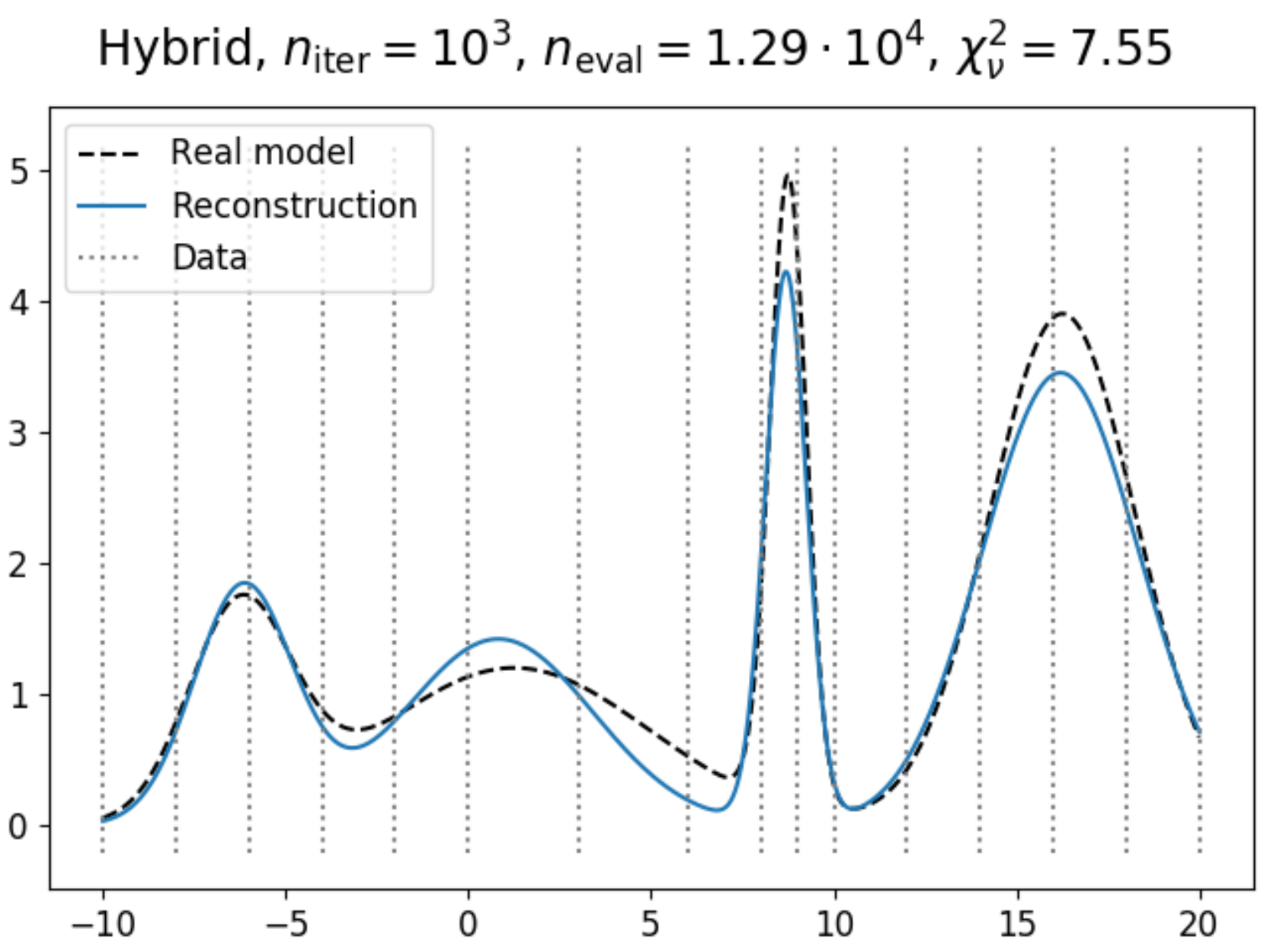}}
    \subfloat{\label{subfig:multi_normal_1000}\includegraphics[width=.37\textwidth]{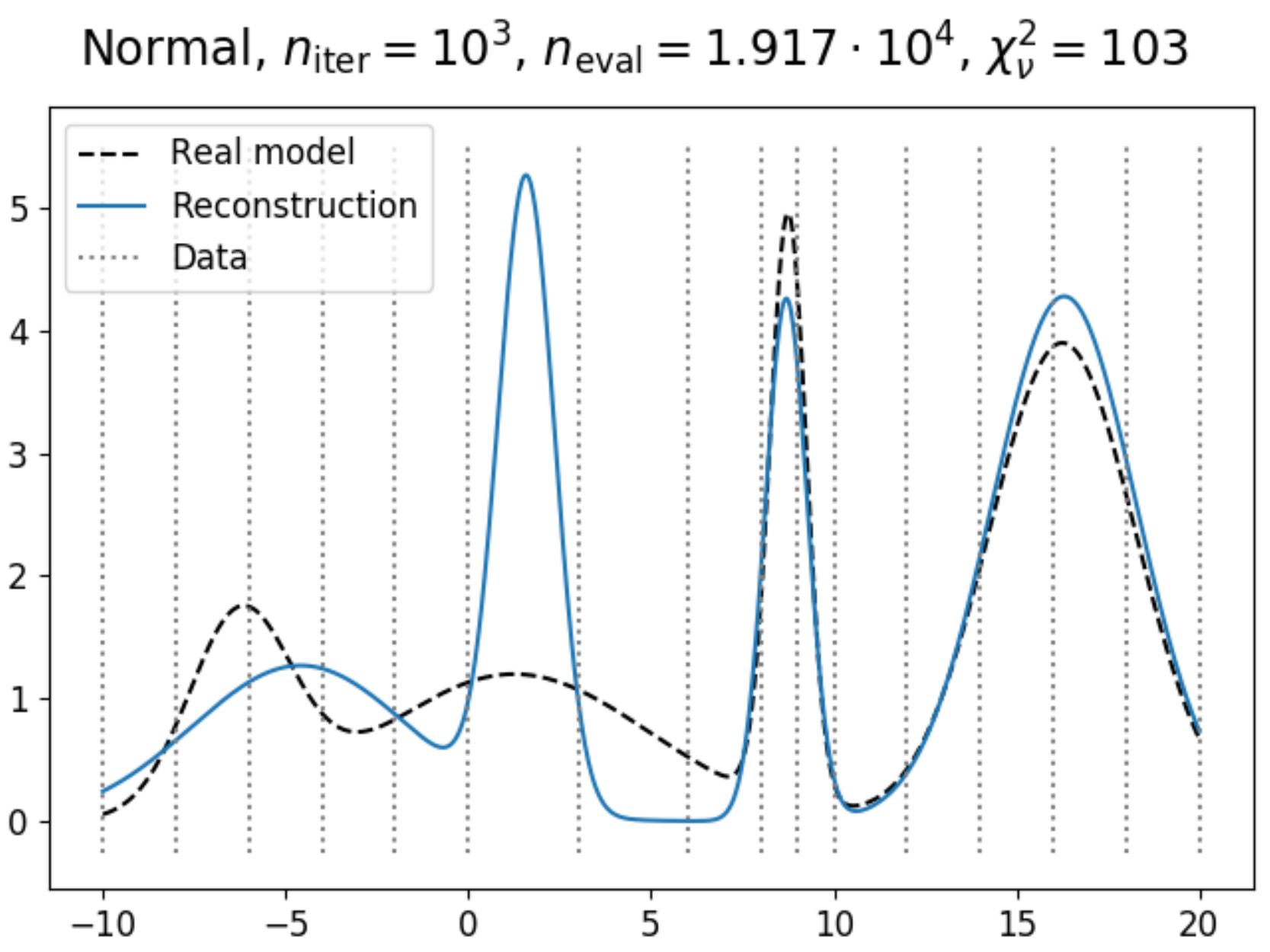}}\\
    \subfloat{\label{subfig:multi_hybrid_10000}\includegraphics[width=.37\textwidth]{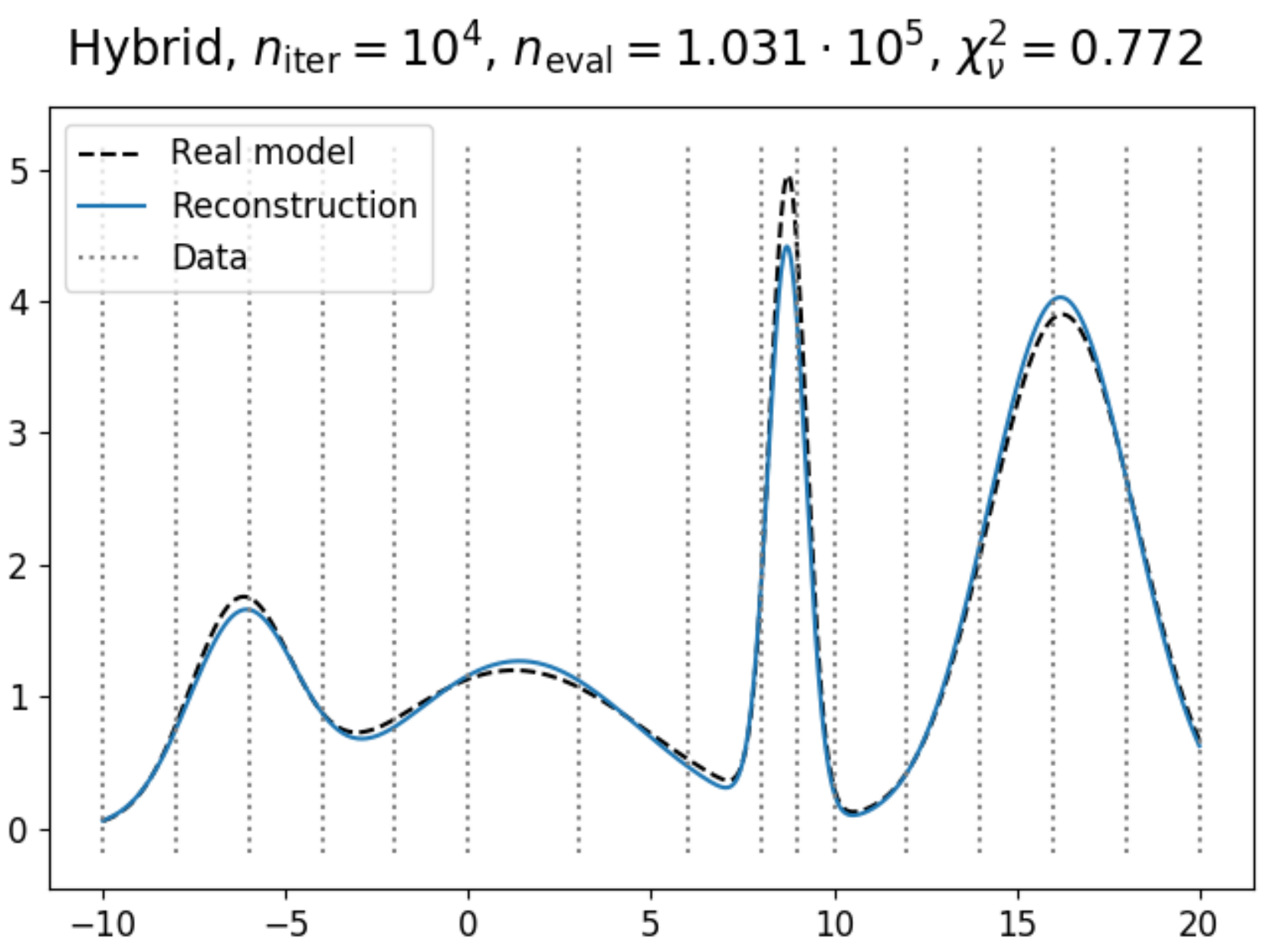}}
    \subfloat{\label{subfig:multi_normal_10000}\includegraphics[width=.37\textwidth]{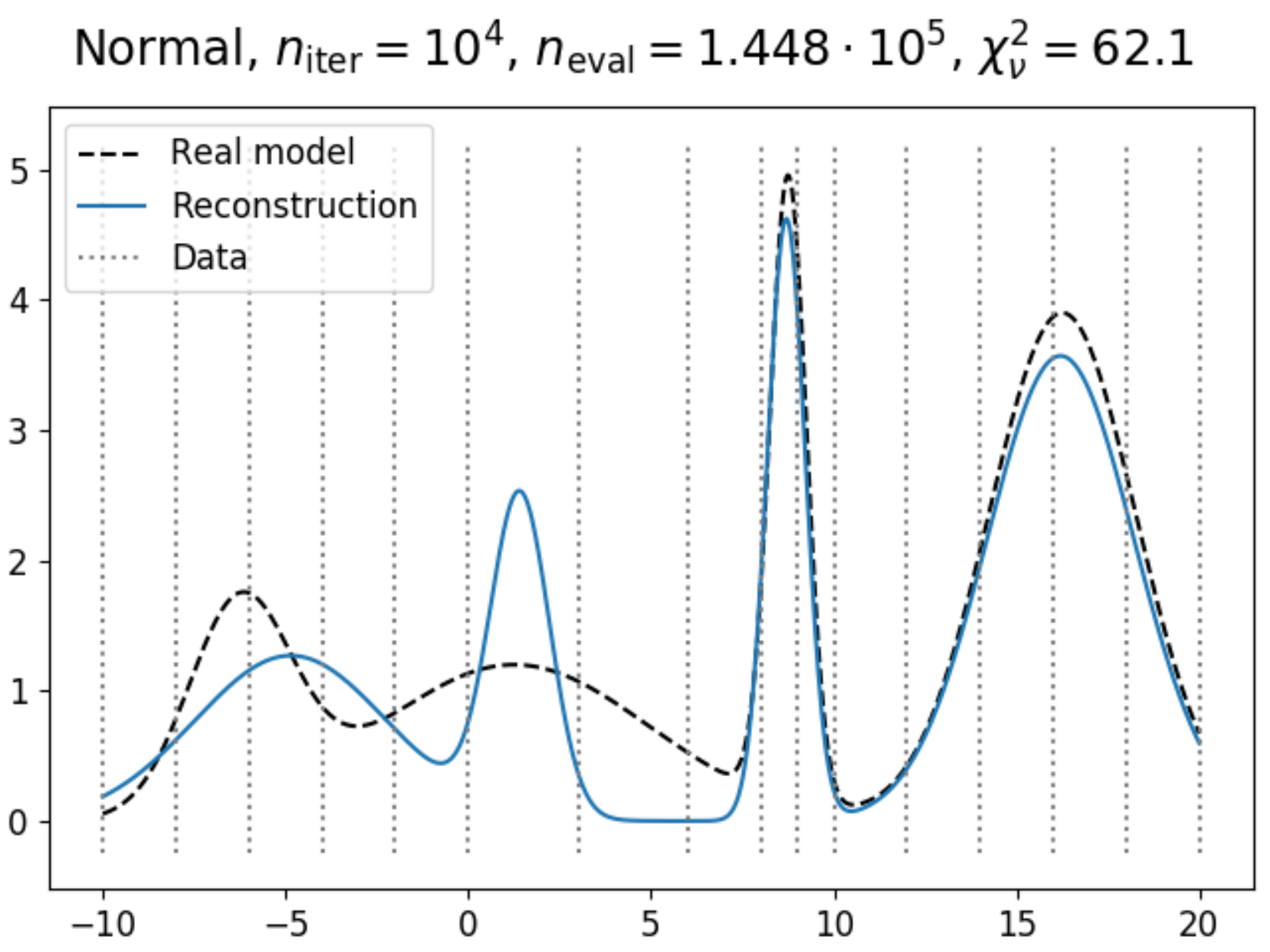}}\\
    \subfloat{\label{subfig:multi_hybrid_100000}\includegraphics[width=.37\textwidth]{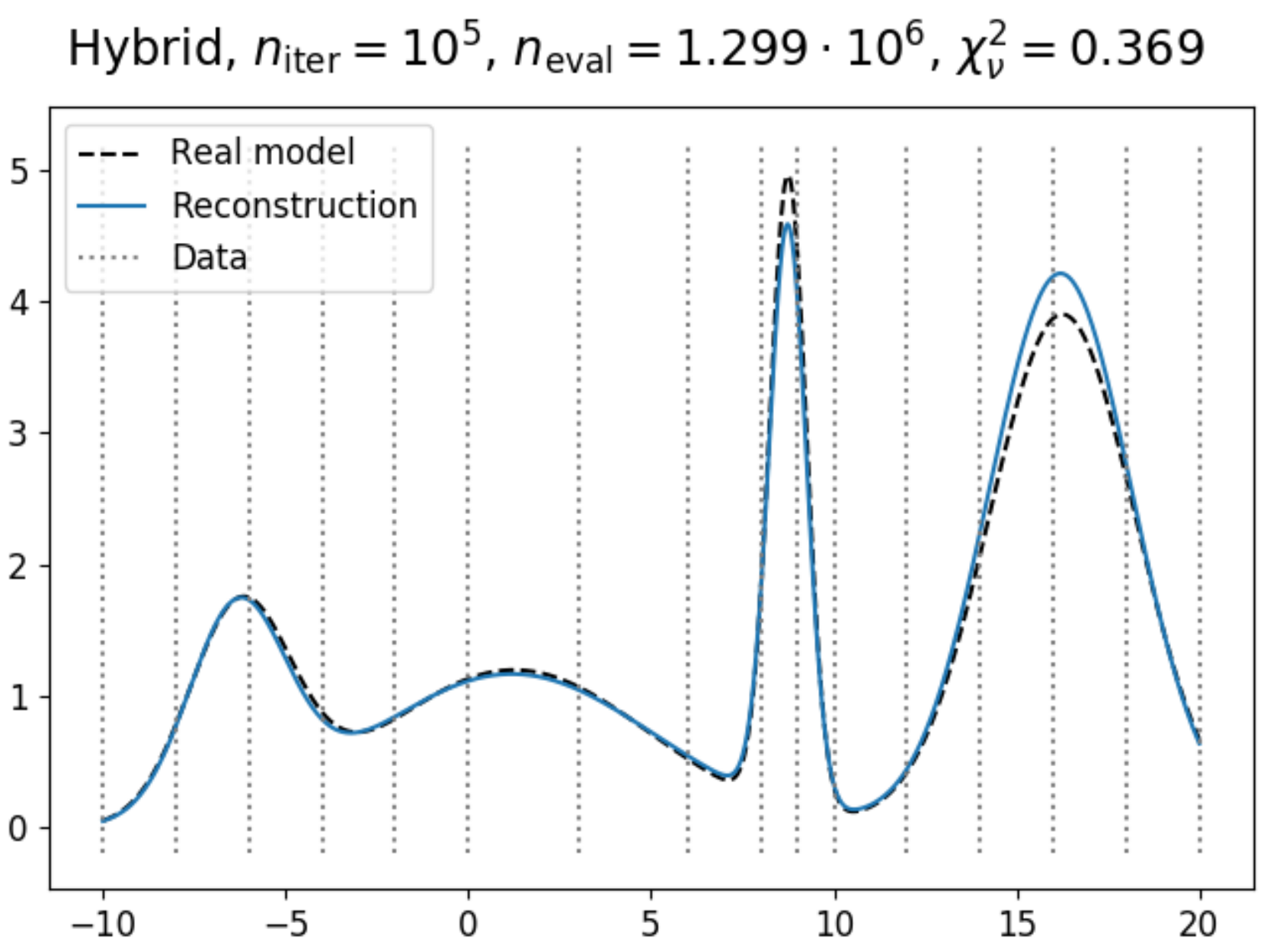}}
    \subfloat{\label{subfig:multi_normal_100000}\includegraphics[width=.37\textwidth]{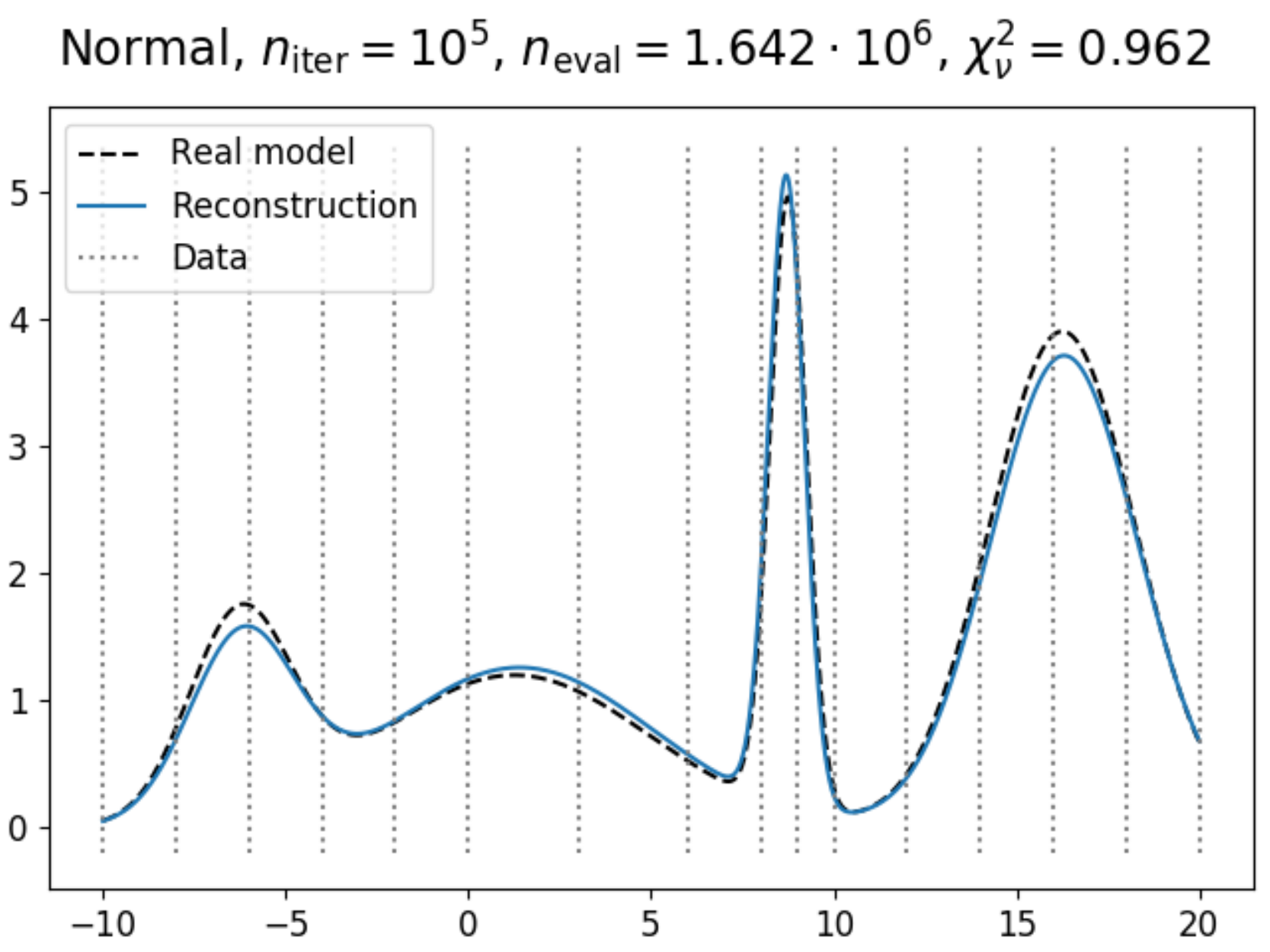}}
    \caption{Parameter estimation results of the reconstruction of a multi-Gaussian model using hybrid sampling and normal sampling, shown for $10^2$, $10^3$, $10^4$ and $10^5$ MCMC iterations.
    The title of every figure states the used sampling type, number of MCMC iterations, number of used model evaluations and the reduced $\chi^2$ of the reconstruction.
    \textbf{Left column:} Reconstruction results using hybrid (\prism\ + MCMC) sampling, which used an emulator with $1,934$ model evaluations and $1.35\cdot10^{-4}\%$ of parameter space remaining.
    Note that the number of used model evaluations mentioned in the title takes the initial number into account.
    \textbf{Right column:} Reconstruction results using normal (MCMC only) sampling.
    Both methods used $26$ MCMC walkers, where the hybrid walkers started in plausible space and the normal walkers started at random.
    The same likelihood calculation was used for both methods, while hybrid sampling used an additional prior based on the information in the emulator.
    All runs were performed from scratch in order to make their results less vulnerable to the burn-in phase.
    See \autoref{tab:multi_gaussian_parameters} in \ref{app:Figures and Tables} for all parameter estimation values and errors.}
    \label{fig:multi_gaussian_results}
\end{center}
\end{figure*}
For comparing hybrid sampling with normal sampling, we make use of a multi-Gaussian model, consisting of four different Gaussians.
In \autoref{fig:multi_gaussian_model}, we show the realization of this model that corresponds to the true parameter values.
To make sure that every Gaussian is distinct (switching parameter labels cannot result in the same realization), the means ($B_i$) each have a parameter range that makes up $25\%$ of the total input domain.
The amplitudes ($A_i$) and the standard deviations ($C_i$) on the other hand, have identical parameter ranges ($[1, 10]$ and $[0, 5]$, respectively).

Given that the model itself has $12$ parameters, we have chosen $16$ data points that are almost equally distributed over the input domain of $[-10, 20]$ in order to try to remain unbiased.
The exception to this is that the expected $x=2$ and $x=4$ points have been replaced by $x=3$ and $x=9$, such that information can be gained about the third Gaussian.
For calculating the posterior probability, we use a flat prior (which returns zero if a sample is not in parameter space and unity otherwise) and the following (Gaussian) likelihood function:
\begin{align*}
    \ln\left(\prob(\vec{z}|\vec{x},I)\right) &= -0.5\cdot\sum_i\frac{\left(f_i(\vec{x})-z_i\right)^2}{\var(\epsilon_{\mathrm{md}, i})+\var(\epsilon_{\mathrm{obs}, i})},
\end{align*}
where $\ln\left(\prob(\vec{z}|\vec{x},I)\right)$ is the natural logarithm of the likelihood, $f_i(\vec{x})$ the model output $i$ for input $\vec{x}$ with corresponding variance $\var(\epsilon_{\mathrm{md}, i})$, and $z_i$ the value of data point $i$ with observational variance  $\var(\epsilon_{\mathrm{obs}, i})$.
Note that this likelihood function may already appear more complex than commonly employed forms.
The same prior and likelihood functions were used for both sampling types, with hybrid sampling using an additional prior as described in \ref{subsubsec:Hybrid sampling}.

By first constructing an emulator of the multi-Gaussian model of \autoref{fig:multi_gaussian_model} and then using either hybrid sampling or normal sampling, we obtain the parameter estimation results in \autoref{fig:multi_gaussian_results}.
In this figure, we show the model reconstructions of both sampling methods for $10^2$, $10^3$, $10^4$ and $10^5$ MCMC iterations/steps.
To make sure that the results are less biased/influenced by the random nature of MCMC and to check for consistency, all runs were performed from scratch (e.g., the $10^5$ chain did not start from the $10^4$ chain).
Given that the used data is artificial and therefore has no physical error, $\chi_{\nu}^2$ is simply an indicator of the quality of the fit (whereas normally, $\chi_{\nu}^2 < 1$ is an indication of over-fitting).

Something that can be noticed immediately, is the best parameter fit that normal sampling returns for $10^3$ and $10^4$ iterations, where it massively overestimates the amplitude of the second Gaussian ($A_2$).
Since every run was done from scratch, these two results are not influenced by each other.
A likely reason for this behavior is that the burn-in phase required a large number of iterations due to the complexity of the problem and had not converged properly yet.
Since the hybrid MCMC walkers started in plausible space, and therefore do not require a burn-in phase, the hybrid sampling results do not exhibit this behavior.

When looking at the various subplots, it is clear that hybrid sampling performs better than normal sampling, scoring much better in terms of $\chi_{\nu}^2$ for the same number of iterations (in addition to using \evnote{fewer} model evaluations).
Additionally, hybrid sampling seems to require less iterations to reach an acceptable result compared to normal sampling, where the third hybrid reconstruction fits the known data better than the final normal reconstruction.
This means that, for this specific model, hybrid sampling required ${\sim}16$ times less model evaluations in comparison to normal sampling.
If one would take into account that the final normal reconstruction could have shown the same behavior as seen for the second and third (analysis of the entire chain showed it did not), this number could potentially have been higher.

We believe that this number is quite significant.
If one assumes that a model evaluation takes significantly more time than anything else (which often is the case for complex models), then using hybrid sampling will be ${\sim}16$ times faster.
Given that this model is (literally) of Gaussian form, it is not unlikely that this number is higher for more complex models.
This, in addition to being built to rapidly analyze a model's behavior, makes \prism\ an excellent alternative to MCMC, while also being able to join forces with it when it comes to constraining models.

\section{Conclusions/Outlook}
\label{sec:Conclusions}
We have introduced a new, publicly available framework for rapid analysis of scientific models \evnote{based on the algorithms described by \citet{Vernon10}}, called \prism.
\prism\ is unique in that it is written with no particular model application in mind, but rather provides a generic and versatile environment for the user.
This makes it modular and allows others to use \prism\ for their own projects with minimal effort.
It has a number of key characteristics:
\begin{itemize}
    \item Written in \python\ for increased versatility and user-friendliness;
    \item Built as a plug-and-play tool where users can adapt it to suit their own needs;
    \item Suited for any type of model;
    \item Capable of reducing relevant parameter space by factors over $100,000$ using only a few thousand model evaluations, as demonstrated in \ref{subsubsec:Parameter estimations};
    \item Can be used alone for analyzing models, or combined with MCMC for efficient model parameter estimations.
\end{itemize}

We have discussed how the \BLA\ and emulation technique can be combined with history matching (\ref{sec:Model analysis}) to efficiently explore the parameter space of a scientific model.
The use of these techniques allows for less information and knowledge to be required in order for a model to be analyzed.
This allows for time spent on \evnote{acquiring} the knowledge and evaluating the model, to be kept to a minimum.

We have described the basic framework of \prism\ (\ref{sec:PRISM}) and the different techniques and methods that it uses.
In \ref{sec:Basic usage}, we showed several application examples of \prism, where we analyzed a Gaussian model's behavior by studying its projections and constrained a multi-Gaussian model.
We also introduced the concept of \textit{hybrid sampling}, where \prism\ and MCMC methods can be combined together to increase the rate at which MCMC optimizations converge.
The comparison between hybrid sampling and normal sampling showed the advantages the former has over the latter, being able to reach an acceptable parameter estimation result ${\sim}16$ times faster than the latter method.

In future work, we will use \prism\ together with the MCMC package \textsw{Mhysa} (Mutch et al. in prep.) to analyze and explore the parameter space of the semi-analytic model \meraxes\ \citep{Meraxes}.
Given that \meraxes\ is designed to be accurate at high redshifts ($z>5$), where qualitative observational data is scarce, \prism\ will be a well-suited choice for performing this task.
Finally, several smaller application projects for \prism\ are currently being planned, as well as adding a low-level MPI implementation and GPU acceleration.

\section*{Acknowledgements}
We would like to thank the referee for the extraordinarily detailed report and constructive feedback.
EV would like to thank Chris Blake, Colin Jacobs and Theo Steininger for fruitful discussions and valuable suggestions.
Parts of this research were supported by the Australian Research Council Centre of Excellence for All Sky Astrophysics in 3 Dimensions (ASTRO 3D), through project number CE170100013.
Parts of this work were performed on the OzSTAR national facility at Swinburne University of Technology. OzSTAR is funded by Swinburne University of Technology and the National Collaborative Research Infrastructure Strategy (NCRIS).



\appendix
\labelformat{section}{App.~#1}
\labelformat{subsection}{App.~#1}
\labelformat{subsubsection}{App.~#1}
\labelformat{lstlisting}{Lst.~#1}
\section{MPI implementation}
\label{app:MPI implementation}
\begin{figure*}
\begin{center}
	\includegraphics[width=\textwidth]{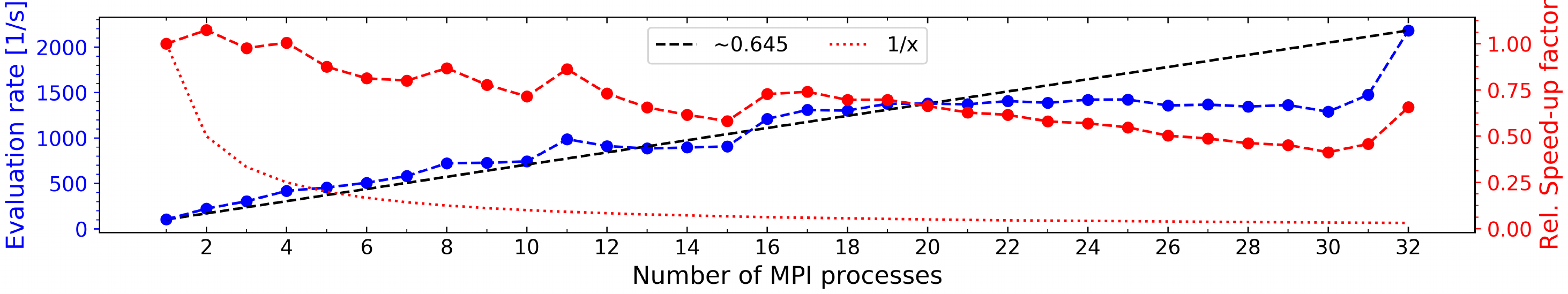}
    \caption{Figure showing the MPI scaling of \prism\ using the emulator of a simple Gaussian model with $32$ emulator systems.
    The tests involved analyzing a Latin-Hypercube design of $3\cdot10^6$ samples in the emulator, determining the average evaluation rate and executing this a total of $20$ times using the same sample set every time.
    The emulator used for this was identical in every instance.
    \textbf{Left axis:} The average evaluation rate of the emulator vs.\ the number of MPI processes it is running on.
    \textbf{Right axis:} The relative speed-up factor vs.\ the number of MPI processes, which is defined as $f(x)/(f(1)\cdot x)$ with $f(x)$ the average evaluation rate and $x$ the number of MPI processes.
    \textbf{Dotted line:} The minimum acceptable relative speed-up factor, which is always $1/x$.
   	\textbf{Dashed line:} A straight line with a slope of ${\sim}0.645$, connecting the lowest and highest evaluation rates.
    The tests were performed using the \textit{OzSTAR computing facility} at the Swinburne University of Technology, Melbourne, Australia.}
    \label{fig:MPI_scaling}
\end{center}
\end{figure*}

Given that most scientific models are either already parallelized or could benefit from parallelization, we had to make sure that \prism\ allows for both MPI and OpenMP coded models to be connected.
Additionally, since individual emulator systems in an emulator iteration are independent of each other, the extra CPUs required for the model should also be usable by the emulator.
For that reason, \prism\ features a high-level MPI implementation for using MPI-coded models, while the \python\ package \numpy\ handles the OpenMP side.
A mixture of both is also possible.

Here, we discuss the MPI scaling tests that were performed on \prism.
For the tests, the same \gaussianlink\ class was used as in \ref{subsec:Minimal example}, but this time with $32$ emulator systems (comparison data points) instead of $3$.
In \prism, all emulator systems are spread out over the available number of MPI processes as much as possible while also trying to balance the number of calculations performed per MPI process.
Since all emulator systems are stored in different HDF5-files, it is possible to reinitialize the \pipeline\ using the same \emulator\ class and \modellink\ subclass on a different number of MPI processes.
To make sure that the results are not influenced by the variation in evaluation rates, we constructed an emulator of the Gaussian model and used the exact same emulator in every test.

The tests were carried out using any number of MPI processes between $1$ and $32$, and using a single OpenMP thread each time for consistency.
We generated a Latin-Hypercube design of $3\cdot10^6$ samples and measured the average evaluation rate of the emulator using the same Latin-Hypercube design each time.
To take into account any variations in the evaluation rate caused by initializations, this test was performed $20$ times.
As a result, this Latin-Hypercube design was evaluated in the emulator a total of $640$ times, giving an absolute total of $1.92\cdot10^9$ emulator evaluations.

In \autoref{fig:MPI_scaling}, we show the results of the performed MPI scaling tests.
On the left $y$-axis, the average evaluation rate vs.\ the number of MPI processes the test ran on is plotted, while the relative speed-up factor vs.\ the number of MPI processes is plotted on the right $y$-axis.
The relative speed-up factor is defined as $f(x)/(f(1)\cdot x)$ with $f(x)$ the average evaluation rate and $x$ the number of MPI processes.
The ideal MPI scaling would correspond to a relative speed-up factor of unity for all $x$.

In this figure, we can see the effect of the high-level MPI implementation.
Because the emulator systems are spread out over the available MPI processes, the evaluation rate is mostly determined by the runtime of the MPI process with the highest number of systems assigned.
Therefore, if the number of emulator systems ($32$ in this case) cannot be divided by the number of available MPI processes, the speed gain is reduced, leading to the plateaus like the one between $x=16$ and $x=31$.
Due to the emulator systems not being the same, their individual evaluation rates are different such that the more MPI processes there are, a different evaluation rate will have a bigger effect on the average evaluation rate of the emulator.
This is shown by the straight dashed line drawn between $f(1)$ and $f(32)$, which has a slope of ${\sim}0.645$.

The relative speed-up factor shows the efficiency of every individual MPI process in a specific run, compared to using a single MPI process.
This also shows the effect of the high-level MPI implementation, giving peaks when the maximum number of emulator systems per MPI process has decreased.
The dotted line shows the minimum acceptable relative speed-up factor, which is always defined as $1/x$.
On this line, the average evaluation rate $f(x)$ for any given number of MPI processes is always equal to $f(1)$.

\section{Writing a ModelLink subclass}
\label{app:Writing a ModelLink subclass}
We have shown in \ref{subsec:Minimal example} how to initialize the \pipeline\ class using a default \modellink\ subclass.
Here, we would like to show the basic steps for making a custom \modellink\ subclass.

\begin{figure*}
\begin{center}
    \begin{minipage}[t]{0.49\textwidth}
    \begin{flushleft}
\begin{lstlisting}[style=defaultpython,basicstyle=\scriptsize\ttfamily\color{text},frame=single,title=\lstname,xleftmargin=0pt,label=lst:ExampleLink,caption={The basic structure of a custom \modellink\ subclass.},captionpos=b]
# -*- coding: utf-8 -*-

# Future imports
from __future__ import absolute_import, division, print_function

# Package imports
import numpy as np

# PRISM imports
from prism.modellink import ModelLink


# ExampleLink class definition
class ExampleLink(ModelLink):
    def __init__(self, *args, **kwargs):
        # Perform any custom operations here
        pass

        # Call ModelLink's __init__()
        super(ExampleLink, self).__init__(*args, **kwargs)

    def get_default_model_parameters(self):
        # Define default parameters (not required)
        par_dict = {}
        return(par_dict)

    def get_default_model_data(self):
        # Define default data (not required)
        data_dict = {}
        return(data_dict)

    # Override call_model method
    def call_model(self, emul_i, par_set, data_idx):
        # Perform operations for obtaining the model output
        # Following is provided:
        # 'emul_i': Requested iteration
        # 'par_set': Requested sample(s)
        # 'data_idx': Requested data point(s)
        pass

    # Override get_md_var method
    def get_md_var(self, emul_i, par_set, data_idx):
        # Perform operations for obtaining the model discrepancy variance
        # Following is provided:
        # 'emul_i': Requested iteration
        # 'par_set': Requested sample
        # 'data_idx': Requested data point(s)
        pass
\end{lstlisting}
    \end{flushleft}
    \end{minipage}~
    \begin{minipage}[t]{0.49\textwidth}
    \begin{flushright}
\begin{lstlisting}[style=defaultpython,basicstyle=\scriptsize\ttfamily\color{text},frame=single,title=\lstname,xleftmargin=0pt,label=lst:LineLink,caption={A custom \modellink\ subclass wrapping a straight line model.},captionpos=b]
# -*- coding: utf-8 -*-

# Future imports
from __future__ import absolute_import, division, print_function

# Package imports
import numpy as np

# PRISM imports
from prism.modellink import ModelLink


# LineLink class definition
class LineLink(ModelLink):
    def __init__(self, *args, **kwargs):
        # No custom operations required
        pass

        # Call ModelLink's __init__()
        super(LineLink, self).__init__(*args, **kwargs)

    def get_default_model_parameters(self):
        # Define default parameters (not required)
        par_dict = {
            'A': [-10, 10, 3],  # Intercept in [-10, 10] with estimate of 3
            'B': [0, 5, 1.5]}   # Slope in [0, 5] with estimate of 1.5
        return(par_dict)

    def get_default_model_data(self):
        # Define default data (not required)
        data_dict = {
            1: [4.5, 0.1],    # f(1) = 4.5 +- 0.1
            2.5: [6.8, 0.1],  # f(2.5) = 6.8 +- 0.1
            -2: [0, 0.1]}     # f(-2) = 0 +- 0.1
        return(data_dict)

    # Override call_model method
    def call_model(self, emul_i, par_set, data_idx):
        # Calculate the value on a straight line for requested data_idx
        vals = par_set['A']+np.array(data_idx)*par_set['B']
        return(vals)

    # Override get_md_var method
    def get_md_var(self, emul_i, par_set, data_idx):
        # Calculate the model discrepancy variance
        # For a straight line, this value can be set to a constant
        return(1e-4*np.ones_like(data_idx))
\end{lstlisting}
    \end{flushright}
    \end{minipage}
\end{center}
\end{figure*}
In \ref{lst:ExampleLink}, we show the basic structure of a \modellink\ subclass, which we used to make a \modellink\ subclass that wraps a straight line model in \ref{lst:LineLink}.

First, we import the packages like before:
\begin{lstlisting}[style=defaultpython]
	|\iin| from prism import Pipeline
    |\iin| from line_link import LineLink
\end{lstlisting}
\prism\ provides a function that allows the user to check if a \modellink\ subclass is properly written, which returns an instance of the subclass if the test passes:
\begin{lstlisting}[style=defaultpython]
    |\iin| from prism.modellink import test_subclass
    |\iin| modellink_obj = test_subclass(LineLink)
\end{lstlisting}
Since no errors were raised, we can initialize the \pipeline\ class and run a single cycle.
To make sure that the results are reproducible, we set \numpy's random seed as well:
\begin{lstlisting}[style=defaultpython]
    |\iin| import numpy as np
	|\iin| np.random.seed(0)
	|\iin| pipe = Pipeline(modellink_obj)
	|\iin| pipe.run()
\end{lstlisting}
We can check the current status of the \pipeline\ with:
\begin{lstlisting}[style=defaultpython]
	|\iin| pipe.details()
\end{lstlisting}
which produces:
\footnotesize
\begin{verbatim}
    PIPELINE DETAILS
    ===============================

    GENERAL
    -------------------------------
    Working directory               'prism_0'
    Emulator type                   'default'
    ModelLink subclass              LineLink
    Emulation method                Regression + Gaussian
    Mock data used?                 No

    ITERATION
    -------------------------------
    Emulator iteration              1
    Construction completed?         Yes
    Plausible regions?              Yes
    Projections available?          Yes (2/2)
    -------------------------------
    # of model evaluation samples   500 ([500])
    # of plausible/analyzed samples 9/1600
    % of parameter space remaining  0.562%
    # of active/total parameters    2/2
    # of emulated data points       3
    # of emulator systems           3
    -------------------------------

    PARAMETER SPACE
    -------------------------------
    *A: [-10.0, 10.0] (3.00000)
    *B: [  0.0,  5.0] (1.50000)
    ===============================
\end{verbatim}
\normalsize
In this overview, we can see that the entire emulator iteration has been constructed successfully, that about $0.562\%$ of parameter space remains, both parameters are considered to be active and three data points were used to constrain the model.
Given the low number of plausible samples, we might have to reanalyze the \pipeline\ with more samples if we would want to construct the second iteration (as warned by the \pipeline).

\begin{figure*}
\begin{center}
	\subfloat[Projection of the intercept value, parameter $A$.]{\label{subfig:1a_proj_A}\includegraphics[width=0.49\textwidth]{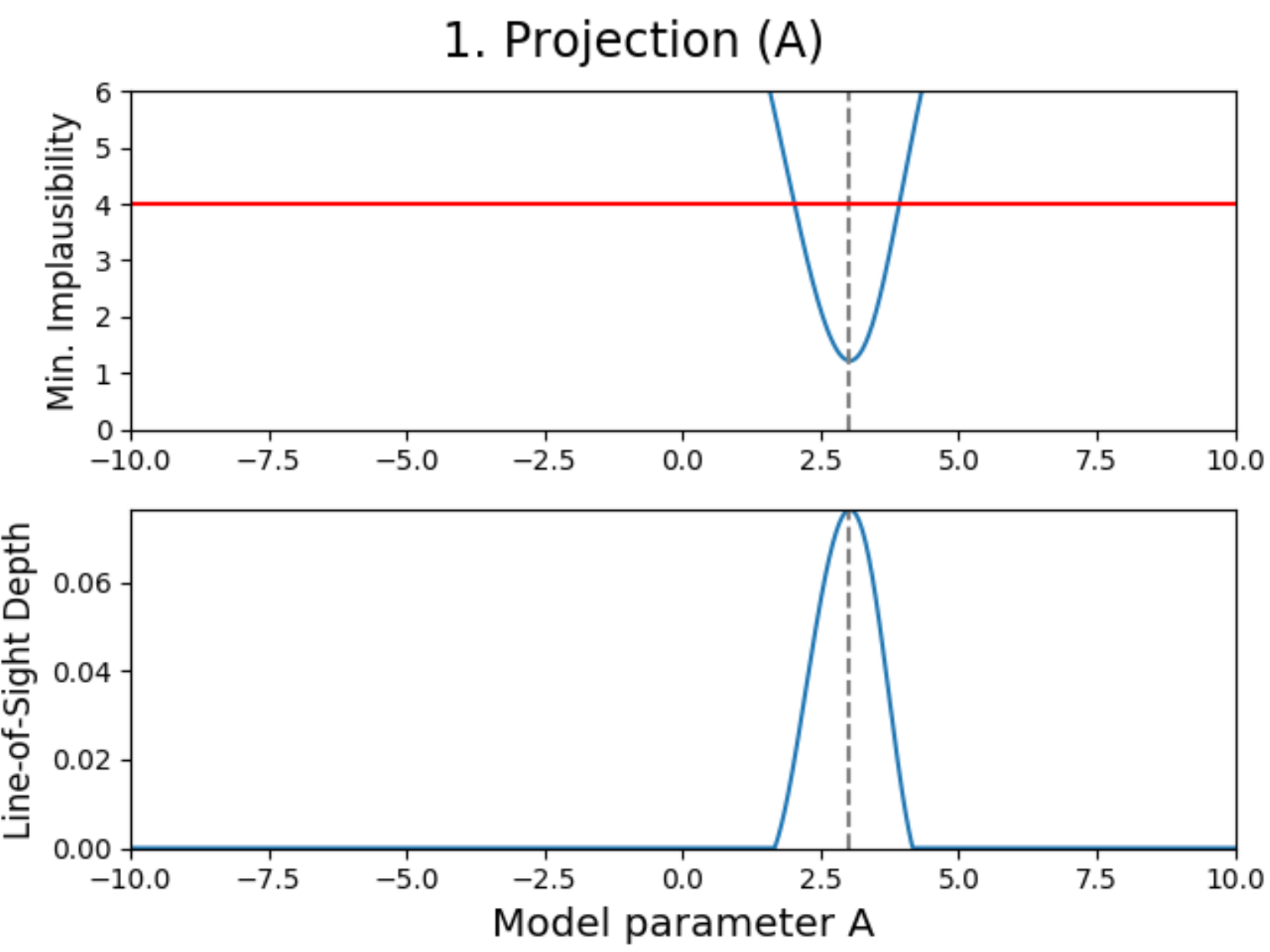}}
    \subfloat[Projection of the slope value, parameter $B$.]{\label{subfig:1a_proj_B}\includegraphics[width=0.49\textwidth]{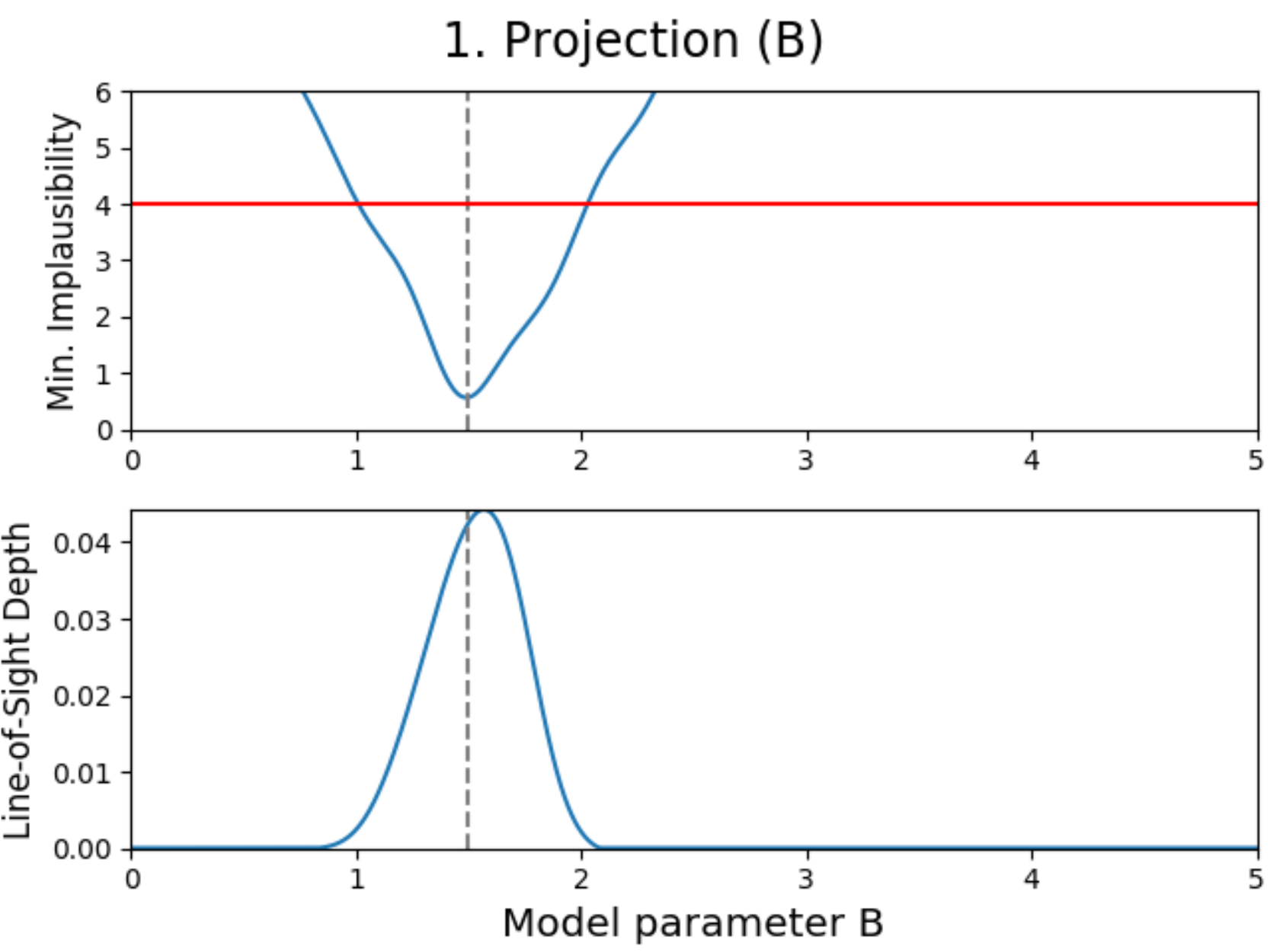}}
	\caption{Projection figures of the emulator of a simple straight line model, defined as $f(x)=A+B\cdot x$.
    These figures are the 2D equivalents of the figures in \autoref{fig:3D_gaussian_projections}.
    \textbf{Top:} The minimum implausibility reached for given parameter value, with the horizontal red line indicating the first non-wildcard implausibility cut-off value.
    \textbf{Bottom:} The fraction of samples that is still plausible for given parameter value.
    The vertical dashed lines indicate the parameter estimates that were given in our definition of the \textsc{LineLink} class.}
    \label{fig:LineLink projections 1a}
\end{center}
\end{figure*}
\begin{figure*}
\begin{center}
	\subfloat[Projection of the intercept value, parameter $A$.]{\label{subfig:1b_proj_A}\includegraphics[width=0.49\textwidth]{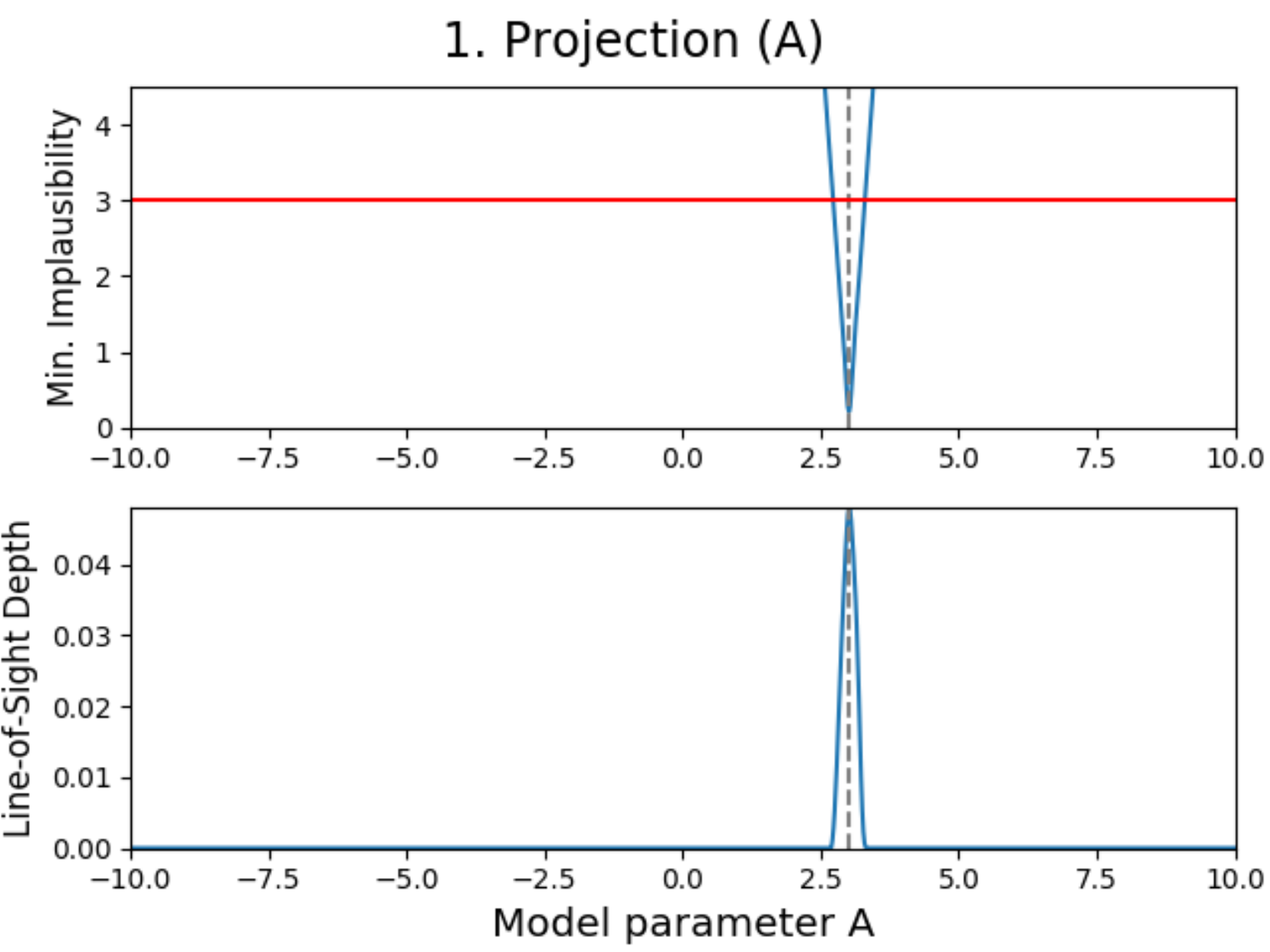}}
    \subfloat[Projection of the slope value, parameter $B$.]{\label{subfig:1b_proj_B}\includegraphics[width=0.49\textwidth]{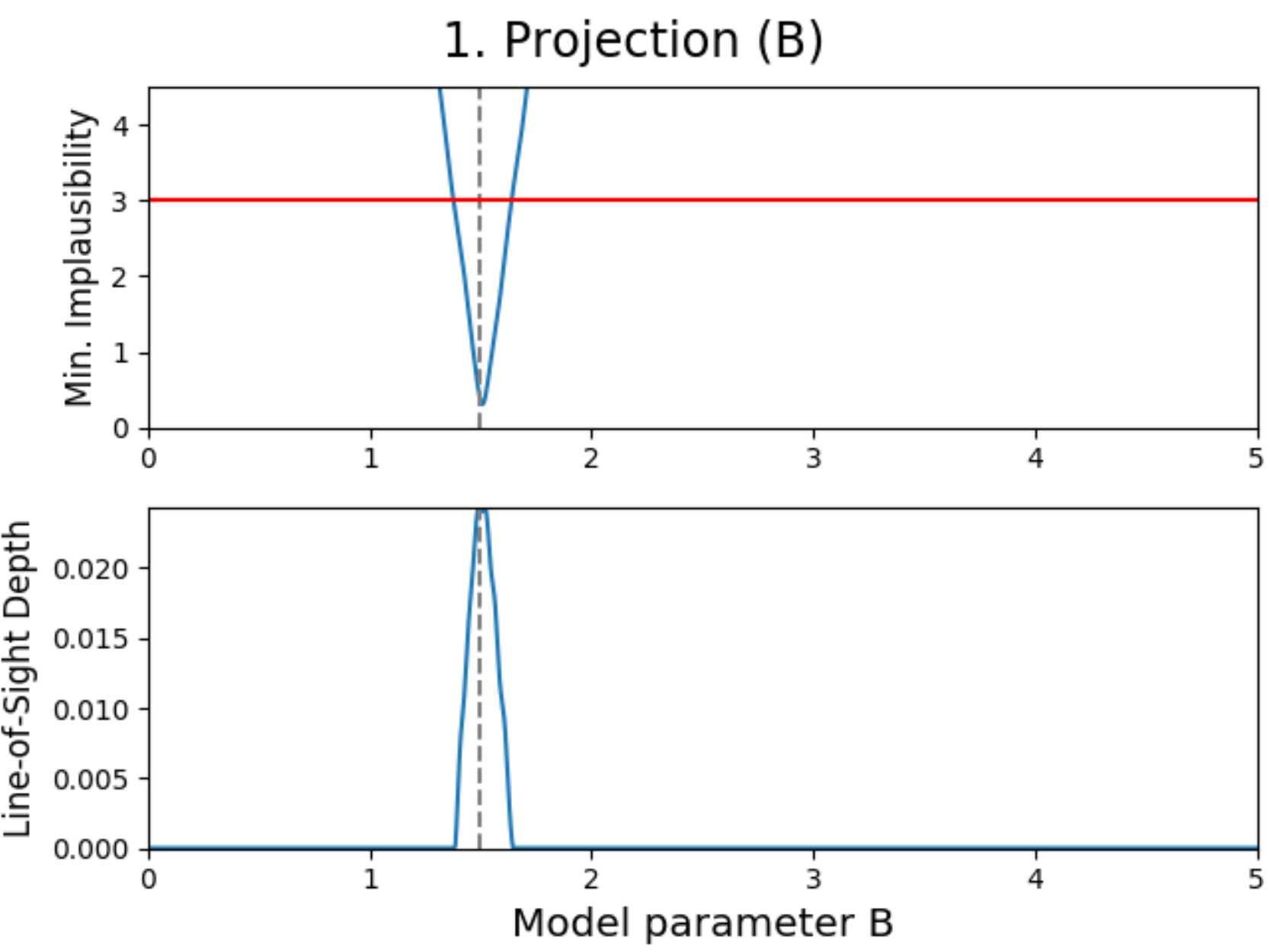}}
	\caption{Projection figures of the emulator of a simple straight line model with more constrained cut-offs.
    \textbf{Top:} The minimum implausibility reached for given parameter value, with the horizontal red line indicating the first non-wildcard implausibility cut-off value.
    \textbf{Bottom:} The fraction of samples that is still plausible for given parameter value.
    The vertical dashed lines indicate the parameter estimates that were given in our definition of the \textsc{LineLink} class.}
    \label{fig:LineLink projections 1b}
\end{center}
\end{figure*}
We can also take a look at the projection figures that were created during the execution of \verb|pipe.run()|, of which two exist according to the details overview shown above.
These figures are shown in \autoref{fig:LineLink projections 1a}.
In these projection figures, we can see that only a very small part of the parameter range for the intercept value ($A$) remains, while the parameter range for the slope value ($B$) is a bit wider.
The minimum implausibility value is the lowest for the estimates that we had provided in the definition of the \textsc{LineLink} class, which corresponded to the values used for generating the data points (excluding added noise).
The line-of-sight plot for both parameters shows that even for those values that have a low minimum implausibility, only a small range of the remaining parameters can yield probable results, which is an indication that both parameters are highly correlated with each other.

When looking at \autoref{fig:LineLink projections 1a}, especially at \ref{subfig:1a_proj_B}, it may look like \prism\ poorly constrained the values of $A$ and $B$, especially as it knows the values of three points on $500$ different parametrizations of a straight line.
The produced regression functions tell a different story however, which can be viewed by checking the determined polynomial coefficients:
\begin{lstlisting}[style=defaultpython]
	|\iin| pipe.emulator.poly_coef[1]
    |\out| [array([0. , 1. , 2.5]),
    		array([ 0.,  1., -2.]),
            array([0., 1., 1.])]
\end{lstlisting}
These values correspond to the coefficients of the constant term, the intercept $A$ and the slope $B$, for the data points $x=\{2.5, -2, 1\}$.
This is exactly the definition of our straight line model $f(x)=A+B\cdot x$.

The reason why the projections look this way, is that \prism\ sets the first implausibility cut-off at $I_{\mathrm{cut}, 2}(\vec{x})=4$ by default for conservative reasons, which means that all samples are included that are expected to be within $4\sigma$ of explaining two out of three data points (since the third is a wildcard).
Given that we are dealing with a simple model here, it is not really necessary to be that conservative.
So, if we now reanalyze the first iteration using $10$ times more samples and $I_{\mathrm{cut}, 1}(\vec{x})=3$, we have $0.100\%$ of parameter space remaining.
Remaking the projections using these new cut-offs gives the figures in \autoref{fig:LineLink projections 1b}.

In these figures, we can see that it is pretty clear what the values of our two model parameters are, especially if one only accepts a minimum implausibility of $1$ or lower (less than $1\sigma$ away from explaining all three data points).
Due to the emulator having perfect accuracy, the data errors (which we set to $0.1$) are dominating in the denominator of \ref{eq:impl_sq} and the emulator cannot be improved anymore.
Using $500$ model evaluations for the emulator was overly conservative and \prism\ could have reached this point with much fewer, but it is generally advised to start with this amount by default.

\section{Picking a prism}
\label{app:Colormaps}
\begin{figure}
\begin{center}
	\subfloat[Jet]{\label{subfig:jet}\includegraphics[width=\linewidth]{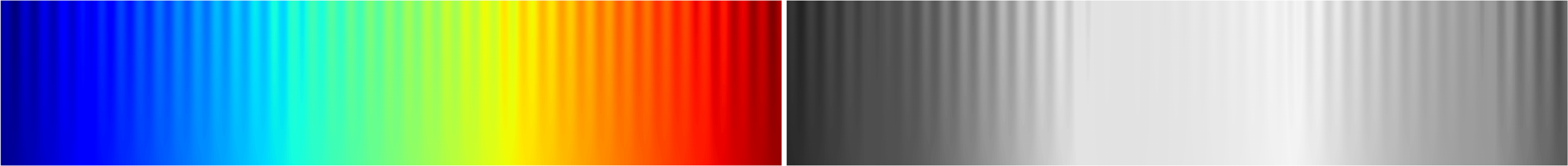}}\\
    \subfloat[Hot]{\label{subfig:hot}\includegraphics[width=\linewidth]{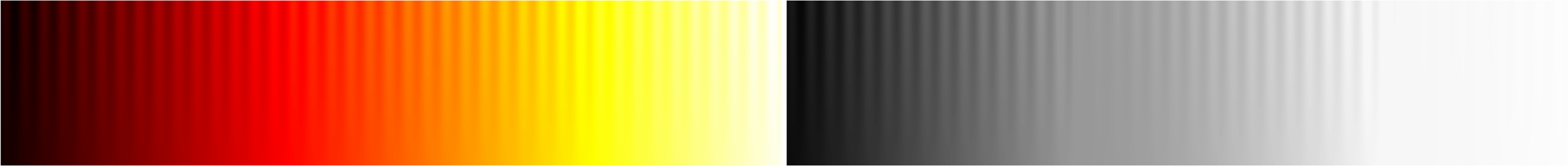}}
	\caption{The \textit{jet} and \textit{hot} colormaps, commonly used in many plotting routines, with color on the left and gray-scale on the right.
    The vertical lines penetrating the colormaps show how well a neighboring color can be distinguished.
    To make a colormap readable and logical, these lines should always be distinct up to roughly the same depth.
    From these images, it is clear that both colormaps, especially \textit{jet}, fail to achieve this.
    Additionally, the \textit{jet} colormap does not monotonically increase in brightness, reaching the brightest point in the center.}
    \label{fig:default_colormaps}
\end{center}
\end{figure}

Since our top priority is to make the emulation technique available to everyone, we had to make sure that this also includes the projections.
There have been many studies on how to properly visualize scientific data, allowing others to interpret them correctly \citep{Rogowitz96,Brychtova16,Szafir18}.
Despite this however, the most commonly used colormaps are the \textit{jet} and \textit{hot} colormaps (see \autoref{fig:default_colormaps}), where \textit{jet} is often the default in many plotting packages.\footnote{Note that some packages are shying away from using this colormap in their latest versions, like \textsw{Matplotlib}.}
Given that we want to ensure that the projections can be understood properly when viewed in gray-scale or by those affected by color vision deficiency (CVD, \citealt{Sharpe99,Birch12}), we had to look for an alternative.

\begin{figure}
\begin{center}
	\subfloat[Viridis]{\includegraphics[width=\linewidth]{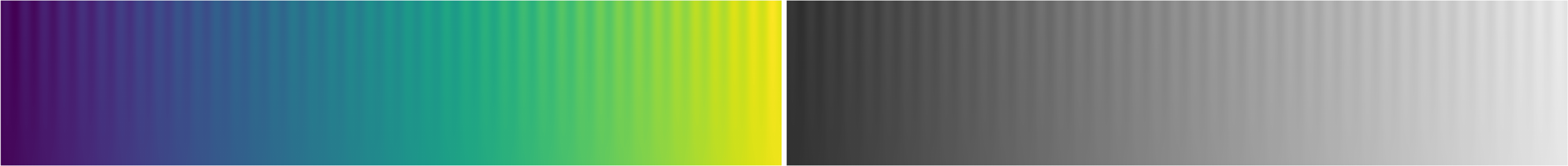}}
	\caption{The \textit{viridis} colormap.
    This colormap solves most of the problems that the \textit{jet} colormap has and \textit{viridis} is used by default now in the newest versions of \textsw{Matplotlib}.
    However, it does have red and green in non-adjacent areas (red at the beginning, green in the middle) which can cause problems for those with CVD.
    Additionally, it only uses three major colors, and we wanted more colors for aesthetic reasons.}
    \label{fig:viridis_colormap}
\end{center}
\end{figure}

A popular alternative colormap that handles most of these problems is the \textit{viridis}\footnote{\url{https://bids.github.io/colormap/}} colormap (see \autoref{fig:viridis_colormap}) made by van der Walt \& Smith (2015) for the \python\ package \textsw{Matplotlib}.
Although this colormap performs well in gray-scale, it can cause problems when viewed by someone with CVD, as demonstrated by \citet{cmaputil}.
Additionally, we felt that \textit{viridis} did not allow for enough fine structure details.
Therefore, we decided to do something similar to the work done by \citet{Kindlmann02}, but instead of using  the six major colors of the rainbow, we wanted to improve the existing ``standards''.

\begin{figure}
\begin{center}
	\subfloat[Rainforest]{\label{subfig:rainforest}\includegraphics[width=\linewidth]{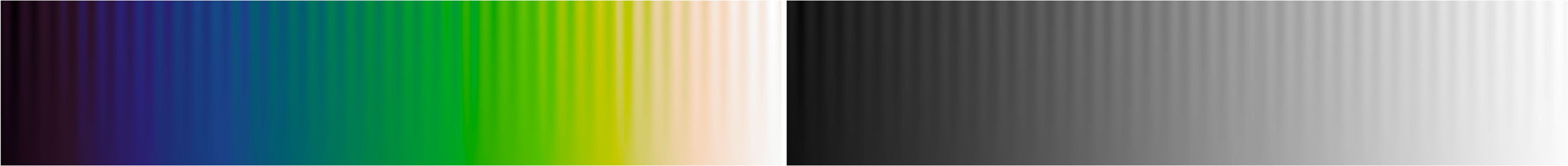}}\\
    \subfloat[Blaze]{\label{subfig:blaze}\includegraphics[width=\linewidth]{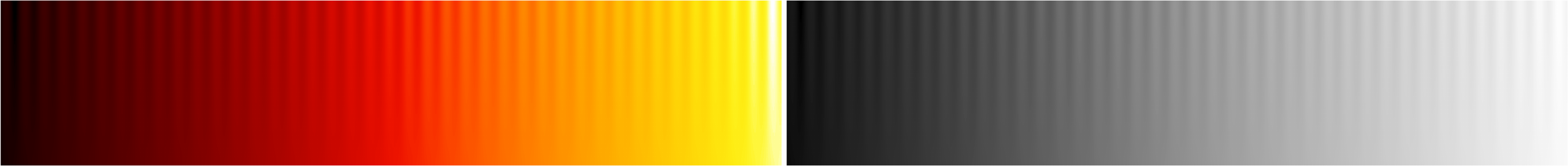}}
	\caption{The improved colormaps \textit{rainforest} (\textit{jet}) and \textit{blaze} (\textit{hot}), which are used by default in \prism.
    As can be seen, these colormaps do not suffer from the same problems mentioned before and are additionally CVD-proof by avoiding the use of red and green in non-adjacent areas (amongst other aspects).}
    \label{fig:new_colormaps}
\end{center}
\end{figure}

By using the \python\ package \textsw{cmaputil} \citep{cmaputil}, we converted the aforementioned \textit{jet} and \textit{hot} colormaps into versions that are CVD-proof and work well in gray-scale (see \autoref{fig:new_colormaps}).
These colormaps, named \textit{rainforest} and \textit{blaze}, are used by default for making \prism's projections.\footnote{Coincidentally, the \textit{rainforest} colormap is very similar to the colormap introduced in Figure 8 of \citet{Kindlmann02}, although the process of obtaining each is completely different. It is also quite similar to the \textit{gist\_earth} colormap in \textsw{Matplotlib}.}
Note that while the colormaps should allow for everybody to interpret the data correctly, those with CVD will see the colormaps differently from those without (unlike the \textit{cividis} colormap introduced by \citealt{cmaputil}).

\section{Figures and Tables}
\label{app:Figures and Tables}
\begin{figure*}
\begin{center}
	\subfloat{\includegraphics[width=0.49\linewidth]{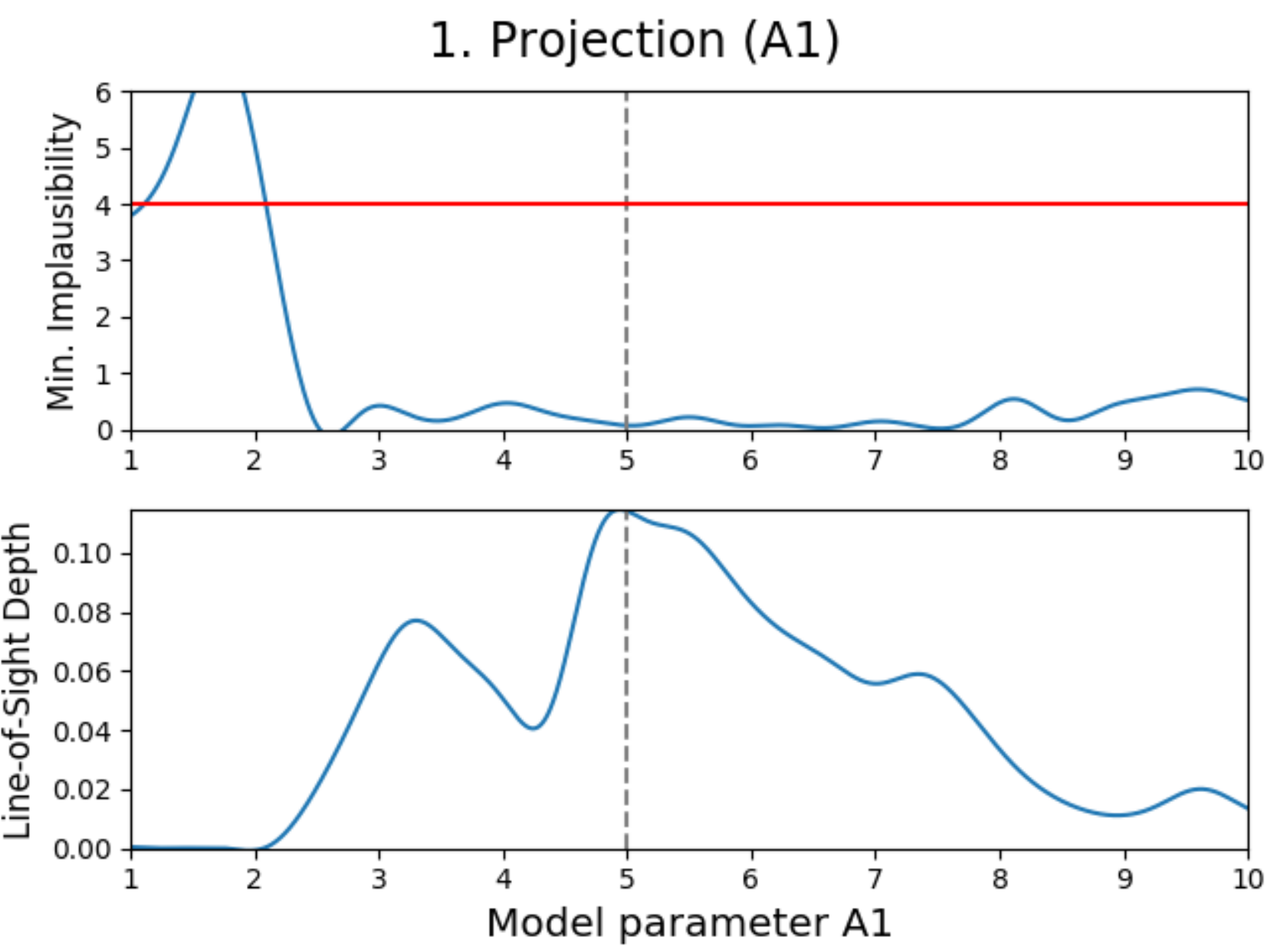}}
    \subfloat{\includegraphics[width=0.49\linewidth]{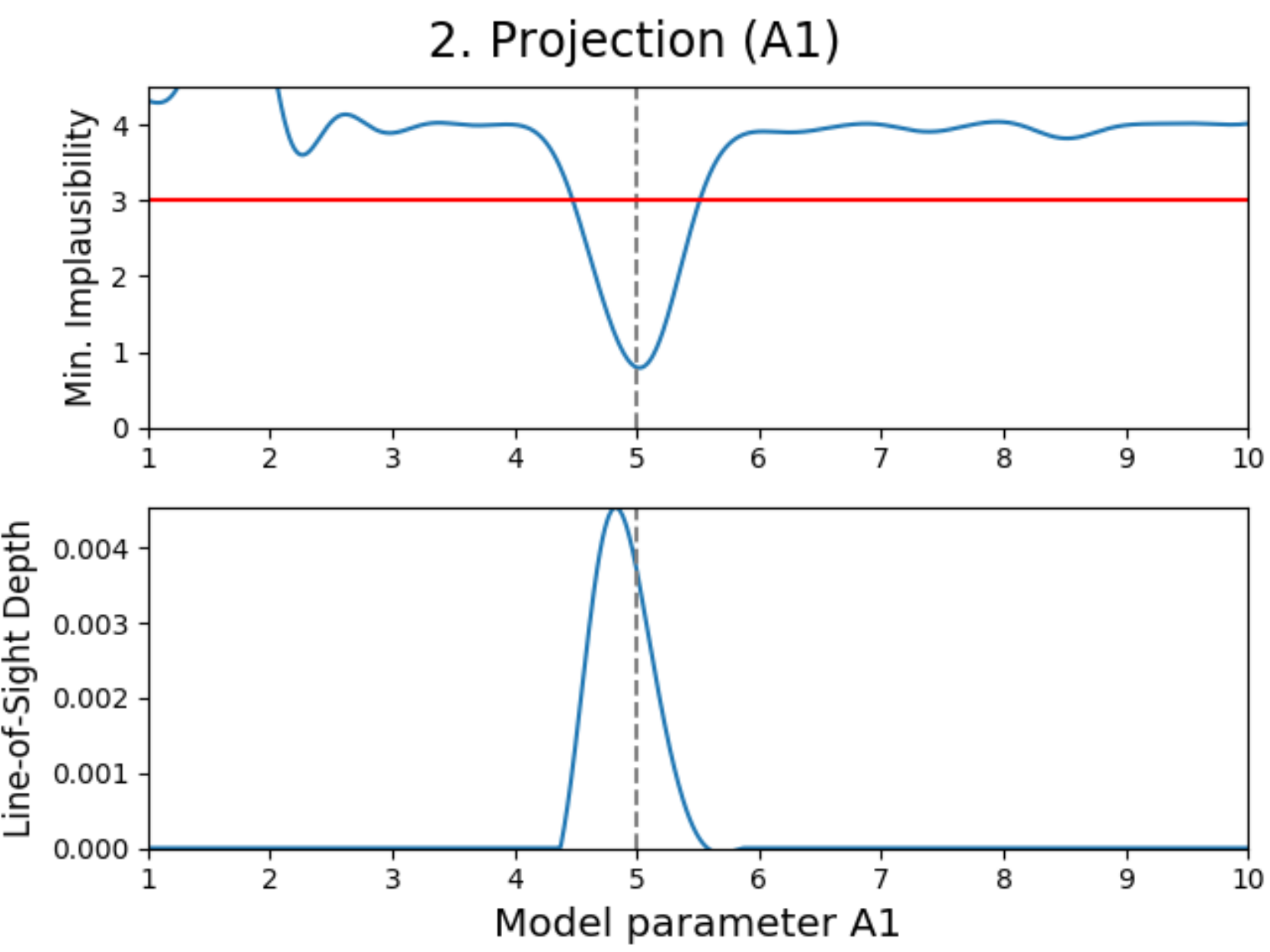}}\\
    \subfloat{\includegraphics[width=0.49\linewidth]{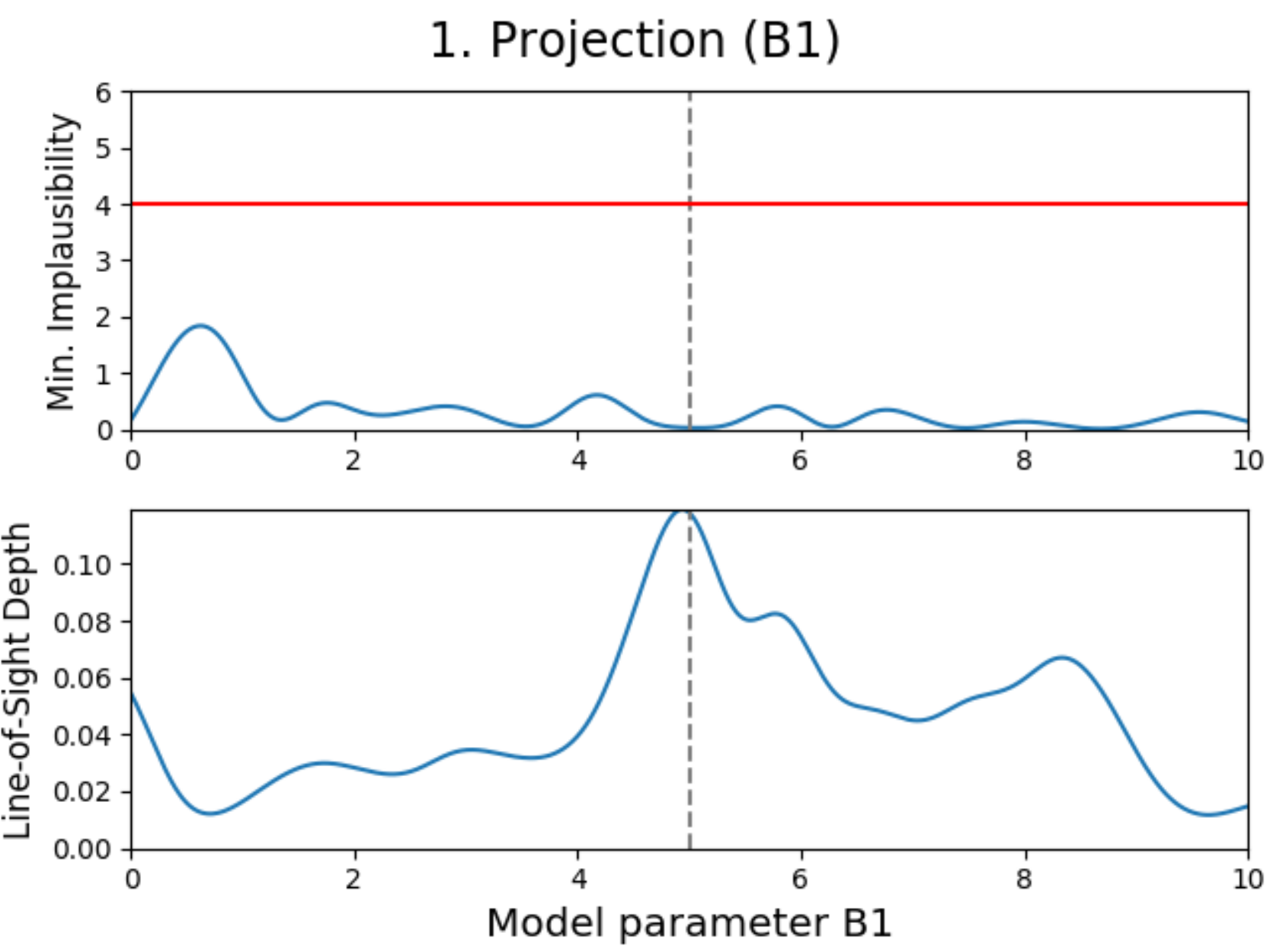}}
    \subfloat{\includegraphics[width=0.49\linewidth]{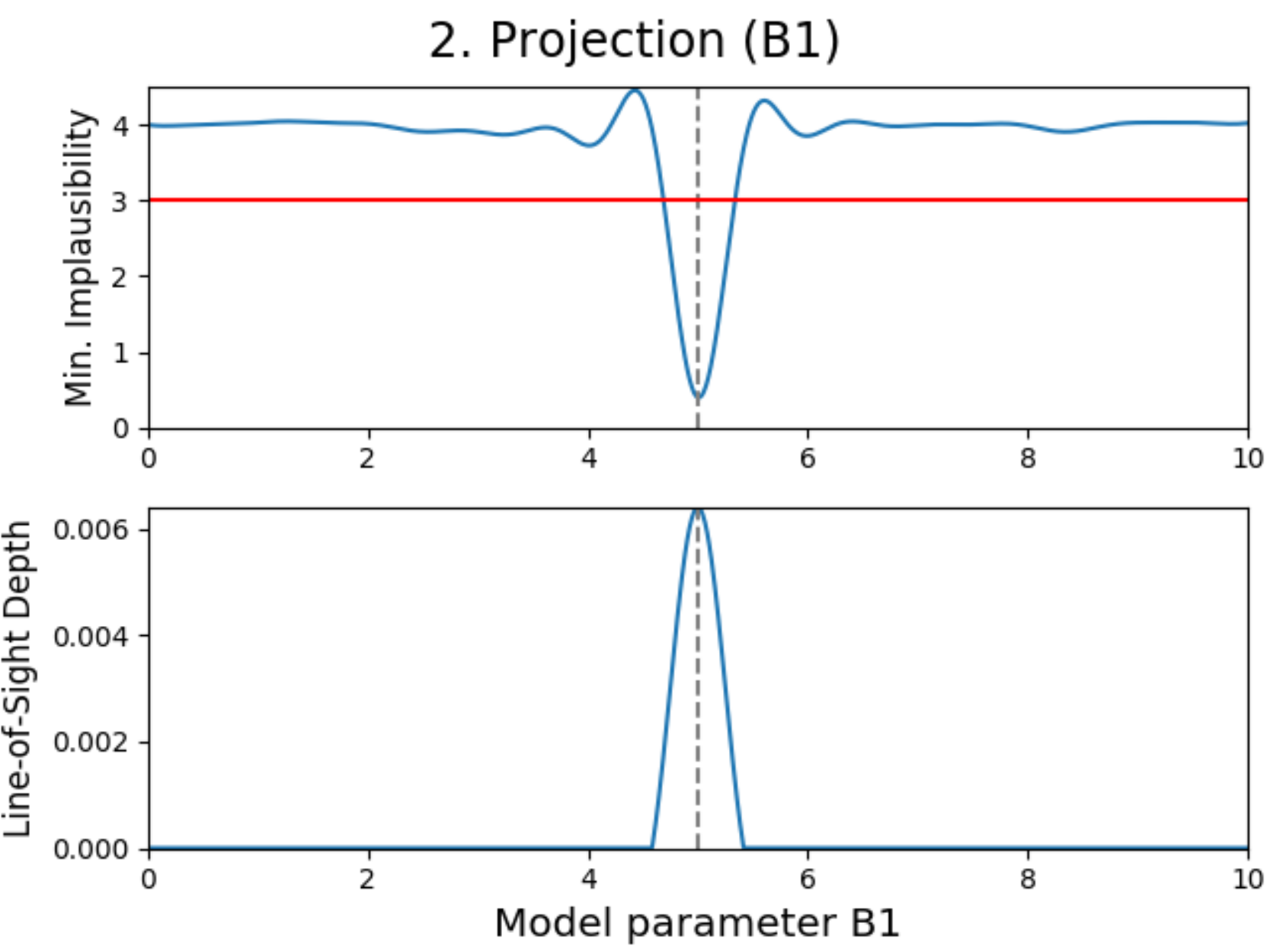}}\\
    \subfloat{\includegraphics[width=0.49\linewidth]{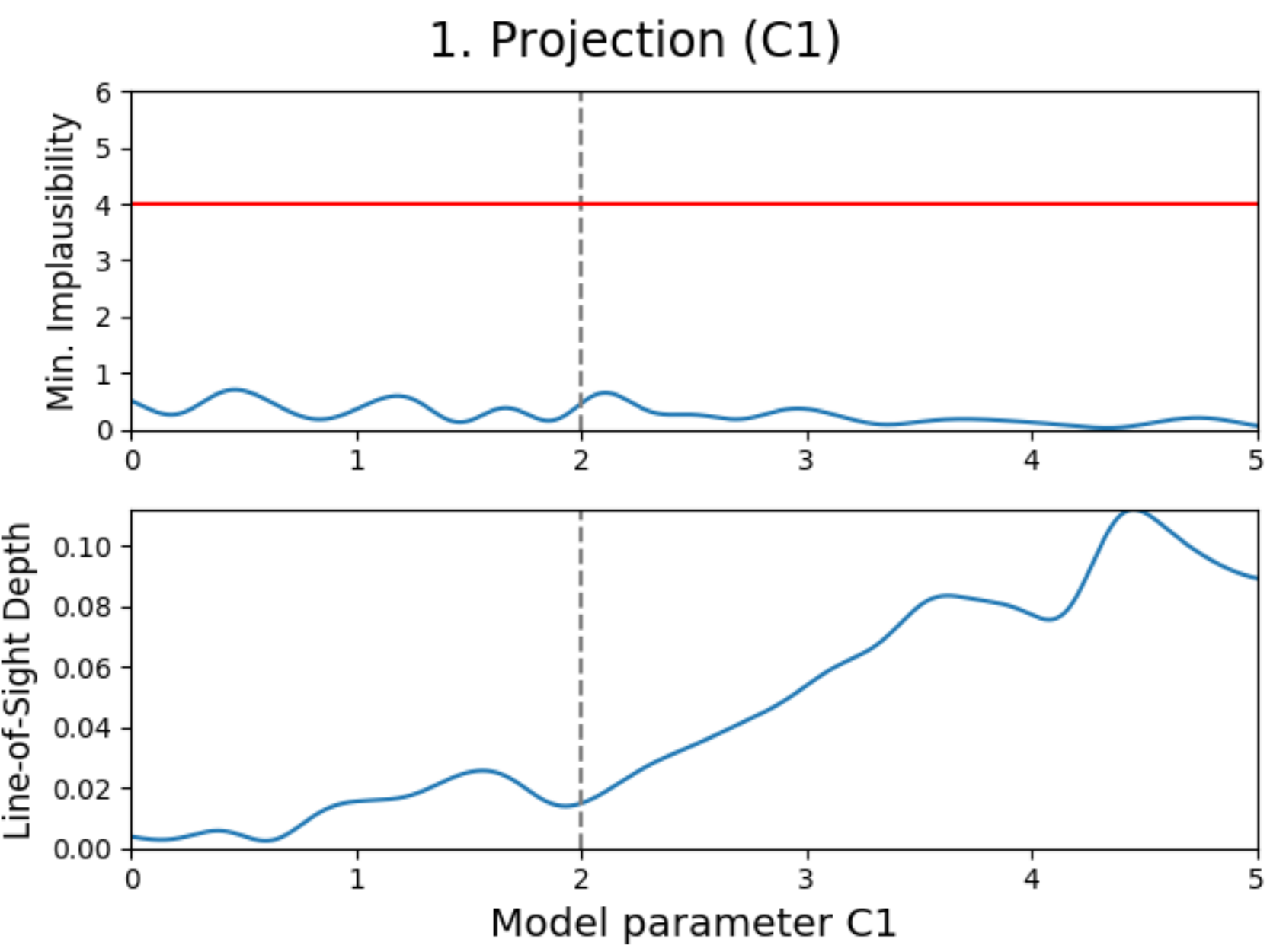}}
    \subfloat{\includegraphics[width=0.49\linewidth]{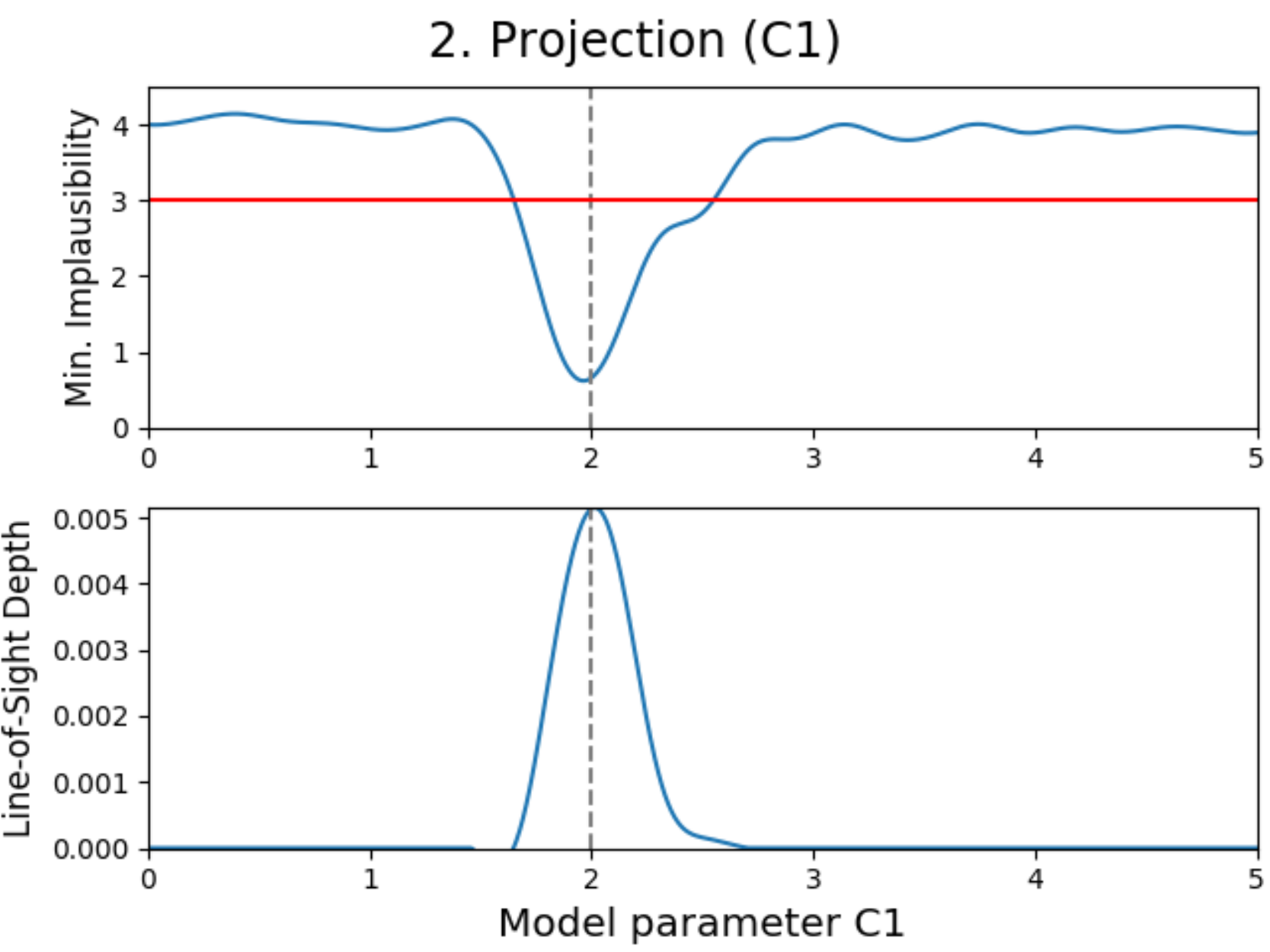}}
	\caption{2D projection figures of the emulator of the \gaussianlink\ class used in \ref{subsec:Minimal example}, where the Gaussian is defined as $f(x)=A_1\cdot\exp\left(-\frac{(x-B_1)^2}{2C_1^2}\right)$.
	These figures are made in the same way as the figures in \autoref{fig:3D_gaussian_projections}, except now only one model parameter is plotted instead of two.
	Whereas the 3D projection figures in \autoref{fig:3D_gaussian_projections} show valuable information for studying model behavior, their 2D variants are useful for parameter estimations.
    \textbf{Left column:} First emulator iteration, $150$ model evaluations, $4.62\%$ of parameter space remaining.
    \textbf{Right column:} Second emulator iteration, $1,110$ model evaluations, $0.0312\%$ of parameter space remaining.
    \textbf{Top subplot:} The minimum implausibility value (at the first cut-off) that can be reached for any given value of the plotted parameter.
    \textbf{Bottom subplot:} The fraction of samples (``line-of-sight depth'') that is plausible for any given value of the plotted parameter.
    \textbf{Gray lines:} Estimates of the plotted parameter, which only show up if the user provided them.
    Note that the first emulator iteration used a wildcard, while the second one did not.}
    \label{fig:2D_gaussian_projections}
\end{center}
\end{figure*}

\begin{table*}
\begin{center}
    \begin{tabular}{V{2.5}r|cV{2.5}c|c|c|cV{2.5}c|c|c|cV{2.5}}
    \clineB{3-10}{2.5}
        \multicolumn{2}{cV{2.5}}{} &
        \multicolumn{4}{cV{2.5}}{Hybrid} &
        \multicolumn{4}{cV{2.5}}{Normal} \\
    \hlineB{2.5}
        Name & Real & $10^2$ & $10^3$ & $10^4$ & $10^5$ & $10^2$ & $10^3$ & $10^4$ & $10^5$ \\
    \hline
    \hline
        $A_1$ & $1.6$ & $1.43^{+1.31}_{-0.35}$ & $1.81^{+0.63}_{-0.45}$ & $1.57^{+1.23}_{-0.37}$ & $1.56^{+0.78}_{-0.34}$ & $2.42^{+1.45}_{-0.78}$ & $1.27^{+0.56}_{-0.20}$ & $1.27^{+5.16}_{-0.19}$ & $1.43^{+0.55}_{-0.28}$ \\
    \hline
        $B_1$ & $-6.25$ & $-5.39^{+0.72}_{-1.40}$ & $-6.16^{+0.99}_{-0.58}$ & $-6.16^{+0.96}_{-0.64}$ & $-6.32^{+0.51}_{-0.44}$ & $-6.40^{+0.99}_{-0.50}$ & $-4.59^{+0.54}_{-0.91}$ & $-4.86^{+2.16}_{-1.28}$ & $-6.19^{+1.05}_{-0.50}$ \\
    \hline
        $C_1$ & $1.4$ & $1.60^{+1.44}_{-0.83}$ & $1.35^{+0.65}_{-0.49}$ & $1.49^{+0.77}_{-0.82}$ & $1.33^{+0.61}_{-0.58}$ & $1.91^{+0.86}_{-0.86}$ & $2.98^{+0.68}_{-0.96}$ & $2.63^{+1.68}_{-2.21}$ & $1.44^{+1.42}_{-0.49}$ \\
    \hline
    \hline
        $A_2$ & $1.2$ & $1.92^{+1.07}_{-0.51}$ & $1.42^{+0.65}_{-0.28}$ & $1.27^{+0.95}_{-0.18}$ & $1.17^{+0.19}_{-0.12}$ & $2.36^{+2.08}_{-0.91}$ & $5.12^{+2.46}_{-3.23}$ & $2.46^{+4.65}_{-1.37}$ & $1.26^{+2.95}_{-0.17}$ \\
    \hline
        $B_2$ & $1.25$ & $1.28^{+0.68}_{-0.74}$ & $0.821^{+0.744}_{-0.538}$ & $1.39^{+0.94}_{-0.93}$ & $1.19^{+0.74}_{-0.99}$ & $0.66^{+1.04}_{-0.94}$ & $1.59^{+0.41}_{-0.20}$ & $1.41^{+0.60}_{-1.93}$ & $1.40^{+0.78}_{-0.87}$ \\
    \hline
        $C_2$ & $3.7$ & $1.79^{+1.74}_{-1.15}$ & $2.59^{+0.79}_{-1.16}$ & $3.27^{+1.42}_{-2.13}$ & $3.91^{+0.80}_{-1.08}$ & $3.03^{+0.56}_{-0.99}$ & $0.768^{+0.594}_{-0.211}$ & $0.808^{+3.22}_{-0.547}$ & $3.68^{+0.88}_{-3.04}$ \\
    \hline
    \hline
        $A_3$ & $4.8$ & $4.07^{+1.70}_{-2.22}$ & $4.20^{+1.33}_{-2.19}$ & $4.31^{+1.32}_{-2.83}$ & $4.40^{+0.92}_{-0.92}$ & $3.44^{+3.59}_{-1.92}$ & $4.26^{+1.28}_{-1.04}$ & $4.62^{+1.07}_{-3.33}$ & $4.95^{+1.42}_{-0.98}$ \\
    \hline
        $B_3$ & $8.75$ & $8.61^{+1.03}_{-2.23}$ & $8.69^{+0.14}_{-1.07}$ & $8.72^{+0.09}_{-1.06}$ & $8.74^{+0.07}_{-0.09}$ & $8.46^{+0.59}_{-0.62}$ & $8.70^{+0.09}_{-0.76}$ & $8.70^{+0.08}_{-1.30}$ & $8.70^{+0.08}_{-0.53}$ \\
    \hline
        $C_3$ & $0.5$ & $0.841^{+1.79}_{-0.337}$ & $0.563^{+2.06}_{-0.079}$ & $0.520^{+2.09}_{-0.067}$ & $0.511^{+0.067}_{-0.059}$ & $1.25^{+1.21}_{-0.53}$ & $0.574^{+0.273}_{-0.076}$ & $0.520^{+1.52}_{-0.077}$ & $0.488^{+0.065}_{-0.084}$ \\
    \hline
    \hline
        $A_4$ & $3.9$ & $3.74^{+1.78}_{-0.93}$ & $3.45^{+0.94}_{-0.64}$ & $4.03^{+1.00}_{-0.78}$ & $4.21^{+0.59}_{-0.62}$ & $3.08^{+2.99}_{-1.46}$ & $4.28^{+1.01}_{-0.85}$ & $3.57^{+0.97}_{-0.74}$ & $3.71^{+0.61}_{-0.69}$ \\
    \hline
        $B_4$ & $16.25$ & $16.2^{+2.0}_{-0.4}$ & $16.2^{+0.2}_{-0.2}$ & $16.2^{+0.2}_{-0.2}$ & $16.2^{+0.2}_{-0.2}$ & $18.5^{+1.1}_{-2.2}$ & $16.3^{+0.2}_{-0.2}$ & $16.2^{+0.2}_{-1.0}$ & $16.3^{+0.2}_{-0.2}$ \\
    \hline
        $C_4$ & $2.0$ & $2.01^{+1.31}_{-0.68}$ & $2.14^{+0.29}_{-0.49}$ & $1.97^{+0.18}_{-0.24}$ & $1.96^{+0.14}_{-0.14}$ & $3.32^{+0.97}_{-1.03}$ & $1.97^{+0.18}_{-0.19}$ & $2.01^{+0.30}_{-0.28}$ & $2.01^{+0.17}_{-0.16}$ \\
    \hlineB{2.5}
        \multicolumn{10}{c}{} \\
    \clineB{2-10}{2.5}
        \multicolumn{1}{cV{2.5}}{} & \multicolumn{1}{rV{2.5}}{$\chi_{\nu}^2$} & $77.7$ & $7.55$ & $0.772$ & $0.369$ & $701$ & $103$ & $62.1$ & $0.962$ \\
    \cline{2-10}
        \multicolumn{1}{cV{2.5}}{} & \multicolumn{1}{rV{2.5}}{$n_{\mathrm{eval}}$} & $2.555\cdot10^3$ & $1.290\cdot10^4$ & $1.031\cdot10^5$ & $1.299\cdot10^6$ & $1.896\cdot10^3$ & $1.917\cdot10^4$ & $1.448\cdot10^5$ & $1.642\cdot10^6$ \\
    \clineB{2-10}{2.5}
    \end{tabular}
    \caption{Overview of the MCMC parameter estimations of the multi-Gaussian described in \ref{subsec:Application}, with $26$ walkers and $1.35\cdot10^{-4}\%$ of parameter space remaining.
    The parameter labeling in the \textbf{first column} corresponds to the four Gaussians shown in \autoref{fig:multi_gaussian_model} and \autoref{fig:multi_gaussian_results} in order, with $A_i$ being its amplitude, $B_i$ its mean and $C_i$ its standard deviation.
    The \textbf{second column} lists the parameter values used to generate the model realization as shown in \autoref{fig:multi_gaussian_model} from which the comparison data was taken.
    All \textbf{remaining columns} show the estimates of all $12$ parameters using hybrid/normal sampling for $10^2$, $10^3$, $10^4$ and $10^5$ MCMC iterations.
    The estimated value is determined by its $0.5$ quantile, with the lower and upper errors being given by the corresponding $0.16$ and $0.84$ quantiles, respectively.
    The errors are rounded to either match the number of significant digits or the number of decimals of the estimated value, whichever comes first.
    For each estimation, the two \textbf{bottom rows} show the corresponding $\chi_{\nu}^2$ and number of required model evaluations $n_{\mathrm{eval}}$.
    Note that all parameter estimations were done from scratch.}
    \label{tab:multi_gaussian_parameters}
\end{center}
\end{table*}
In \autoref{fig:2D_gaussian_projections}, we show the 2D versions of the projection figures shown in \autoref{fig:3D_gaussian_projections}.
In \autoref{tab:multi_gaussian_parameters}, we list the parameter estimation values used for all plots in \autoref{fig:multi_gaussian_results}.


\newpage
\clearpage
\bibliographystyle{aasjournal}
\bibliography{bibliography}

\begin{thebibliography}{}
\expandafter\ifx\csname natexlab\endcsname\relax\def\natexlab#1{#1}\fi
\providecommand{\url}[1]{\href{#1}{#1}}

\bibitem[{Andrianakis {et~al.}(2017)Andrianakis, McCreesh, Vernon, J.~McKinley,
  Oakley, Nsubuga, Goldstein, \& White}]{Andrianakis17}
Andrianakis, I., McCreesh, N., Vernon, I., {et~al.} 2017, SIAM/ASA Journal on
  Uncertainty Quantification, 5, 694

\bibitem[{Andrianakis {et~al.}(2016)Andrianakis, Vernon, McCreesh, McKinley,
  Oakley, Nsubuga, Goldstein, \& White}]{Andrianakis16}
Andrianakis, I., Vernon, I., McCreesh, N., {et~al.} 2016, Journal of the Royal
  Statistical Society: Series C (Applied Statistics), 66, 717

\bibitem[{{Andrianakis} {et~al.}(2015){Andrianakis}, {Vernon}, {McCreesh},
  {McKinley}, {Oakley}, {Nsubuga}, {Goldstein}, \& {White}}]{Andrianakis15}
{Andrianakis}, I., {Vernon}, I.~R., {McCreesh}, N., {et~al.} 2015, PLoS
  Computational Biology, 11, e1003968

\bibitem[{{Betancourt}(2017)}]{Betancourt}
{Betancourt}, M. 2017, ArXiv e-prints, arXiv:1701.02434

\bibitem[{Birch(2012)}]{Birch12}
Birch, J. 2012, J. Opt. Soc. Am. A, 29, 313

\bibitem[{{Bower} {et~al.}(2010){Bower}, {Vernon}, {Goldstein}, {Benson},
  {Lacey}, {Baugh}, {Cole}, \& {Frenk}}]{Bower10}
{Bower}, R.~G., {Vernon}, I., {Goldstein}, M., {et~al.} 2010, \mnras, 407, 2017

\bibitem[{Brooks {et~al.}(2011)Brooks, Gelman, Jones, \&
  Meng}]{brooks2011handbook}
Brooks, S., Gelman, A., Jones, G., \& Meng, X. 2011, {Handbook of Markov Chain
  Monte Carlo}, Chapman \& Hall/CRC Handbooks of Modern Statistical Methods
  (CRC Press)

\bibitem[{Brychtov{\'a} \& \c{C}{\"o}ltekin(2016)}]{Brychtova16}
Brychtov{\'a}, A., \& \c{C}{\"o}ltekin, A. 2016, Cartography and Geographic
  Information Science, 44, 229

\bibitem[{Cawley \& Talbot(2010)}]{Cawley10}
Cawley, G.~C., \& Talbot, N.~L. 2010, J. Mach. Learn. Res., 11, 2079

\bibitem[{Collette(2013)}]{h5py}
Collette, A. 2013, {Python and HDF5: Unlocking scientific data}, 1st edn.
  (O'Reilly Media)

\bibitem[{Craig {et~al.}(1996)Craig, Goldstein, Seheult, \& Smith}]{Craig96}
Craig, P.~S., Goldstein, M., Seheult, A.~H., \& Smith, J.~A. 1996, in Bayesian
  Statistics 5, ed. J.~M. Bernardo, A.~P. Berger, A.~P. Dawid, \& A.~F.~M.
  Smith (Oxford, UK: Clarendon Press), 69--95

\bibitem[{Craig {et~al.}(1997)Craig, Goldstein, Seheult, \& Smith}]{Craig97}
Craig, P.~S., Goldstein, M., Seheult, A.~H., \& Smith, J.~A. 1997, in Case
  Studies in Bayesian Statistics, ed. C.~Gatsonis, J.~S. Hodges, R.~E. Kass,
  R.~McCulloch, P.~Rossi, \& N.~D. Singpurwalla (New York, NY: Springer New
  York), 37--93

\bibitem[{{Croton} {et~al.}(2016){Croton}, {Stevens}, {Tonini}, {Garel},
  {Bernyk}, {Bibiano}, {Hodkinson}, {Mutch}, {Poole}, \& {Shattow}}]{SAGE}
{Croton}, D.~J., {Stevens}, A.~R.~H., {Tonini}, C., {et~al.} 2016, \apjs, 222,
  22

\bibitem[{Currin {et~al.}(1991)Currin, Mitchell, Morris, \&
  Ylvisaker}]{Currin91}
Currin, C., Mitchell, T., Morris, M., \& Ylvisaker, D. 1991, Journal of the
  American Statistical Association, 86, 953

\bibitem[{Dalc\'in {et~al.}(2005)Dalc\'in, Paz, \& Storti}]{mpi4py}
Dalc\'in, L., Paz, R., \& Storti, M. 2005, Journal of Parallel and Distributed
  Computing, 65, 1108

\bibitem[{{De Finetti}(1974)}]{DeFinetti74}
{De Finetti}, B. 1974, Probability and Statistics, Vol.~1, Theory of
  Probability (Wiley)

\bibitem[{{De Finetti}(1975)}]{DeFinetti75}
---. 1975, Probability and Statistics, Vol.~2, Theory of Probability (Wiley)

\bibitem[{{Foreman-Mackey} {et~al.}(2013){Foreman-Mackey}, {Hogg}, {Lang}, \&
  {Goodman}}]{emcee}
{Foreman-Mackey}, D., {Hogg}, D.~W., {Lang}, D., \& {Goodman}, J. 2013, \pasp,
  125, 306

\bibitem[{Gelman {et~al.}(2014)Gelman, Carlin, Stern, Dunson, Vehtari, \&
  Rubin}]{gelman2014bayesian}
Gelman, A., Carlin, J.~B., Stern, H.~S., {et~al.} 2014, {Bayesian data
  analysis}, 3rd edn. (Taylor \& Francis Group, LLC)

\bibitem[{Geman \& Geman(1984)}]{Gibbs}
Geman, S., \& Geman, D. 1984, IEEE Transactions on Pattern Analysis and Machine
  Intelligence, PAMI-6, 721

\bibitem[{Goldstein(1999)}]{Goldstein99}
Goldstein, M. 1999, in Encyclopedia of Statistical Sciences (Wiley), 29--34

\bibitem[{Goldstein(2006)}]{Goldstein06}
Goldstein, M. 2006, Bayesian Anal., 1, 403

\bibitem[{Goldstein \& Rougier(2006)}]{Goldsteinetal06}
Goldstein, M., \& Rougier, J. 2006, Journal of the American Statistical
  Association, 101, 1132

\bibitem[{Goldstein \& Wilkinson(2000)}]{Goldstein00}
Goldstein, M., \& Wilkinson, D.~J. 2000, Statistics and Computing, 10, 311

\bibitem[{Goldstein \& Wooff(2007)}]{BLA}
Goldstein, M., \& Wooff, D. 2007, {Bayes Linear Statistics: Theory and
  Methods}, 1st edn. (John Wiley \& Sons Ltd.)

\bibitem[{{Goodman} \& {Weare}(2010)}]{affine_invariant}
{Goodman}, J., \& {Weare}, J. 2010, Communications in Applied Mathematics and
  Computational Science, 5, 65

\bibitem[{Hastings(1970)}]{Hastings}
Hastings, W.~K. 1970, Biometrika, 57, 97

\bibitem[{{Hoffman} \& {Gelman}(2011)}]{NUTS}
{Hoffman}, M.~D., \& {Gelman}, A. 2011, ArXiv e-prints, arXiv:1111.4246

\bibitem[{Iman \& Conover(1982)}]{IC82}
Iman, R.~L., \& Conover, W.~J. 1982, Communications in Statistics - Simulation
  and Computation, 11, 311

\bibitem[{{Jaffe} {et~al.}(2010){Jaffe}, {Leahy}, {Banday}, {Leach}, {Lowe}, \&
  {Wilkinson}}]{Jaffe10}
{Jaffe}, T.~R., {Leahy}, J.~P., {Banday}, A.~J., {et~al.} 2010, \mnras, 401,
  1013

\bibitem[{{Jaffe} {et~al.}(2013){Jaffe}, {Ferri{\`e}re}, {Banday}, {Strong},
  {Orlando}, {Mac{\'{\i}}as-P{\'e}rez}, {Fauvet}, {Combet}, \&
  {Falgarone}}]{Jaffe13}
{Jaffe}, T.~R., {Ferri{\`e}re}, K.~M., {Banday}, A.~J., {et~al.} 2013, \mnras,
  431, 683

\bibitem[{{Jansson} \& {Farrar}(2012{\natexlab{a}})}]{JF12a}
{Jansson}, R., \& {Farrar}, G.~R. 2012{\natexlab{a}}, \apj, 757, 14

\bibitem[{{Jansson} \& {Farrar}(2012{\natexlab{b}})}]{JF12b}
---. 2012{\natexlab{b}}, \apjl, 761, L11

\bibitem[{Johnson {et~al.}(1990)Johnson, Moore, \& Ylvisaker}]{JMY90}
Johnson, M., Moore, L., \& Ylvisaker, D. 1990, Journal of Statistical Planning
  and Inference, 26, 131

\bibitem[{Joseph \& Hung(2008)}]{JH08}
Joseph, V.~R., \& Hung, Y. 2008, Statistica Sinica, 18, 171

\bibitem[{Kennedy \& O'Hagan(2001)}]{Kennedy01}
Kennedy, M.~C., \& O'Hagan, A. 2001, Journal of the Royal Statistical Society:
  Series B (Statistical Methodology), 63, 425

\bibitem[{Kindlmann {et~al.}(2002)Kindlmann, Reinhard, \& Creem}]{Kindlmann02}
Kindlmann, G., Reinhard, E., \& Creem, S. 2002, in Proceedings of the
  Conference on Visualization '02, VIS '02 (Washington, DC, USA: IEEE Computer
  Society), 299--306

\bibitem[{{Lagos} {et~al.}(2018){Lagos}, {Tobar}, {Robotham}, {Obreschkow},
  {Mitchell}, {Power}, \& {Elahi}}]{Shark}
{Lagos}, C.~d.~P., {Tobar}, R.~J., {Robotham}, A.~S.~G., {et~al.} 2018, \mnras,
  481, 3573

\bibitem[{McKay {et~al.}(1979)McKay, Beckman, \& Conover}]{LHS}
McKay, M.~D., Beckman, R.~J., \& Conover, W.~J. 1979, Technometrics, 21, 239

\bibitem[{{Message Passing Interface Forum}(1994)}]{mpi-1}
{Message Passing Interface Forum}. 1994, International Journal of Supercomputer
  Applications, 8, 159

\bibitem[{{Message Passing Interface Forum}(1998)}]{mpi-2}
---. 1998, High Performance Computing Applications, 12, 1

\bibitem[{{Metropolis} {et~al.}(1953){Metropolis}, {Rosenbluth}, {Rosenbluth},
  {Teller}, \& {Teller}}]{Metropolis}
{Metropolis}, N., {Rosenbluth}, A.~W., {Rosenbluth}, M.~N., {Teller}, A.~H., \&
  {Teller}, E. 1953, \jcp, 21, 1087

\bibitem[{Morris \& Mitchell(1995)}]{MM95}
Morris, M.~D., \& Mitchell, T.~J. 1995, Journal of Statistical Planning and
  Inference, 43, 381

\bibitem[{{Mutch} {et~al.}(2016){Mutch}, {Geil}, {Poole}, {Angel}, {Duffy},
  {Mesinger}, \& {Wyithe}}]{Meraxes}
{Mutch}, S.~J., {Geil}, P.~M., {Poole}, G.~B., {et~al.} 2016, \mnras, 462, 250

\bibitem[{{Nu{\~n}ez} {et~al.}(2018){Nu{\~n}ez}, {Anderton}, \&
  {Renslow}}]{cmaputil}
{Nu{\~n}ez}, J.~R., {Anderton}, C.~R., \& {Renslow}, R.~S. 2018, PLoS ONE, 13,
  e0199239

\bibitem[{Oakley \& O'Hagan(2002)}]{Oakley02}
Oakley, J., \& O'Hagan, A. 2002, Biometrika, 89, 769

\bibitem[{O'Hagan(2006)}]{O'Hagan06}
O'Hagan, A. 2006, Reliability Engineering \& System Safety, 91, 1290

\bibitem[{Oliphant(2006)}]{NumPy}
Oliphant, T. 2006, {NumPy}: A guide to {NumPy},  USA: Trelgol Publishing.
\newblock \url{http://www.numpy.org/}

\bibitem[{Owen(1994)}]{Owen94}
Owen, A.~B. 1994, Journal of the American Statistical Association, 89, 1517

\bibitem[{Pedregosa {et~al.}(2011)Pedregosa, Varoquaux, Gramfort, Michel,
  Thirion, Grisel, Blondel, Prettenhofer, Weiss, Dubourg, Vanderplas, Passos,
  Cournapeau, Brucher, Perrot, \& Duchesnay}]{Sklearn}
Pedregosa, F., Varoquaux, G., Gramfort, A., {et~al.} 2011, Journal of Machine
  Learning Research, 12, 2825

\bibitem[{{Pshirkov} {et~al.}(2011){Pshirkov}, {Tinyakov}, {Kronberg}, \&
  {Newton-McGee}}]{Pshirkov11}
{Pshirkov}, M.~S., {Tinyakov}, P.~G., {Kronberg}, P.~P., \& {Newton-McGee},
  K.~J. 2011, \apj, 738, 192

\bibitem[{Pukelsheim(1994)}]{Pukelsheim94}
Pukelsheim, F. 1994, The American Statistician, 48, 88

\bibitem[{Raftery {et~al.}(1995)Raftery, Givens, \& Zeh}]{Raftery95}
Raftery, A.~E., Givens, G.~H., \& Zeh, J.~E. 1995, Journal of the American
  Statistical Association, 90, 402

\bibitem[{Raschka(2018)}]{Mlxtend}
Raschka, S. 2018, JOSS, 3, doi:10.21105/joss.00638

\bibitem[{{Rodrigues} {et~al.}(2017){Rodrigues}, {Vernon}, \&
  {Bower}}]{Rodriques17}
{Rodrigues}, L.~F.~S., {Vernon}, I., \& {Bower}, R.~G. 2017, \mnras, 466, 2418

\bibitem[{Rogowitz {et~al.}(1996)Rogowitz, Treinish, \& Bryson}]{Rogowitz96}
Rogowitz, B.~E., Treinish, L.~A., \& Bryson, S. 1996, Computers in Physics, 10,
  268

\bibitem[{Sacks {et~al.}(1989)Sacks, Welch, Mitchell, \& Wynn}]{Sacks89}
Sacks, J., Welch, W.~J., Mitchell, T.~J., \& Wynn, H.~P. 1989, Statistical
  Science, 4, 409

\bibitem[{Sharpe {et~al.}(1999)Sharpe, Stockman, Jaegle, \& Nathans}]{Sharpe99}
Sharpe, L.~T., Stockman, A., Jaegle, H., \& Nathans, J. 1999, in Color Vision:
  From Genes to Perception (Cambridge University Press), 3--51

\bibitem[{{Sivia} \& {Skilling}(2012)}]{BayesianBook}
{Sivia}, D.~S., \& {Skilling}, J. 2012, Data Analysis: A Bayesian Tutorial, 2nd
  edn. (Oxford Science Publications)

\bibitem[{Skilling(2006)}]{Skilling}
Skilling, J. 2006, Bayesian Anal., 1, 833.
\newblock \url{https://doi.org/10.1214/06-BA127}

\bibitem[{{Steininger} {et~al.}(2018){Steininger}, {En{\ss}lin}, {Greiner},
  {Jaffe}, {van der Velden}, {Wang}, {Haverkorn}, {H{\"o}randel}, {Jasche}, \&
  {Rachen}}]{IMAGINE}
{Steininger}, T., {En{\ss}lin}, T.~A., {Greiner}, M., {et~al.} 2018, ArXiv
  e-prints, arXiv:1801.04341

\bibitem[{Stone(1974)}]{Stone74}
Stone, M. 1974, Journal of the Royal Statistical Society. Series B
  (Methodological), 36, 111

\bibitem[{{Sun} {et~al.}(2008){Sun}, {Reich}, {Waelkens}, \&
  {En{\ss}lin}}]{Sun08}
{Sun}, X.~H., {Reich}, W., {Waelkens}, A., \& {En{\ss}lin}, T.~A. 2008, \aap,
  477, 573

\bibitem[{Szafir(2018)}]{Szafir18}
Szafir, D.~A. 2018, IEEE Transactions on Visualization and Computer Graphics,
  24, 392

\bibitem[{Tang(1998)}]{Tang98}
Tang, B. 1998, Statistica Sinica, 8, 965

\bibitem[{{Terral} \& {Ferri{\`e}re}(2017)}]{Terral16}
{Terral}, P., \& {Ferri{\`e}re}, K. 2017, \aap, 600, A29

\bibitem[{{Unger} \& {Farrar}(2017)}]{UngerFarrar17}
{Unger}, M., \& {Farrar}, G.~R. 2017, International Cosmic Ray Conference, 301,
  558

\bibitem[{{van der Velden}(2019)}]{PRISM_JOSS}
{van der Velden}, E. 2019, JOSS, 4, doi:10.21105/joss.01229

\bibitem[{{Van Eck} {et~al.}(2011){Van Eck}, {Brown}, {Stil}, {Rae}, {Mao},
  {Gaensler}, {Shukurov}, {Taylor}, {Haverkorn}, {Kronberg}, \&
  {McClure-Griffiths}}]{VanEck11}
{Van Eck}, C.~L., {Brown}, J.~C., {Stil}, J.~M., {et~al.} 2011, \apj, 728, 97

\bibitem[{{Vernon} {et~al.}(2010){Vernon}, {Goldstein}, \& {Bower}}]{Vernon10}
{Vernon}, I., {Goldstein}, M., \& {Bower}, R.~G. 2010, Bayesian Anal., 5, 619

\bibitem[{{Vernon} {et~al.}(2014){Vernon}, {Goldstein}, \& {Bower}}]{Vernon14}
---. 2014, Statist. Sci., 29, 81.
\newblock \url{https://doi.org/10.1214/12-STS412}

\bibitem[{Vernon {et~al.}(2018)Vernon, Liu, Goldstein, Rowe, Topping, \&
  Lindsey}]{Vernon18}
Vernon, I., Liu, J., Goldstein, M., {et~al.} 2018, BMC Systems Biology, 12, 1

\end{thebibliography}


\end{document}